\documentclass[12pt]{article}
\usepackage{enumitem}
\def\ArxivCombined{1}

\PassOptionsToPackage{unicode}{hyperref}
\PassOptionsToPackage{hyphens}{url}
\PassOptionsToPackage{dvipsnames,svgnames,x11names}{xcolor}
\ifdefined\ArxivCombined\else
\documentclass[
  12pt]{article}
\fi

\usepackage[letterpaper,margin=1in]{geometry}
\usepackage{setspace}
\usepackage{amsmath,amssymb}
\usepackage{iftex}
\ifPDFTeX
  \usepackage[T1]{fontenc}
  \usepackage[utf8]{inputenc}
  \usepackage{textcomp} 
\else 
  \usepackage{unicode-math}
  \defaultfontfeatures{Scale=MatchLowercase}
  \defaultfontfeatures[\rmfamily]{Ligatures=TeX,Scale=1}
\fi
\usepackage{lmodern}
\ifPDFTeX\else  
\fi
\IfFileExists{upquote.sty}{\usepackage{upquote}}{}
\IfFileExists{microtype.sty}{
  \usepackage[]{microtype}
  \UseMicrotypeSet[protrusion]{basicmath} 
}{}
\makeatletter
\@ifundefined{KOMAClassName}{
  \IfFileExists{parskip.sty}{%
    \usepackage{parskip}
  }{
    \setlength{\parindent}{0pt}
    \setlength{\parskip}{6pt plus 2pt minus 1pt}}
}{
  \KOMAoptions{parskip=half}}
\makeatother
\makeatletter
\ifx\paragraph\undefined\else
  \let\oldparagraph\paragraph
  \renewcommand{\paragraph}{
    \@ifstar
      \xxxParagraphStar
      \xxxParagraphNoStar
  }
  \newcommand{\xxxParagraphStar}[1]{\oldparagraph*{#1}\mbox{}}
  \newcommand{\xxxParagraphNoStar}[1]{\oldparagraph{#1}\mbox{}}
\fi
\ifx\subparagraph\undefined\else
  \let\oldsubparagraph\subparagraph
  \renewcommand{\subparagraph}{
    \@ifstar
      \xxxSubParagraphStar
      \xxxSubParagraphNoStar
  }
  \newcommand{\xxxSubParagraphStar}[1]{\oldsubparagraph*{#1}\mbox{}}
  \newcommand{\xxxSubParagraphNoStar}[1]{\oldsubparagraph{#1}\mbox{}}
\fi
\makeatother

\usepackage{longtable,booktabs,array}
\usepackage{calc} 
\usepackage{etoolbox}
\makeatletter
\patchcmd\longtable{\par}{\if@noskipsec\mbox{}\fi\par}{}{}
\makeatother
\IfFileExists{footnotehyper.sty}{\usepackage{footnotehyper}}{\usepackage{footnote}}
\makesavenoteenv{longtable}

\makeatletter
\def\maxwidth{\ifdim\Gin@nat@width>\linewidth\linewidth\else\Gin@nat@width\fi}
\def\maxheight{\ifdim\Gin@nat@height>\textheight\textheight\else\Gin@nat@height\fi}
\usepackage{algorithmicx}%
\usepackage{algpseudocode}%
\makeatother
\usepackage{graphicx}

\makeatletter
\def\maxwidth{%
  \ifdim\Gin@nat@width>\linewidth
    \linewidth
  \else
    \Gin@nat@width
  \fi
}
\def\maxheight{%
  \ifdim\Gin@nat@height>\textheight
    \textheight
  \else
    \Gin@nat@height
  \fi
}
\makeatother

\setkeys{Gin}{
  width=\maxwidth,
  height=\maxheight,
  keepaspectratio
}

\usepackage{algorithmicx}
\usepackage{algpseudocode}
\makeatletter
\def\fps@figure{htbp}
\makeatother

\makeatletter
\@ifpackageloaded{caption}{}{\usepackage{caption}}
\AtBeginDocument{%
\ifdefined\contentsname
  \renewcommand*\contentsname{Table of contents}
\else
  \newcommand\contentsname{Table of contents}
\fi
\ifdefined\listfigurename
  \renewcommand*\listfigurename{List of Figures}
\else
  \newcommand\listfigurename{List of Figures}
\fi
\ifdefined\listtablename
  \renewcommand*\listtablename{List of Tables}
\else
  \newcommand\listtablename{List of Tables}
\fi
\ifdefined\figurename
  \renewcommand*\figurename{Figure}
\else
  \newcommand\figurename{Figure}
\fi
\ifdefined\tablename
  \renewcommand*\tablename{Table}
\else
  \newcommand\tablename{Table}
\fi
}
\@ifpackageloaded{float}{}{\usepackage{float}}
\floatstyle{ruled}
\@ifundefined{c@chapter}{\newfloat{codelisting}{h}{lop}}{\newfloat{codelisting}{h}{lop}[chapter]}
\floatname{codelisting}{Listing}

\makeatother
\makeatletter
\@ifpackageloaded{caption}{}{\usepackage{caption}}
\@ifpackageloaded{subcaption}{}{\usepackage{subcaption}}
\makeatother

\usepackage{multirow}%
\usepackage{amsmath,amssymb,amsfonts}%
\usepackage{amsthm}%
\usepackage{eucal}%
\usepackage[title]{appendix}%

\usepackage{textcomp}%
\usepackage{manyfoot}%
\usepackage{booktabs}%
\usepackage{algorithm}%
\usepackage{algorithmicx}%
\usepackage{algpseudocode}%
\usepackage{listings}%
\usepackage{caption}
\usepackage{subcaption}%
\usepackage{tikz}%

\usepackage{graphicx}
\usepackage{xcolor}
\usepackage{tikz}
\usetikzlibrary{decorations.pathreplacing}
\tikzset{
  itemdot/.style={circle,draw=black!35,line width=.3pt,inner sep=0pt},
  paneltitle/.style={font=\scriptsize\bfseries},
  matrixcell/.style={draw=black!60,line width=.35pt},
  setline/.style={font=\scriptsize},
  note/.style={font=\scriptsize,align=center,draw=black!25,dashed,
               rounded corners=1pt,inner sep=3pt}
}
\usepackage{array}%
\usepackage{tabularx}%

\newcolumntype{Y}{>{\raggedright\arraybackslash}X}
\newcommand{\ind}{\mathbf{1}}
\newcommand{\Ga}{\mathrm{Gamma}}
\newcommand{\GN}{\mathrm{GN}}
\newcommand{\suppsectionref}[2]{%
  \ifdefined\ArxivCombined Section~\ref{#1}\else Section~S#2\fi}
\newcommand{\suppsectionsref}[4]{%
  \ifdefined\ArxivCombined Sections~\ref{#1}--\ref{#2}\else Sections~S#3--S#4\fi}

\newcommand{\PctExactKept}{43.9}

\newcommand{\ChatSalso}{19}
\newcommand{\KhatSalso}{259}
\newcommand{\CmodePost}{19}
\newcommand{\KmodePost}{259}
\newcommand{\PctCMode}{50.4}
\newcommand{\PctKMode}{87.2}

\newcommand{\EssK}{245}

\newcommand{\SmXAccept}{0.00505}

\providecommand{\ChatSalso}{19}
\providecommand{\KhatSalso}{259}
\providecommand{\CmodePost}{19}
\providecommand{\KmodePost}{259}
\providecommand{\PctCMode}{50.5}
\providecommand{\PctKMode}{87.2}

\providecommand{\EssK}{245}

\providecommand{\SmXAccept}{0.00505}

\providecommand{\PctExactKept}{43.9}

\tikzset{
  card/.style={
    draw=black!55,
    line width=.45pt,
    rounded corners=1.5pt
  },
  panel/.style={
    draw=black!70,
    line width=.55pt,
    rounded corners=1.5pt
  },
  paneltitle/.style={
    font=\bfseries\scriptsize
  },
  itemdot/.style={
    circle,
    draw=black!45,
    line width=.35pt,
    minimum size=3.8mm,
    inner sep=0pt
  },
  note/.style={
    draw=black!45,
    dashed,
    rounded corners=1.5pt,
    inner sep=3pt,
    font=\scriptsize,
    align=center
  },
  setline/.style={
    font=\scriptsize,
    align=left
  },
  matrixcell/.style={
    draw=black!60,
    line width=.4pt
  }
}
\theoremstyle{plain}

\theoremstyle{definition}

\theoremstyle{remark}

\ifLuaTeX
  \usepackage{selnolig}  
\fi
\usepackage[]{natbib}
\usepackage{bookmark}

\IfFileExists{xurl.sty}{\usepackage{xurl}}{} 
\hypersetup{
  pdftitle={Bayesian Plackett--Luce latent block models for ranked data},
  pdfauthor={Lapo Santi; Nial Friel; Valeria Vitelli},
  pdfkeywords={Bayesian nonparametrics; co-clustering; gene expression; Gnedin process; Markov chain Monte Carlo},
  colorlinks=true,
  linkcolor={blue},
  filecolor={Maroon},
  citecolor={Blue},
  urlcolor={Blue},
  pdfcreator={LaTeX via pandoc}}

\newcommand{\anon}{1}

\begin{document}

\def\spacingset#1{\renewcommand{\baselinestretch}%
{#1}\small\normalsize} \spacingset{1}

\if1\anon
{
  \title{\bf Bayesian Plackett--Luce latent block models for ranked data}
  \author{
    Lapo Santi\thanks{
      School of Mathematics and Statistics, University College Dublin,
      Dublin, Ireland.
      Corresponding author: \texttt{lapo.santi@ucdconnect.ie}.
      ORCID: 0009-0005-9363-3353}
    \quad Nial Friel\thanks{
      School of Mathematics and Statistics, University College Dublin,
      Dublin, Ireland; Rinn Artificial Intelligence.
      Email: \texttt{nial.friel@ucd.ie}.
      ORCID: 0000-0003-4778-0254}
    \quad Valeria Vitelli\thanks{
      Oslo Centre for Biostatistics and Epidemiology, University of Oslo,
      Oslo, Norway.
      Email: \texttt{valeria.vitelli@medisin.uio.no}.
      ORCID: 0000-0002-6746-0453}
  }
  \date{}
  \maketitle
} \fi

\if0\anon
{
  \bigskip
  \bigskip
  \bigskip
  \begin{center}
    {\LARGE\bf Bayesian Plackett--Luce latent block models for ranked data}
  \end{center}
  \medskip
} \fi
\bigskip
\begin{abstract}
We introduce a Bayesian latent block model that jointly partitions assessors and items under a Plackett--Luce observation model. Assessors are assigned to $C$ clusters and items to $K$ blocks; items in a block share a common strength parameter within each assessor cluster, yielding a parsimonious $C\times K$ co-clustering representation. Independent Gnedin priors infer $C$ and $K$. Data augmentation gives conjugate Gibbs updates and a tractable MCMC sampler with split--merge moves. Simulations characterize recovery and posterior uncertainty as signal, ranking depth, and group balance vary. Applied to the cancer gene atlas (TCGA) pan-cancer top-500 gene-expression rankings, the model reveals tissue-driven sample structure while compressing gene-level heterogeneity into interpretable blocks. Rank-based GSEA of posterior gene scores supports biological interpretation.
\end{abstract}

\noindent%


\spacingset{1}
\setlength{\abovedisplayskip}{4pt plus 2pt minus 2pt}
\setlength{\belowdisplayskip}{4pt plus 2pt minus 2pt}
\setlength{\abovedisplayshortskip}{0pt plus 2pt}
\setlength{\belowdisplayshortskip}{4pt plus 2pt minus 2pt}

\section{Introduction}\label{sec:intro}

Ranking data record how assessors order a common set of \(n\) items according to preference, relevance, performance, or another criterion, and arise in fields such as voting, consumer research, sports, and genomics. They are particularly useful when scores lack natural units, represent intangible quantities, or cannot be meaningfully compared across items, since recording relative positions largely bypasses the need for a common measurement scale. Rankings can therefore preserve the information of interest in settings where imposing a common numerical scale might instead obscure it. These advantages may partly explain the renewed interest in ranking data, as reflected in several recent contributions, including \citet{vitelli2018bayesianmallows}, \citet{piancastelli_clustered_2024}, \citet{PEARCE_2025}, and \citet{Henderson2025}.

A widely used probabilistic model for ranking data is the Plackett--Luce (PL) model \citep{luce_possible_1959,plackett_analysis_1975}.
It assigns each item \(i\) a positive strength \(\lambda_i\) and generates a ranking through a sequence of choices, with larger strengths favoring higher positions. If \(A_t\) denotes the set of items available before position \(t\), the probability of selecting \(i\in A_t\) is
\begin{equation}\label{eq:pl-basic}
\Pr(i\text{ is chosen at position }t\mid A_t,\boldsymbol\lambda)
=
\frac{\lambda_i}{\sum_{j\in A_t}\lambda_j}.
\end{equation}
Once selected, item \(i\) is removed from the available set, so that \(A_{t+1}=A_t\setminus{i}\), and the process continues until the top \(m\) positions have been filled, where \(m\leq n\). The resulting ranking is complete when \(m=n\) and partial when \(m<n\). At the limiting case \(n=2\), the model reduces to a pairwise comparison, with the Bradley--Terry model arising as a special case of the PL model \citep{SanFri2026}.

The standard Plackett--Luce model assumes agreement among assessors, although assessors may rank the same items systematically differently. Ranking mixtures model this heterogeneity through groups with distinct ranking behavior
\citep{gormley_mixture_2008,mollica_bayesian_2017}, but retain a separate strength for every item in every group. This captures assessor variation but, with many items, can be difficult to estimate and interpret and cannot directly identify items with similar roles across assessor populations.

To address these limitations, we let similar items share strength values. In the proposed Bayesian Plackett--Luce latent block model, assessors belong to \(C\) clusters and items to \(K\) blocks. Items in a block share a strength within an assessor cluster, although it can vary across clusters. The blocks are shared across all assessor clusters, providing a common basis for comparison and replacing an assessor-only \(C\times n\) strength matrix with a potentially much smaller \(C\times K\) representation. Throughout, we use \emph{cluster} for an assessor group and \emph{block} for an item group; \emph{co-clustering} and \emph{latent block model} retain their standard two-way meanings.

The paper makes three methodological contributions. First, it formulates a latent block model directly for ranked data, jointly modelling assessor clusters and item blocks. Standard Plackett--Luce mixtures partition assessors but retain an item-specific strength vector in each component. The nested Bayesian model of \citet{johnson2020wand} partitions both margins but permits the item grouping to vary across assessor clusters. We instead estimate one global item partition: the resulting \(C\times K\) strength matrix is a common coordinate system for comparing assessor populations and separates heterogeneity in assessor rankings from heterogeneity in item treatment, which amounts to distinguishing which assessors rank similarly from which items are treated similarly.

Second, we infer the complexity of this two-way structure rather than fixing it. Independent Gnedin partition priors on the assessor and item allocations
\citep{gnedin_species_2010,DeBlasiEtAl2015} assign positive probability to a finite but unknown number of groups. This is appropriate in contexts such as the present one, where the numbers of underlying assessor groups and item profiles are not expected to grow without bound. Applying the priors independently allows the data to determine both (C) and (K), while retaining item-only and assessor-only models as special cases.

Third, exponential augmentation of the Plackett--Luce likelihood
\citep{caronEfficientBayesianInference2012} lets the block structure restore Gamma conjugacy and integrate out the \(C\times K\) strengths in either partition update. We combine the resulting collapsed allocation probabilities with split--merge proposals. For fixed \(C\), \(K\), and maximum ranking length \(m\), one Markov chain Monte Carlo sweep costs \(O(LmK+LCK+nCK)\), hence is linear in assessors and items; when \(K\ll n\), the same representation replaces the assessor-only \(C\times n\) parameters with a smaller \(C\times K\) table.

We evaluate the model in a structured simulation study and on gene-expression rankings from The Cancer Genome Atlas (TCGA). \citet{vitelli2023transcriptomic} represented gene expression as rankings, with the highest-expressed gene first. This reduces measurement-scale differences across samples caused by technical variation and biological heterogeneity associated with sex, environmental exposures, or disease status, giving a common representation for heterogeneous tumor samples.

Our goal is to identify tumor groups and their characterizing gene blocks. Unlike the Bayesian Mallows analysis of \citet{vitelli2023transcriptomic}, which gives every sample cluster a separate full-gene ranking, our model estimates one gene partition shared across clusters. A block thus represents the same genes in every group, so its importance is directly comparable. In the application for instance, the same block has the highest estimated strength in both a lung squamous-cell and a head-and-neck squamous-cell cluster; this can be checked directly using model quantities, rather than inferred post-hoc, for example by computing the overlap among separate lists. We analyze top-500 rankings from \(2{,}617\) tumor samples over \(1{,}247\) genes, estimating tissue-related sample structure and a compressed set of shared blocks. Rank-based gene set enrichment analysis supports biological interpretation while retaining posterior uncertainty in both partitions. The joint model also improves conditional sample-level predictive fit relative to the assessor-only Plackett--Luce mixture used as an internal benchmark.

The remainder of the paper is organized as follows. Section~\ref{sec:literature} reviews related work on latent block models and probabilistic models for rankings. Section~\ref{sec:model} introduces the proposed model and prior specification. Section~\ref{sec:computation} describes posterior computation and summarization. Section~\ref{sec:sim-results} evaluates partition recovery and computational behavior in simulated data, and Section~\ref{sec:tcga} presents the pan-cancer analysis.

\section{Background and related work}\label{sec:literature}

The proposed model combines two established lines of work: methods that jointly group rows and columns of a data matrix, and probabilistic ranking models. The former task is variously called co-clustering \citep{govaert_nadif_2013,biernacki_jacques_keribin_2023}, biclustering \citep{madeira_oliveira_2004}, two-mode clustering \citep{van_mechelen_bock_de_boeck_2004}, or latent block modelling \citep{govaert_nadif_2003}. These labels describe closely related two-way grouping tasks, but usage varies by application; we use \emph{co-clustering} for the generic task and \emph{latent block model} for its model-based formulation \citep{govaert_nadif_2013,biernacki_jacques_keribin_2023}. Our key step is to model rankings as a two-way latent structure, with assessor clusters on one margin and item blocks on the other, under a Plackett--Luce likelihood.

The latent block model (LBM) of \citet{govaert_nadif_2003} is a foundational model-based formulation: conditional on row and column allocations, observations in each row--column cell group share a distributional parameter. \citet{govaert_nadif_2008} give the canonical binary Bernoulli example, and \citet{keribin_brault_celeux_govaert_2014} develop estimation and selection for categorical latent block models.

From a Bayesian perspective, \citet{wyse_friel_2012} integrated out cell-specific parameters in a collapsed latent block model. Its posterior is joint over row and column memberships and their occupied-group counts, providing a direct precedent for our collapsed partition updates.

For ranked data, the Plackett--Luce model represents rankings as sequential choices governed by positive item strengths \citep{luce_possible_1959,plackett_analysis_1975}. The augmentation of \citet{caronEfficientBayesianInference2012} restores Gamma conjugacy, while extensions accommodate partial rankings and ties \citep{turner_plackettluce_2020,Henderson2025}.

Assessor heterogeneity is commonly modelled through mixtures of Plackett--Luce, related multistage, or Mallows models \citep{gormley_mixture_2008,mollica_bayesian_2017}. Existing approaches include finite Bayesian Plackett--Luce mixtures for partial rankings \citep{mollica_bayesian_2017}, a nonparametric construction for an unbounded collection of items \citep{caron_bayesian_2014}, and results on mixture identifiability and estimation \citep{liu_zhao_liao_lu_xia_2019}. These models identify assessor populations but retain component-specific item strengths rather than a shared item partition. A close precedent is the Bayesian WAND model of \citet{johnson2020wand}, whose nested prior clusters assessors and groups items with exchangeable strengths within assessor clusters. By contrast, our global item partition compares all assessor clusters through the same finite \(C\times K\) strength table.
\section{The joint Plackett--Luce latent block model}\label{sec:model}

The data consist of $L$ observed rankings, each reported by one assessor over a common set of items $V=\{1,\dots,n\}$. We denote the ranking reported by assessor $\ell \in \{1,\dots,L\}$ by $\rho^{(\ell)}$, so that the observed data are given by the collection $\{\rho^{(1)},\dots,\rho^{(L)}\}$, which we assume to be a stochastic realization of latent quantities. Figure~\ref{fig:PL-LBM-chessboard} provides an illustrative example of this data format. Each $\rho^{(\ell)}$ is a top-$m_\ell$ ordering, \[ \rho^{(\ell)} = (\rho^{(\ell)}_1,\dots,\rho^{(\ell)}_{m_\ell}), \] where $m_\ell\leq n$, $\rho^{(\ell)}_t \in V$ denotes the item placed at position~$t$, and no item appears more than once. A complete ranking is obtained as the special case $m_\ell=n$. In many applications, however, asking every assessor to rank all $n$ items is impractical or uninformative. The observed data are therefore more naturally represented by partial rankings, with $m_\ell<n$ and potentially varying across assessors. The formulation below accommodates both partial and complete rankings. We first introduce the two one-dimensional models for item blocks and assessor clusters, respectively, before presenting the joint latent block model.

\subsection*{One-dimensional item-block model}

As the number of ranked items increases, the assumption in
\eqref{eq:pl-basic} that every item has a distinct strength may become
implausible. Instead, suppose that each item~$i$ belongs to one of $K$
blocks, with allocation $x_i\in\{1,\dots,K\}$, and that all items in
block~$k$ share a common strength $\lambda_k>0$. Item~$i$ then inherits
the strength $\lambda_{x_i}$, and the likelihood becomes
\begin{equation}\label{eq:lik-item}
  p(\rho^{(\ell)} \mid \mathbf{x}, \boldsymbol\lambda)
  =
  \prod_{t=1}^{m_\ell}
  \frac{\lambda_{x_{\rho_t^{(\ell)}}}}
  {\sum_{j\in A_t^{(\ell)}}\lambda_{x_j}}.
\end{equation}

All assessors share the same block-strength vector
$(\lambda_1,\dots,\lambda_K)$, so that items within the same block are
treated as having the same underlying ranking strength. Although highly
parsimonious, this model assumes homogeneous assessors
(Fig.~\ref{fig:PL-LBM-chessboard}a).

\subsection*{One-dimensional assessor clustering}

Assessor clustering considers the complementary setting in which $K=n$ and
$x_i=i$, that is, items retain distinct strength parameters, while groups of assessors may express
different ranking preferences. Let each assessor~$\ell$ belong to one of
$C$ clusters, with $w_\ell\in\{1,\dots,C\}$, and let cluster~$c$ have the
item-specific strength vector
\[
  \boldsymbol\lambda_c =
  (\lambda_{c1},\dots,\lambda_{cn}).
\]
The resulting strength matrix $\Lambda=(\lambda_{ci})$ has dimension
$C\times n$, with $\lambda_{ci}>0$, and the likelihood for assessor~$\ell$
in cluster~$c$ is
\begin{equation}\label{eq:lik-assessor}
  p(\rho^{(\ell)} \mid w_\ell=c,\Lambda)
  =
  \prod_{t=1}^{m_\ell}
  \frac{\lambda_{c,\rho_t^{(\ell)}}}
  {\sum_{j\in A_t^{(\ell)}}\lambda_{cj}}.
\end{equation}
This model captures heterogeneity across assessors but provides no
compression on the item side. Consequently, $\Lambda$ may be large when
$n$ is in the hundreds or thousands
(Fig.~\ref{fig:PL-LBM-chessboard}b).

\subsection*{Joint PL latent block model (PL-LBM)}
The joint model combines the two one-dimensional constructions. It retains
the two sets of latent allocation variables presented above, namely
$w_\ell \in \{1,\dots,C\}$ for assessor clusters, with
$\ell = 1,\dots,L$, and $x_i \in \{1,\dots,K\}$ for item blocks, with
$i = 1,\dots,n$. The resulting latent strength parameter is a
$C \times K$ matrix
\begin{equation}\label{eq:Lambda-def}
  \Lambda = [\lambda_{ck}]_{\substack{c=1,\dots,C \\ k=1,\dots,K}},
  \qquad \lambda_{ck} > 0.
\end{equation}

Conditional on the allocations and the strength matrix, the
likelihood is
\begin{equation}\label{eq:lik-joint}
  p(\boldsymbol\rho \mid \mathbf{w}, \mathbf{x}, \Lambda) =
  \prod_{\ell=1}^{L}\prod_{t=1}^{m_\ell}
  \frac{\lambda_{w_\ell,\,x_{\rho_t^{(\ell)}}}}
  {\sum_{j \in A_t^{(\ell)}}
  \lambda_{w_\ell,\,x_j}}.
\end{equation}

Crucially, two items in the same block share the same strength within each
assessor cluster, yet different assessor clusters may weight that same block
differently (see Fig.~\ref{fig:PL-LBM-chessboard}c). All assessor clusters
share the same number of item blocks. The item-only and assessor-only models
are recovered by setting $C=1$ and by setting $K=n$ with $x_i=i$,
respectively.

\definecolor{itemred}{HTML}{C94C4C}
\definecolor{itemorange}{HTML}{D98B2B}
\definecolor{itemblue}{HTML}{4B74B8}
\definecolor{itemgreen}{HTML}{5B9654}
\definecolor{panelhead}{gray}{0.94}
\definecolor{lambdaLow}{gray}{0.96}
\definecolor{lambdaMid}{gray}{0.78}
\definecolor{lambdaHigh}{gray}{0.48}
\begin{figure}[t]
\centering
\resizebox{\textwidth}{!}{%
\begin{tikzpicture}[x=1cm,y=1cm, font=\small]
  \def\pw{4.95}
  \def\ph{4.85}
  \def\py{-2.90}
  \def\figcentre{7.625}
  \def\rankshift{2.95}

  \draw[black!25,line width=.35pt] (0.20,-0.03)--(15.05,-0.03);
  \node[fill=white,inner xsep=5pt,font=\scriptsize\bfseries] at (\figcentre,-0.03)
    {Illustrative ranking data};

  \begin{scope}[xshift=\rankshift cm]
  \draw[black!45,line width=.3pt] (2.25,-0.31)--(2.25,-1.86);
  \draw[black!45,line width=.3pt] (0.10,-0.78)--(9.25,-0.78);
  \node[anchor=west,font=\tiny\bfseries] at (.2,-0.57)
    {Rankings};
  \foreach \xx/\ord in {2.55/1st,4.20/2nd,5.85/3rd,7.50/4th}
    \node[font=\tiny] at (\xx,-0.57) {\ord};
  \foreach \xx/\col/\name in {2.65/itemred/Red,4.20/itemorange/Orange,5.85/itemblue/Blue,7.50/itemgreen/Green}{
    
  }
  \foreach \yy/\lab in {-0.96/{$\rho^{(1)}=$},-1.32/{$\rho^{(2)}=$},-1.68/{$\rho^{(3)}=$}}{
    \node[anchor=east,font=\scriptsize] at (2.15,\yy) {\lab};
  }

  \foreach \yy/\a/\b/\c/\d in {
    -0.96/itemred/itemorange/itemblue/itemgreen,
    -1.32/itemblue/itemgreen/itemred/itemorange,
    -1.68/itemorange/itemred/itemblue/itemgreen}{
    \foreach \xx/\col in {2.55/\a,4.20/\b,5.85/\c,7.50/\d}
      \node[itemdot,fill=\col,minimum size=3.0mm] at (\xx,\yy) {};
  }
  \end{scope}
  \node[font=\scriptsize\itshape,align=center] at (5.05,-2.17)
    {};
  \draw[black!25,line width=.35pt] (0.20,-2.62)--(15.05,-2.62);
  \node[fill=white,inner xsep=5pt,font=\scriptsize\bfseries] at (\figcentre,-2.62)
    {PL model representations};

  \begin{scope}[shift={(0,\py)}]
    \fill[panelhead] (0,0) rectangle (\pw,-0.46);
    \node[paneltitle] at (2.48,-0.23) {(a) Item-block model ($C=1$)};
    \node[font=\scriptsize] at (2.00,-0.78) {Block 1};
    \node[font=\scriptsize] at (3.60,-0.78) {Block 2};
    \draw[decorate,decoration={brace,amplitude=2.5pt}] (1.20,-1.05)--(2.80,-1.05);
    \draw[decorate,decoration={brace,amplitude=2.5pt}] (2.80,-1.05)--(4.40,-1.05);
    \foreach \xx/\col in {1.60/itemred,2.40/itemorange,3.20/itemblue,4.00/itemgreen}
      \node[itemdot,fill=\col,minimum size=3.0mm] at (\xx,-1.32) {};
    \node[font=\tiny,align=center,text width=1.05cm] at (0.62,-1.98)
      {All assessors\\$\{1,2,3\}$};
    \fill[lambdaHigh] (1.20,-1.60) rectangle (2.80,-2.35);
    \fill[lambdaLow]  (2.80,-1.60) rectangle (4.40,-2.35);
    \draw[matrixcell] (1.20,-1.60) rectangle (4.40,-2.35);
    \draw[black!60,line width=.35pt] (2.80,-1.60)--(2.80,-2.35);
    \node at (2.00,-1.98) {$\lambda_1$};
    \node at (3.60,-1.98) {$\lambda_2$};
    \node[setline,anchor=west] at (0.45,-2.96)
      {$x_{\mathrm{Red}}=x_{\mathrm{Orange}}=1$};
    \node[setline,anchor=west] at (0.45,-3.34)
      {$x_{\mathrm{Blue}}=x_{\mathrm{Green}}=2$};
    \node[note,text width=3.85cm] at (2.48,-4.25)
      {Items are partitioned into two blocks, each sharing the same latent strength};
  \end{scope}

  \begin{scope}[shift={(5.15,\py)}]
    \fill[panelhead] (0,0) rectangle (\pw,-0.46);
    \node[paneltitle] at (2.48,-0.23) {(b) Assessor clustering ($K=n$)};
    \node[font=\tiny] at (2.66,-0.78) {Items (columns)};
    \foreach \xx/\col in {1.55/itemred,2.29/itemorange,3.03/itemblue,3.77/itemgreen}
      \node[itemdot,fill=\col,minimum size=3.0mm] at (\xx,-1.32) {};
    \node[font=\tiny,rotate=90] at (4.62,-2.35) {Assessors (rows)};
    \node[font=\tiny,anchor=east,align=right] at (1.08,-1.98)
      {Cluster 1\\$\{1,3\}$};
    \node[font=\tiny,anchor=east,align=right] at (1.08,-2.73)
      {Cluster 2\\$\{2\}$};
    \fill[lambdaHigh] (1.18,-1.60) rectangle ++(.74,-.75);
    \fill[lambdaMid]  (1.92,-1.60) rectangle ++(.74,-.75);
    \fill[lambdaLow]  (2.66,-1.60) rectangle ++(.74,-.75);
    \fill[lambdaLow]  (3.40,-1.60) rectangle ++(.74,-.75);
    \fill[lambdaLow]  (1.18,-2.35) rectangle ++(.74,-.75);
    \fill[lambdaLow]  (1.92,-2.35) rectangle ++(.74,-.75);
    \fill[lambdaMid]  (2.66,-2.35) rectangle ++(.74,-.75);
    \fill[lambdaHigh] (3.40,-2.35) rectangle ++(.74,-.75);
    \draw[matrixcell] (1.18,-1.60) rectangle (4.14,-3.10);
    \draw[black!60,line width=.35pt] (1.18,-2.35)--(4.14,-2.35);
    \foreach \xx in {1.92,2.66,3.40} \draw[black!60,line width=.35pt] (\xx,-1.60)--(\xx,-3.10);
    \node[font=\tiny,text=white] at (1.55,-1.98) {$\lambda_{1,1}$};
    \node[font=\tiny] at (2.29,-1.98) {$\lambda_{1,2}$};
    \node[font=\tiny] at (3.03,-1.98) {$\lambda_{1,3}$};
    \node[font=\tiny] at (3.77,-1.98) {$\lambda_{1,4}$};
    \node[font=\tiny] at (1.55,-2.73) {$\lambda_{2,1}$};
    \node[font=\tiny] at (2.29,-2.73) {$\lambda_{2,2}$};
    \node[font=\tiny] at (3.03,-2.73) {$\lambda_{2,3}$};
    \node[font=\tiny,text=white] at (3.77,-2.73) {$\lambda_{2,4}$};
    \node[setline,anchor=west] at (0.45,-3.43)
      {$w_1=w_3=1,\qquad w_2=2$};
    \node[note,text width=3.85cm] at (2.48,-4.25)
      {Assessors are partitioned into two clusters, each with a full latent strength vector};
  \end{scope}

  \begin{scope}[shift={(10.30,\py)}]
    \fill[panelhead] (0,0) rectangle (\pw,-0.46);
    \node[paneltitle] at (2.48,-0.23) {(c) Joint co-clustering};
    \node[font=\tiny] at (2.65,-0.56) {Items (columns)};
    \node[font=\scriptsize] at (1.93,-0.78) {Block 1};
    \node[font=\scriptsize] at (3.38,-0.78) {Block 2};
    \draw[decorate,decoration={brace,amplitude=2.5pt}] (1.20,-1.05)--(2.65,-1.05);
    \draw[decorate,decoration={brace,amplitude=2.5pt}] (2.65,-1.05)--(4.10,-1.05);
    \foreach \xx/\col in {1.56/itemred,2.29/itemorange,3.01/itemblue,3.74/itemgreen}
      \node[itemdot,fill=\col,minimum size=3.0mm] at (\xx,-1.32) {};
    \node[font=\tiny,rotate=90] at (4.62,-2.35) {Assessors (rows)};
    \node[font=\tiny,anchor=east,align=right] at (1.08,-1.98)
      {Cluster 1\\$\{1,3\}$};
    \node[font=\tiny,anchor=east,align=right] at (1.08,-2.73)
      {Cluster 2\\$\{2\}$};
    \fill[lambdaHigh] (1.20,-1.60) rectangle (2.65,-2.35);
    \fill[lambdaLow]  (2.65,-1.60) rectangle (4.10,-2.35);
    \fill[lambdaLow]  (1.20,-2.35) rectangle (2.65,-3.10);
    \fill[lambdaHigh] (2.65,-2.35) rectangle (4.10,-3.10);
    \draw[matrixcell] (1.20,-1.60) rectangle (4.10,-3.10);
    \draw[black!60,line width=.35pt] (2.65,-1.60)--(2.65,-3.10);
    \draw[black!60,line width=.35pt] (1.20,-2.35)--(4.10,-2.35);
    \node[font=\scriptsize,text=white] at (1.93,-1.98) {$\lambda_{1,1}$};
    \node[font=\scriptsize] at (3.38,-1.98) {$\lambda_{1,2}$};
    \node[font=\scriptsize] at (1.93,-2.73) {$\lambda_{2,1}$};
    \node[font=\scriptsize,text=white] at (3.38,-2.73) {$\lambda_{2,2}$};
    \node[font=\scriptsize,align=center] at (2.65,-3.43)
      {$\mathbf{w}=(1,2,1),\qquad \mathbf{x}=(1,1,2,2)$};
    \node[note,text width=3.85cm] at (2.65,-4.25)
      {Assessors are partitioned into two clusters and items into two blocks};
  \end{scope}

\end{tikzpicture}%
}
\caption{From observed rankings to joint partitions. The top panel shows three complete toy rankings from $L=3$ assessors, with colours identifying $n=m=4$ items. The set beneath each row label lists the indices of the assessors assigned to that cluster; thus $\{1,3\}$ denotes assessors 1 and 3, while $\{1,2,3\}$ in panel~(a) denotes the single cluster containing all assessors. Panel~(a) groups items into shared blocks; panel~(b) clusters assessors while retaining item-specific strengths; and panel~(c) combines both structures in the PL-LBM, with a common item partition and cluster-specific block-strength profiles. Darker gray shading indicates greater relative strength.}
\label{fig:PL-LBM-chessboard}
\end{figure}
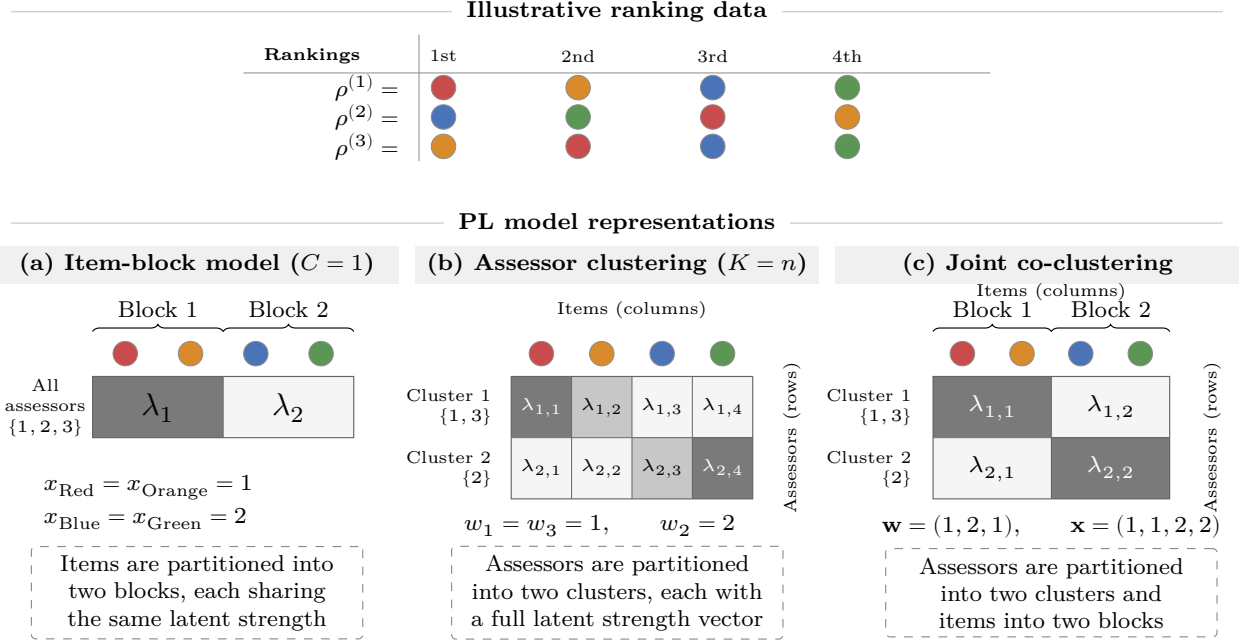

\subsection{Partition prior: Gnedin process}\label{subsec:gnedin-prior}

The priors on the two partitions play a structural role in the proposed model,
as they determine how strongly the data are compressed into assessor clusters
and item blocks. We adopt two independent Gnedin processes,
$p_{\GN}(\mathbf{w}\mid\gamma_w)$ and
$p_{\GN}(\mathbf{x}\mid\gamma_x)$, with scalar hyperparameters
$\gamma_w,\gamma_x\in(0,1)$. The Gnedin process is the $\sigma=-1$ member
of the Gibbs-type family and belongs to its negative-discount subclass
\citep{gnedin_species_2010,DeBlasiEtAl2015}. Under this prior, the total
number of groups in the underlying population is random but finite almost
surely, and therefore does not grow indefinitely as more units are observed.

This property is particularly suitable for the PL-LBM. Conditional on
$(C,K)$, the model is represented by a finite $C\times K$ strength matrix,
but its dimensions are not known in advance. Fixing $C$ or $K$, or imposing
an arbitrary deterministic upper bound, would therefore make the inference
unnecessarily restrictive.

Formally, consider an exchangeable partition of
$[N]=\{1,\dots,N\}$, and let $H_N$ denote its number of occupied groups.
Under the projective extension of the Gnedin process,
\[
  H_\infty=\lim_{N\rightarrow\infty}H_N
\]
exists and is finite almost surely. In the one-parameter formulation used
here, its distribution is
\begin{equation}\label{eq:gnedin-terminal-count}
  \Pr(H_\infty=h\mid\gamma)
  =
  \frac{\gamma(1-\gamma)_{h-1}}{h!}
  =
  \frac{\gamma\,\Gamma(h-\gamma)}
       {\Gamma(1-\gamma)\Gamma(h+1)},
  \qquad h=1,2,\dots,
\end{equation}
where $(a)_r=a(a+1)\cdots(a+r-1)$ denotes the ascending factorial, with
$(a)_0=1$ \citep{gnedin_species_2010}. Hence,
\begin{equation}\label{eq:gnedin-terminal-tail}
  \Pr(H_\infty=h\mid\gamma)
  \sim
  \frac{\gamma}{\Gamma(1-\gamma)}h^{-(1+\gamma)}
  \qquad\text{as }h\rightarrow\infty.
\end{equation}
The prior on the terminal number of groups is therefore heavy-tailed:
although $H_\infty$ is finite almost surely,
$\mathrm{E}(H_\infty)=\infty$ for every $\gamma\in(0,1)$. The prior can
thus favor parsimonious partitions while retaining non-negligible support
for substantially larger dimensions when required by the likelihood. This
property concerns the limit population count; for a data set of size $N$,
the occupied count $H_N$ is necessarily bounded by $N$.

The parameter $\gamma$ controls this balance. Since
$\Pr(H_\infty=1\mid\gamma)=\gamma$ and the tail exponent in
\eqref{eq:gnedin-terminal-tail} increases with $\gamma$, larger values
impose stronger prior parsimony, whereas smaller values place more mass on
larger numbers of groups. This allows different prior resolutions on the two
margins of the PL-LBM: $\gamma_w$ controls the number of assessor populations,
while $\gamma_x$ controls the compression of item-specific strengths into
shared blocks.

For a generic partition of $[N]$ into $H$ nonempty groups of sizes
$n_1,\dots,n_H$, the corresponding exchangeable partition probability
function is
\begin{equation}\label{eq:gnedin-eppf-corrected}
  p_{\GN}(n_1,\dots,n_H\mid\gamma)
  =
  \psi_{N,H}(\gamma)
  \prod_{h=1}^{H}n_h!,
\end{equation}
where
\begin{equation}\label{eq:gnedin-psi-corrected}
  \psi_{N,H}(\gamma)
  =
  \frac{(H-1)!}{(N-1)!}
  \frac{(1-\gamma)_{H-1}(\gamma)_{N-H}}
       {(1+\gamma)_{N-1}},
  \qquad \gamma\in(0,1).
\end{equation}
For the item partition, $N=n$ and $H=K$; for the assessor partition,
$N=L$ and $H=C$.

The corresponding predictive probabilities for the item partition are
\begin{align}
  p(x_{n+1}=k\mid\mathbf{x},\gamma_x)
  &=
  \frac{(n_k^{(x)}+1)(n-K+\gamma_x)}
       {n(n+\gamma_x)},
  \qquad k=1,\dots,K,
  \label{eq:gnedin-old-corrected}\\
  p(x_{n+1}=K+1\mid\mathbf{x},\gamma_x)
  &=
  \frac{K(K-\gamma_x)}
       {n(n+\gamma_x)}.
  \label{eq:gnedin-new-corrected}
\end{align}
Here, $n_k^{(x)}=\#\{i:x_i=k\}$ is the current size of item block $k$.
The analogous formulas for the assessor partition follow by replacing
$(n,K,\gamma_x)$ with $(L,C,\gamma_w)$.

The first probability displays the reinforcement property of Gibbs-type
partitions: items are more likely to join blocks that are already well
represented. Meanwhile, the probability of creating a new block decreases
as the sample grows and, for fixed $(n,K)$, as $\gamma_x$ increases. In the
PL-LBM, reinforcement discourages the unnecessary creation of nearly
duplicate strength profiles, while the heavy-tailed prior allows a finer
partition when supported by the rankings.
\subsection{Model posterior}\label{subsec:bayesian}

The remaining prior to be specified is that on the item-block strengths, which
are taken to be independent:
\[
  \lambda_{ck} \mid \mathbf{x},\mathbf{w} \sim \Ga(a,b),
  \qquad c=1,\dots,C,\quad k=1,\dots,K,
\]
where $\Ga(a,b)$ denotes a Gamma distribution with shape $a$ and rate $b$.
The Gamma prior is conjugate to the augmented PL likelihood
(Section~\ref{subsec:augmentation}), yielding closed-form full conditional
distributions for $\Lambda$.

By Bayes' theorem, the full posterior is
\begin{equation}\label{eq:full-posterior}
  p(\mathbf{w}, \mathbf{x}, \Lambda \mid \boldsymbol\rho) \propto
  \underbrace{p(\boldsymbol\rho \mid \mathbf{w}, \mathbf{x},
  \Lambda)}_{\text{PL likelihood}}
  \underbrace{p(\Lambda \mid
  \mathbf{w}, \mathbf{x})}_{\text{strength prior}}
  \underbrace{p_{\GN}(\mathbf{w}\mid\gamma_w)\,
  p_{\GN}(\mathbf{x}\mid\gamma_x)}_{\text{partition priors}}.
\end{equation}

\section{Posterior computation}\label{sec:computation}
Our goal is to sample from the full posterior~\eqref{eq:full-posterior}
using Markov chain Monte Carlo (MCMC). The PL likelihood~\eqref{eq:lik-joint}
involves denominators
$\sum_{j \in A_t^{(\ell)}} \lambda_{w_\ell,\,x_j}$ that couple the
strength parameters, thereby preventing direct conjugate Gibbs updates.
The following data augmentation restores conditional Gamma conjugacy.

\subsection{Data augmentation and collapsed model}
\label{subsec:augmentation}

Following \citet{caronEfficientBayesianInference2012}, we introduce
exponentially distributed latent variables $Z_{\ell t}>0$ and define the
augmented density
\begin{equation}\label{eq:aug-joint-corrected}
  p(\boldsymbol\rho,\mathbf Z \mid \mathbf w,\mathbf x,\Lambda)
  =
  \prod_{\ell=1}^{L}\prod_{t=1}^{m_\ell}
  \lambda_{w_\ell,x_{\rho_t^{(\ell)}}}
  \exp\!\left\{
    -Z_{\ell t}
    \sum_{j\in A_t^{(\ell)}}
    \lambda_{w_\ell,x_j}
  \right\}.
\end{equation}
To see why this augmentation recovers the Plackett--Luce factor, let
$r_{\ell t}=\rho_t^{(\ell)}$ and write
$R_{\ell t}=\sum_{j\in A_t^{(\ell)}}\lambda_{w_\ell,x_j}$. Then
\begin{equation}\label{eq:augmentation-integral}
  \int_0^\infty
  \lambda_{w_\ell,x_{r_{\ell t}}}
  \exp\{-zR_{\ell t}\}\,dz
  =
  \frac{\lambda_{w_\ell,x_{r_{\ell t}}}}{R_{\ell t}},
\end{equation}
which is precisely the probability of the observed choice at stage $t$.
Thus, $Z_{\ell t}$ is an auxiliary exponential waiting time with rate
$R_{\ell t}$; conditioning on it separates the strength parameters in the
augmented likelihood.

The same representation can be written at the item-block level using the following quantities:
\begin{align}
  w_{ck}
  &=
  \sum_{\ell=1}^{L}
  \ind(w_\ell=c)
  \sum_{t=1}^{m_\ell}
  \ind\!\bigl(x_{\rho_t^{(\ell)}}=k\bigr),
  \label{eq:wck-corrected}\\
  S_{ck}
  &=
  \sum_{\ell=1}^{L}
  \ind(w_\ell=c)
  \sum_{t=1}^{m_\ell}
  Z_{\ell t}
  \sum_{j\in A_t^{(\ell)}}
  \ind(x_j=k).
  \label{eq:Sck-corrected}
\end{align}
Here, $w_{ck}$ counts how many times an item from block~$k$ is selected by
an assessor in cluster~$c$, whereas $S_{ck}$ weights each latent waiting time by the number
of items from block~$k$ that remain available at that stage. The augmented
likelihood consequently factorizes as
\begin{equation}\label{eq:aug-factorised-corrected}
  p(\boldsymbol\rho,\mathbf Z \mid \mathbf w,\mathbf x,\Lambda)
  \propto
  \prod_{c=1}^{C}\prod_{k=1}^{K}
  \lambda_{ck}^{w_{ck}}
  \exp(-\lambda_{ck}S_{ck}).
\end{equation}
Thus, conditional on the latent waiting times, the likelihood contribution
for each $\lambda_{ck}$ has a Gamma kernel, with shape contribution $w_{ck}$
and rate contribution $S_{ck}$.

Combining \eqref{eq:aug-factorised-corrected} with the independent
$\Ga(a,b)$ prior and integrating out $\Lambda$ yields the collapsed posterior
\begin{equation}\label{eq:collapsed-augmented-corrected}
  p(\mathbf w,\mathbf x \mid \boldsymbol\rho,\mathbf Z)
  \propto
  p_{\GN}(\mathbf w \mid \gamma_w)\,
  p_{\GN}(\mathbf x \mid \gamma_x)\,
  \prod_{c=1}^{C}\prod_{k=1}^{K}
  \frac{b^a}{\Gamma(a)}
  \frac{\Gamma(a+w_{ck})}
       {(b+S_{ck})^{a+w_{ck}}},
\end{equation}
which is analogous to the collapsed posterior used in Gaussian and Bernoulli
latent block models, where cell-specific parameters are integrated out
analytically under conjugate priors.

\subsection{Gibbs sampler with split--merge moves}
\label{subsec:mcmc}

We employ a partially collapsed sampler targeting the augmented posterior
\begin{equation}\label{eq:aug-posterior-corrected}
  p(\mathbf w,\mathbf x,\Lambda,\mathbf Z \mid \boldsymbol\rho)
  \propto
  p(\boldsymbol\rho,\mathbf Z \mid \mathbf w,\mathbf x,\Lambda)\,
  p(\Lambda \mid \mathbf w,\mathbf x)\,
  p_{\GN}(\mathbf w \mid \gamma_w)\,
  p_{\GN}(\mathbf x \mid \gamma_x).
\end{equation}
Within each partition sweep, the allocation probabilities are evaluated
using the collapsed posterior~\eqref{eq:collapsed-augmented-corrected}, in
which $\Lambda$ has been integrated out analytically. The MCMC (see Alg.~\ref{alg:compact-mcmc})
refreshes $\Lambda$ after every complete collapsed allocation sweep and after
every accepted split--merge proposal \citep{vanDykPark2008}. Each refresh is followed immediately by
row-wise geometric-mean normalization, with the removed positive row scale
retained for the subsequent updates (See Section~\ref{sec:identifiability}). Rejected split--merge proposals leave both the
partition and this strength state unchanged. The overall procedure is
summarized in Algorithm~\ref{alg:compact-mcmc}.

\begin{algorithm}[t]
  \caption{Compact partially collapsed MCMC sampler. Full allocation
  probabilities, split--merge proposals, and implementation details are
  provided in \suppsectionsref{sec:full-conditionals}{sec:split-merge}{1}{2}
  of the online Supplementary Material.}
  \label{alg:compact-mcmc}

  \begin{spacing}{1.0}
  \begin{algorithmic}[1]
    \Require Rankings $\boldsymbol\rho$, hyperparameters
    $(a,b,\gamma_w,\gamma_x)$, and number of iterations $S$
    \Ensure Posterior draws of $(\mathbf w,\mathbf x,\widetilde\Lambda)$ and
    identified summaries

    \State Initialize $\mathbf w$, $\mathbf x$, and the stored pair
    $(\widetilde\Lambda,\mathbf s)$

    \For{$s=1,\ldots,S$}
      \State Reconstruct $\Lambda=\operatorname{diag}(\mathbf s)
      \widetilde\Lambda$ and sample $\mathbf Z$ from
      \eqref{eq:Z-cond-corrected}

      \For{each configured assessor-allocation sweep}
        \State Update $\mathbf w$ using the collapsed probabilities in
        \eqref{eq:w-existing-corrected}--\eqref{eq:w-new-corrected}
        \State Draw $\Lambda$ from \eqref{eq:lambda-cond-corrected} and
        decompose it into $(\widetilde\Lambda,\mathbf s)$
      \EndFor

      \State Apply the configured collapsed split--merge proposals to
      $\mathbf w$; after each accepted proposal, redraw and decompose $\Lambda$

      \For{each configured item-allocation sweep}
        \State Update $\mathbf x$ using the analogous collapsed probabilities
        in \suppsectionref{sec:full-conditionals}{1} of the online Supplementary Material
        \State Draw $\Lambda$ from \eqref{eq:lambda-cond-corrected} and
        decompose it into $(\widetilde\Lambda,\mathbf s)$
      \EndFor

      \State Apply the configured collapsed split--merge proposals to
      $\mathbf x$; after each accepted proposal, redraw and decompose $\Lambda$

      \State Store $(\mathbf w,\mathbf x,\widetilde\Lambda,\mathbf s)$
    \EndFor
  \end{algorithmic}
  \end{spacing}
\end{algorithm}

\subsubsection{Step 1: update latent times
\texorpdfstring{$\mathbf Z$}{Z}}

Given $(\mathbf w,\mathbf x,\Lambda)$, the latent waiting times
$Z_{\ell t}$ are sampled independently from
\begin{equation}\label{eq:Z-cond-corrected}
  Z_{\ell t}\mid
  \mathbf w,\mathbf x,\Lambda,\boldsymbol\rho
  \sim
  \mathrm{Exp}\!\left(
    \sum_{j\in A_t^{(\ell)}}
    \lambda_{w_\ell,x_j}
  \right),
  \qquad
  \ell=1,\dots,L,\;\;
  t=1,\dots,m_\ell.
\end{equation}
For computational purposes, define $c_{t,k}^{(\ell)}
  =
  \#\{j\in A_t^{(\ell)}:x_j=k\},$
the number of items from block~$k$ that remain available before position
$t$. The exponential rate can then be evaluated directly at the item-block
level by weighting each strength $\lambda_{w_\ell,k}$ by
$c_{t,k}^{(\ell)}$, avoiding a separate summation over all available items.

\subsubsection{Step 2: update assessor allocations
\texorpdfstring{$\mathbf w$}{w}}

For assessor~$\ell$, define its item-block-wise contributions
\begin{align}
  w_{\ell k}
  &=
  \sum_{t=1}^{m_\ell}
  \ind\!\bigl(x_{\rho_t^{(\ell)}}=k\bigr),
  \label{eq:wlk-corrected}\\
  S_{\ell k}
  &=
  \sum_{t=1}^{m_\ell}
  Z_{\ell t}
  \sum_{j\in A_t^{(\ell)}}\ind(x_j=k).
  \label{eq:Slk-corrected}
\end{align}
Let $w_{-\ell,ck}$ and $S_{-\ell,ck}$ denote the item-block sufficient
statistics computed with assessor~$\ell$ removed. Let $C_{-\ell}$ be the
number of occupied assessor clusters after removing assessor~$\ell$, and
let $n_{-\ell,c}^{(w)}$ be the size of cluster~$c$ in
$\mathbf w_{-\ell}$.

For an existing cluster $c=1,\dots,C_{-\ell}$,
\begin{align}
  p(w_\ell=c \mid \mathbf w_{-\ell},\mathbf x,\mathbf Z,\boldsymbol\rho)
  &\propto
  p_{\GN}(w_\ell=c \mid \mathbf w_{-\ell},\gamma_w)
  \prod_{k=1}^{K}
  \frac{\Gamma(a+w_{-\ell,ck}+w_{\ell k})}
       {\Gamma(a+w_{-\ell,ck})}
  \\[-0.25em]
  &\quad\times
  \frac{(b+S_{-\ell,ck})^{a+w_{-\ell,ck}}}
       {(b+S_{-\ell,ck}+S_{\ell k})
       ^{a+w_{-\ell,ck}+w_{\ell k}}},
  \label{eq:w-existing-corrected}
\end{align}
where the Gnedin predictive term is obtained from
\eqref{eq:gnedin-old-corrected} by replacing
$(n,K,\gamma_x)$ with $(L-1,C_{-\ell},\gamma_w)$.

For a new cluster,
\begin{align}
  p(w_\ell=C_{-\ell}+1 \mid
  \mathbf w_{-\ell},\mathbf x,\mathbf Z,\boldsymbol\rho)
  &\propto
  p_{\GN}(\text{new} \mid \mathbf w_{-\ell},\gamma_w)
  \prod_{k=1}^{K}
  \frac{b^a}{\Gamma(a)}
  \frac{\Gamma(a+w_{\ell k})}
       {(b+S_{\ell k})^{a+w_{\ell k}}},
  \label{eq:w-new-corrected}
\end{align}
where the Gnedin predictive term is obtained analogously from
\eqref{eq:gnedin-new-corrected}.

\subsubsection{Step 3: update item allocations
\texorpdfstring{$\mathbf x$}{x}}

The item allocations are updated analogously to the assessor allocations.
The item-specific sufficient statistics and the probabilities of assignment
to existing and new blocks are given in \suppsectionref{sec:full-conditionals}{1} of the online
Supplementary Material.

\subsubsection{Step 4: refresh strength parameters
\texorpdfstring{$\Lambda$}{Lambda}}
\label{subsubsec:lambda-update}

Given $(\mathbf w,\mathbf x,\mathbf Z,\boldsymbol\rho)$, the conditional
posterior of each item-block strength is
\begin{equation}\label{eq:lambda-cond-corrected}
  \lambda_{ck}\mid \mathbf w,\mathbf x,\mathbf Z,\boldsymbol\rho
  \sim
  \Ga(a+w_{ck},\,b+S_{ck}),
  \qquad c=1,\dots,C,\;\;k=1,\dots,K.
\end{equation}
Each draw is row-normalized using~\eqref{eq:norm-geomean-rewrite}; the removed
positive row scale is retained. In the implemented kernel, this conditional
Gamma refresh is made after every complete assessor or item allocation sweep
and immediately after every accepted split--merge proposal, using statistics
for the resulting partition. A rejected proposal retains the current draw and
row scales. Their product reconstructs the unconstrained Gamma matrix used in
the next latent-time update, while the normalized component provides the
identified posterior summary. These intermediate conditional draws are Gibbs
refreshes of the same augmented target; they are not additional model layers.

\subsubsection{Split--merge updates via restricted Gibbs proposals}
\label{subsubsec:split-merge}

Single-entity Gibbs updates make local adjustments but can become trapped in
local posterior modes. To help the sampler traverse low-density regions, we
supplement each partition sweep with configurable split--merge
Metropolis--Hastings updates that move several assessors or items at a time.
This follows the restricted
split--merge construction of \citet{jain_neal_2004}; see also
\citet{bouchard_cote_particle_gibbs_2015} for a description of its
anchor-and-restricted-allocation form. More generally, state-dependent and
similarity-guided split--merge proposals have been studied for difficult
partition spaces \citep{peixoto_merge_split_2020,luo_shrivastava_2019}.

For the assessor partition, two assessors are selected as anchors. If they
belong to the same cluster, a split is proposed; otherwise, a merge is
proposed. The candidate partition is constructed through a restricted Gibbs
scan over the cluster or pair of clusters containing the anchors, while all
allocations outside this affected set remain fixed. The proposal is accepted
using the collapsed posterior ratio together with the corresponding forward
and reverse proposal probabilities. The item-partition update is defined
analogously. Full details of the anchor proposal, restricted scan, and
Metropolis--Hastings ratio are provided in \suppsectionref{sec:split-merge}{2} of the online
Supplementary Material.

\subsection{Identifiability}
\label{sec:identifiability}

As in any Plackett--Luce model, the likelihood identifies strengths only
up to a common positive scale
\citep{caronEfficientBayesianInference2012,johnson2020wand}. In the
PL-LBM, this invariance holds separately within each assessor cluster:
multiplying row~$c$ of $\Lambda$ by any $\alpha_c>0$ leaves all choice
probabilities unchanged. The likelihood therefore identifies only the
relative strengths within each row. To preserve the conjugate Gamma updates, the sampler draws the
unconstrained strengths and then decomposes each row into a positive
scale $s_c=(\prod_{k=1}^{K}\lambda_{ck})^{1/K}$ and normalized strengths
\begin{equation}\label{eq:norm-geomean-rewrite}
  \widetilde{\lambda}_{ck}
  =
  \frac{\lambda_{ck}}{s_c}
  =
  \frac{\lambda_{ck}}
       {\left(\prod_{k'=1}^{K}\lambda_{ck'}\right)^{1/K}}.
\end{equation}
Each normalized row has unit geometric mean, equivalently
$K^{-1}\sum_{k=1}^{K}\log\widetilde{\lambda}_{ck}=0$. Thus,
$\widetilde{\lambda}_{ck}>1$ indicates above-geometric-mean strength
within assessor cluster~$c$, whereas $\widetilde{\lambda}_{ck}<1$
indicates below-geometric-mean strength. No information is lost because
$\Lambda=\operatorname{diag}(\mathbf{s})\widetilde{\Lambda}$. The row
scales are retained to reconstruct the unconstrained Gamma draw for the
next latent-time update, while $\widetilde{\Lambda}$ is used for
posterior interpretation and summarization
\citep{caronEfficientBayesianInference2012,newman_ranking_2022}.

Under the shape--rate parameterization,
$\lambda_{ck}\sim\operatorname{Gamma}(a,b)$ gives
$\mathbb{E}(\log\lambda_{ck})=\psi(a)-\log b$ and
$\operatorname{Var}(\log\lambda_{ck})=\psi'(a)$. Hence,
$b=\exp\{\psi(a)\}$ centres the unconstrained prior at zero on the log
scale. 

The second invariance arises from permutations of the assessor-cluster and
item-block labels. The VI partition estimates introduced below are
label-invariant. Strength matrices with the VI dimensions are aligned to the
two VI partitions before they are summarized, as described below and in
\suppsectionref{sec:supp-identifiability}{3} of the Supplementary Material.

\subsection{Point estimation and convergence diagnostics}
\label{subsec:diagnostics}

The posterior distribution over partitions is the primary inferential object.
To obtain the related point estimates, we minimize the VI loss of
\citet{Meila2003} separately on each partition space,
\[
  \widehat{\mathbf w}_{\mathrm{VI}}
  =
  \arg\min_{\mathbf w^\prime}
  \mathrm{E}\!\left[
    \mathrm{VI}(\mathbf w^\prime,\mathbf w)
    \mid \boldsymbol\rho
  \right],
\]
which we compute using \texttt{salso}
\citep{DahlJohnsonMuller2022,Dahl2025salso}. The item partition
$\mathbf x$ is handled analogously.

Because the numbers of occupied assessor clusters and item blocks vary across iterations,
we also summarize their posterior distributions through
\[
  \widehat C_{\mathrm{mode}}
  =
  \operatorname{mode}\{C^{(s)}\},
  \qquad
  \widehat K_{\mathrm{mode}}
  =
  \operatorname{mode}\{K^{(s)}\}.
\]
Alternative summaries are
$\widehat C_{\mathrm{VI}}$ and $\widehat K_{\mathrm{VI}}$, the numbers of
unique labels in the corresponding VI point estimates.

\paragraph*{Posterior summary of
\texorpdfstring{$\Lambda$}{Lambda}.}

An entrywise posterior mean of $\Lambda$ requires matrices with a common
dimension. We therefore retain draws satisfying
$(C^{(s)},K^{(s)})=(\widehat C_{\mathrm{VI}},\widehat K_{\mathrm{VI}})$ and
align both allocation margins to the corresponding VI partitions using ECR
relabeling from the \texttt{ R label.switching} package \citep{
papastamoulis_labelswitching_2016}. If $\sigma_s$ and $\tau_s$ denote the
resulting row and column permutations, the aligned normalized draw is
\[
  \widetilde\Lambda^{(s)}_{\mathrm{align}}
  =
  \widetilde\Lambda^{(s)}[\sigma_s,\tau_s],
\]
and $\widehat\Lambda$ is its Monte Carlo mean over the same-dimensions draws. \suppsectionref{sec:supp-identifiability}{3} of the Supplementary Material gives the full
construction.
\section{Simulation study}\label{sec:sim-results}

The simulations assess recovery of both latent partitions and where rankings no longer distinguish them. Primary data are generated exactly from the PL-LBM with $L=120$ assessors, $n=160$ items, and $(C^\star,K^\star)=(3,4)$. Adjacent assessor clusters and item blocks have log-strength separations $\delta\in\{0.50,0.70,1.10\}$ (weak, moderate, and strong signal). We observe $m=16$ or $64$ positions (10\% or 40\% of items) and use equal-sized or imbalanced groups, the latter with smallest assessor cluster and item block of 15 and 16 members. The $3\times2\times2$ design has eight independently generated data sets per setting. 

Each data set is fitted with two chains initialized at $(C,K)=(2,2)$ and $(7,9)$, each run for 4,000 iterations with 1,200 discarded and split--merge proposals using one exact restricted-allocation pass. We assess the 2,800 retained $C$ and $K$ values separately with rank-normalized split $\widehat R$ \citep{vehtari_gelman_simpson_2021,gelman_rubin_1992}; values near one indicate that both starts explore the same marginal distribution. A data set passes only if $\widehat R_C\leq1.01$ and $\widehat R_K\leq1.01$; in this case, the two independent chains are pooled. Bulk ESS is the approximate number of independent draws represented by the 5,600 correlated two-chain draws. In Table~\ref{tab:main-simulation-summary}, $\widehat R$ and ESS average all eight data sets in a setting, whereas recovery uses only those passing this screen.

The Co-clustering Adjusted Rand Index (CARI) compares estimated and true assessor--item cells \citep{robert_vasseur_brault_2020}: values near one indicate nearly perfect joint recovery and values near zero chance-level agreement. The Supplement reports side-specific ARIs, rare-group F1 scores, posterior count probabilities, and the full design.

\begin{figure}[htbp]
  \centering
  \includegraphics[width=\textwidth]{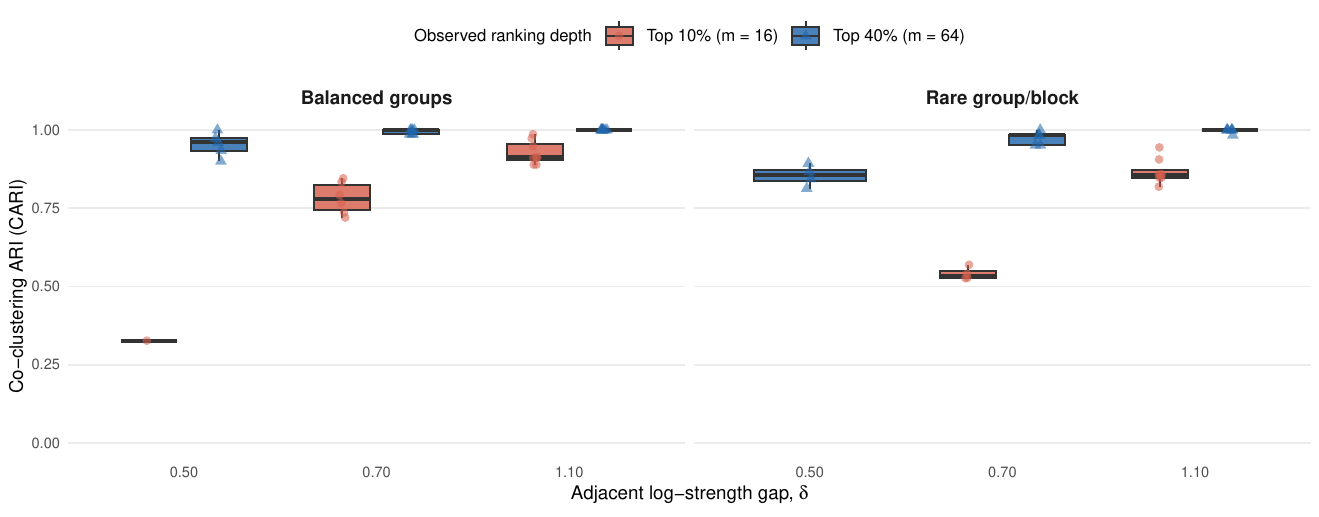}
  \caption{Joint partition recovery among data sets satisfying
  $\widehat R_C\leq1.01$ and $\widehat R_K\leq1.01$. Boxes summarize CARI
  across independently generated data sets, and points show individual data
  sets. A missing box means that none of the eight data sets in that condition
  produced a pair of chains that could be pooled.}
  \label{fig:sim-cari-main}
\end{figure}

\begin{table}[htbp]
  \centering
  \caption{Selected simulation results. ``Passed'' counts data sets with both $\widehat R_C\leq1.01$ and $\widehat R_K\leq1.01$. Mean CARI uses only passing data sets; mean rank-normalized $\widehat R$ and bulk ESS use all eight. Each ESS is based on 5,600 retained draws from two chains.}
  \label{tab:main-simulation-summary}
  \footnotesize
  \setlength{\tabcolsep}{4pt}
  \begin{tabular}{lrrrrrr}
    \toprule
    Regime & Passed & Mean CARI & Mean $\widehat R_C$ & Mean $\widehat R_K$ & Mean ESS$_C$ & Mean ESS$_K$ \\
    \midrule
    Weak, shallow, balanced & 1/8 & 0.33 & 1.070 & 1.105 & 146 & 86 \\
    Moderate, shallow, balanced & 6/8 & 0.78 & 1.002 & 1.006 & 1802 & 497 \\
    Moderate, deep, balanced & 7/8 & 0.99 & 1.004 & 1.003 & 1767 & 1864 \\
    Moderate, deep, unbalanced & 5/8 & 0.97 & 1.003 & 1.007 & 1408 & 1187 \\
    Strong, deep, balanced & 8/8 & 1.00 & 1.000 & 1.001 & 4926 & 3537 \\
    \bottomrule
  \end{tabular}
\end{table}

Figure~\ref{fig:sim-cari-main} and Table~\ref{tab:main-simulation-summary} show two clear patterns. Under balanced, moderate signal, observing 64 rather than 16 positions raises mean CARI from 0.78 to 0.99 among qualifying data sets and mean item-count ESS from 497 to 1,864. With strong signal and deep rankings, all eight data sets pass and mean CARI is 1.00. Overall, 90 of 128 data sets pass the common-budget diagnostic, including 31 of 32 strong-signal data sets. Only 1 of 16 weak-signal, shallow-ranking data sets passes, rising to 9 of 16 with deeper rankings. This weak regime is an information boundary: posterior mass remains spread over plausible partitions and occupied counts, so shallow rankings need not uniquely recover the generating partition.

A fixed-data check with three chains started at $(C,K)=(1,1)$, $(8,12)$, and $(20,30)$ gives overlapping 7,000-draw traces, $\widehat R_C=1.0004$, $\widehat R_K=1.0001$, and bulk ESS 3,640 and 8,249, supporting satisfactory occupied-count mixing under informative rankings.

\suppsectionref{sec:supp-simulation}{4} provides complete tables, side-specific recovery plots, within-block heterogeneity analysis, and additional convergence diagnostics.
\section{Application: TCGA pan-cancer gene expression rankings}\label{sec:tcga}

We consider the Pan-Cancer-12 data set previously analyzed by
\citet{ciriello2013emerging} and \citet{hoadley2014multiplatform}, comprising
12 diverse cancer types from The Cancer Genome Atlas. Preprocessed RNA
sequencing data were available for a subset of $L=2{,}617$ tumor samples and
were obtained from the Synapse repository\footnote{Synapse ID
\texttt{syn1715755}; DOI: \url{https://doi.org/10.7303/syn2468297}}.
For computational tractability, and to focus the analysis on genes with an
established relevance to cancer, we retained the expression variables
corresponding to genes affected by at least one of the selected functional
events identified by \citet{ciriello2013emerging}. Genes with more than
$50\%$ missing expression values were removed. The resulting data set
contains $n=1{,}247$ genes. Each sample contributes its top 500 ranked
positions, while all 1,247 genes remain in every Plackett--Luce choice set.

We fitted the following PL models:
\begin{enumerate}
  \item \textbf{Item-block-only model.} We set $C=1$, so the strength table is
        $1\times K$ (Eq.~\ref{eq:lik-item}). We use
        $\lambda_k\sim\Ga(50,\exp(\psi(50)))$ and a Gnedin process on
        $\mathbf x$ with $\gamma_x=0.99$, which favors a parsimonious number
        of gene blocks.
  \item \textbf{Assessor-cluster-only model.} There are no item blocks, so the
        strength table is $C\times n$ (Eq.~\ref{eq:lik-assessor}). We use the
        same Gamma prior on the cluster-specific strengths and a Gnedin process
        on $\mathbf w$ with $\gamma_w=0.5$. This is the PL analog of the
        Bayes Mallows sample-clustering model of
        \citet{vitelli2023transcriptomic}.
  \item \textbf{Joint PL-LBM.} This is the full model with both assessor
        clusters and item blocks, so the strength table is $C\times K$. We set
        $\gamma_x=0.99$ and $\gamma_w=0.5$. The key structural point is that
        the gene blocks are shared across assessor clusters: block composition
        is common to all clusters, and only the within-cluster block strengths
        change.
\end{enumerate}

\subsection{MCMC configuration.}
All three models above are run for $10{,}000$ Gibbs iterations with a $2{,}000$
iteration burn-in, retaining $8{,}000$ post-burn-in draws.
The archived fits are initialized with $C=3$ and $K=3$ where relevant for
the model being fitted.
Each joint-model Gibbs sweep is augmented with 5 split--merge
proposals for the assessor partition and 2 for the item partition. The
item-only fit uses 2 item proposals per sweep, while the assessor-only
fit uses one assessor proposal. Further details on this are specified in
\suppsectionref{sec:split-merge}{2} of the Supplementary Material. The joint model ran for
$40{,}458$~s; the assessor-only model took $17{,}563$~s.

Marginal posterior distributions and occupied-count traces are given in
\suppsectionref{sec:supp-tcga-results}{6} of the online Supplementary Material. Posterior summaries are
defined in \suppsectionref{sec:supp-identifiability}{3}.

\subsection{Predictive comparison}

Table~\ref{tab:tcga-waic-comparison} compares the three fitted models using WAIC \citep{vehtariPracticalBayesianModel2016}.
The calculation uses post-burn-in draws from each fit to compute the point-wise log-likelihood of 2,617 sample rankings. 
The joint model improves conditional repeat-ranking fit relative to the
assessor-cluster-only model, while the item-block-only model is less favorable.
Moreover, the joint model uses $19\times259=4{,}921$ PL strength parameters,
compared with $12\times1{,}247=14{,}964$ for the assessor-only fit. It therefore
combines improved conditional fit with an interpretable compression of the
gene dimension.

\begin{table}[htbp]
  \centering
  \caption{Sample-level conditional WAIC comparison for the three TCGA models. Lower WAIC
  is preferred; standard errors are computed over the 2,617 sample-level
  contributions.}
  \label{tab:tcga-waic-comparison}
  \begin{tabular}{lrrr}
    \toprule
    Model & WAIC & SE(WAIC) & $p_{\mathrm{WAIC}}$ \\
    \midrule
    Item-block-only & 18,001,773 & 1,894 & 1,226 \\
    Assessor-cluster-only & 16,978,028 & 11,111 & 8,111 \\
    Joint PL-LBM & 16,935,942 & 14,029 & 6,907 \\
    \bottomrule
  \end{tabular}
\end{table}

\subsection{Posterior distributions of \texorpdfstring{$C$}{C} and \texorpdfstring{$K$}{K} of the joint model}

\begin{figure}[htbp]
  \centering
  \includegraphics[width=0.62\textwidth]{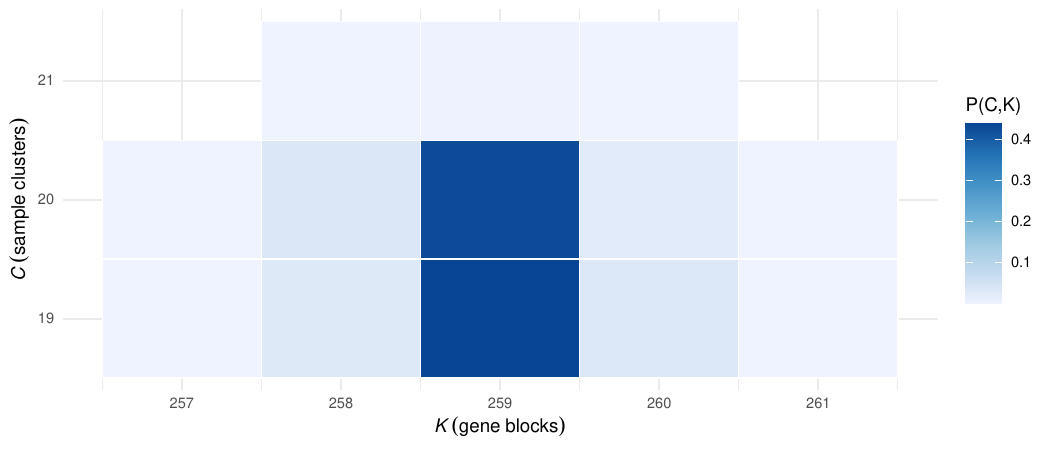}
  \caption{Joint posterior $P(C,K\mid\mathrm{data})$ for the TCGA joint
  PL-LBM, shown over the occupied-count pairs visited by the chain. Posterior
  mass is concentrated at $C\in\{19,20\}$ and $K=259$.}
  \label{fig:tcga-posterior-CK}
\end{figure}

We now turn to inspect the results specifically for the PL-LBM. Its posterior mass for $C$
lies almost entirely on 19 and 20, with $\PctCMode\%$ at its mode
$C_{\mathrm{mode}}=\CmodePost$. The posterior of $K$ spans $257$--$261$,
with $\PctKMode\%$ of post-burn-in draws at its mode
$K_{\mathrm{mode}}=\KmodePost$ and effective sample size $\EssK$; the item-side
split--merge acceptance rate is $\SmXAccept$. Only \PctExactKept\% of retained
draws have exactly the $19\times259$ reporting dimension (See Figure~\ref{fig:tcga-posterior-CK} for the joint posterior over the two
occupied counts). Thus the VI
partition ($\widehat C_{\mathrm{VI}}=\ChatSalso$,
$\widehat K_{\mathrm{VI}}=\KhatSalso$) is a useful snapshot of a concentrated, but non-degenerate,
posterior distribution that still quantifies uncertainty around the point estimate considered here.
The joint model has more sample clusters than the assessor-only PL model
($19$ rather than $12$), and both are reasonably aligned with the 16 clusters found by the Bayesian Mallows analysis of
\citet{vitelli2023transcriptomic}.

\subsection{Sample clusters and cancer-type composition}

In the 19-cluster VI reporting partition, most clusters are strongly aligned
with tissue of origin: KIRC, LAML, GBM, and three BRCA clusters are essentially
single-tissue clusters. At the same time, several clusters are cross-tissue,
notably a colorectal cluster (COAD with substantial READ) and mixed LUAD/LUSC
or HNSC/LUSC clusters. This pattern mirrors the findings from previous
pan-cancer studies \citep{Hoadley2018CellOrigin, vitelli2023transcriptomic}:
strong tissue-of-origin organization with a smaller number of cross-tissue
groupings. The full 19-sample-cluster summary table, including the strongest block for
each sample cluster, is given in \suppsectionref{sec:supp-tcga-results}{6} of the online Supplementary Material.

\begin{figure}[htbp]
  \centering
  \includegraphics[width=0.88\textwidth]{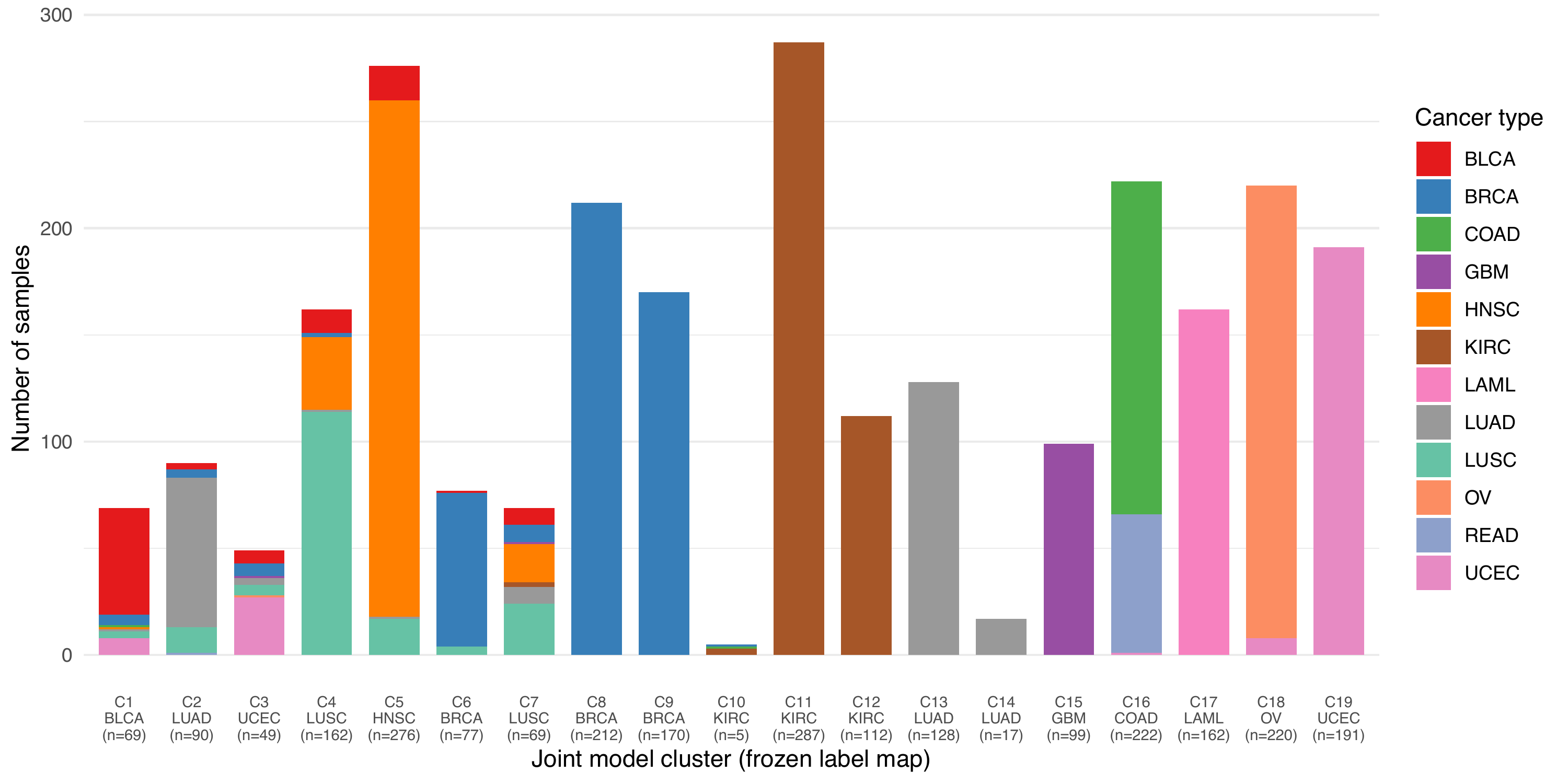}
  \caption{Composition of the 19 sample clusters in the VI reporting partition
  of the joint PL-LBM, shown as stacked barplots of cancer-type composition.
  Each cluster is annotated by its dominant cancer type and size.}
  \label{fig:tcga-cluster-composition}
\end{figure}

Figure~\ref{fig:tcga-cluster-composition} shows that samples cluster primarily
by organ site, consistent with the finding of \citet{vitelli2023transcriptomic} that
tissue of origin is the main driver of gene-expression rankings in the TCGA
pan-cancer data. LAML, GBM, and one UCEC cluster are single-tissue, while the
OV cluster is nearly pure. KIRC has two large pure clusters and one small mixed
cluster; LUAD has two pure clusters and one mixed LUAD/LUSC cluster; and BRCA
has two pure clusters and a third that is 94\% BRCA. COAD and READ cluster
together, as expected for colorectal cancers, while the mixed squamous
clusters connect HNSC and LUSC.

The $\widehat K_{\mathrm{VI}}=259$ gene blocks in this reporting partition
have median size 4, range 1--14, and 17 singleton blocks. Thus the block
representation is a substantial but non-uniform compression of the 1,247-gene
universe; small blocks retain gene-specific ranking patterns where the
posterior supports them. The shared-block structure also gives a direct
description of recurrent programmes: for example, the same 10-gene block
containing FGFR3, HRAS, GPC1, PARD6G, and SEMA4B is strongest for the C4 LUSC
and C5 HNSC clusters, while the five-gene FOXA1--GATA3--N4BP3--TBX3--ZNF552
block is strongest for both pure-BRCA clusters C8 and C9. Such sharing is
represented directly in the fitted model rather than reconstructed from
overlap among separate cluster-specific gene lists.

\subsection{Posterior latent strength matrix \texorpdfstring{$\Lambda$}{Lambda}}
\label{subsec:strength-heatmap}

The joint fit gives an ECR-aligned $\widehat C_{\mathrm{VI}}\times
\widehat K_{\mathrm{VI}}=19\times259$ conditional strength summary, computed
from the 3,513 of 8,000 retained draws having exactly these dimensions. 

\begin{figure}[htbp]
  \centering
  \includegraphics[width=0.78\textwidth]{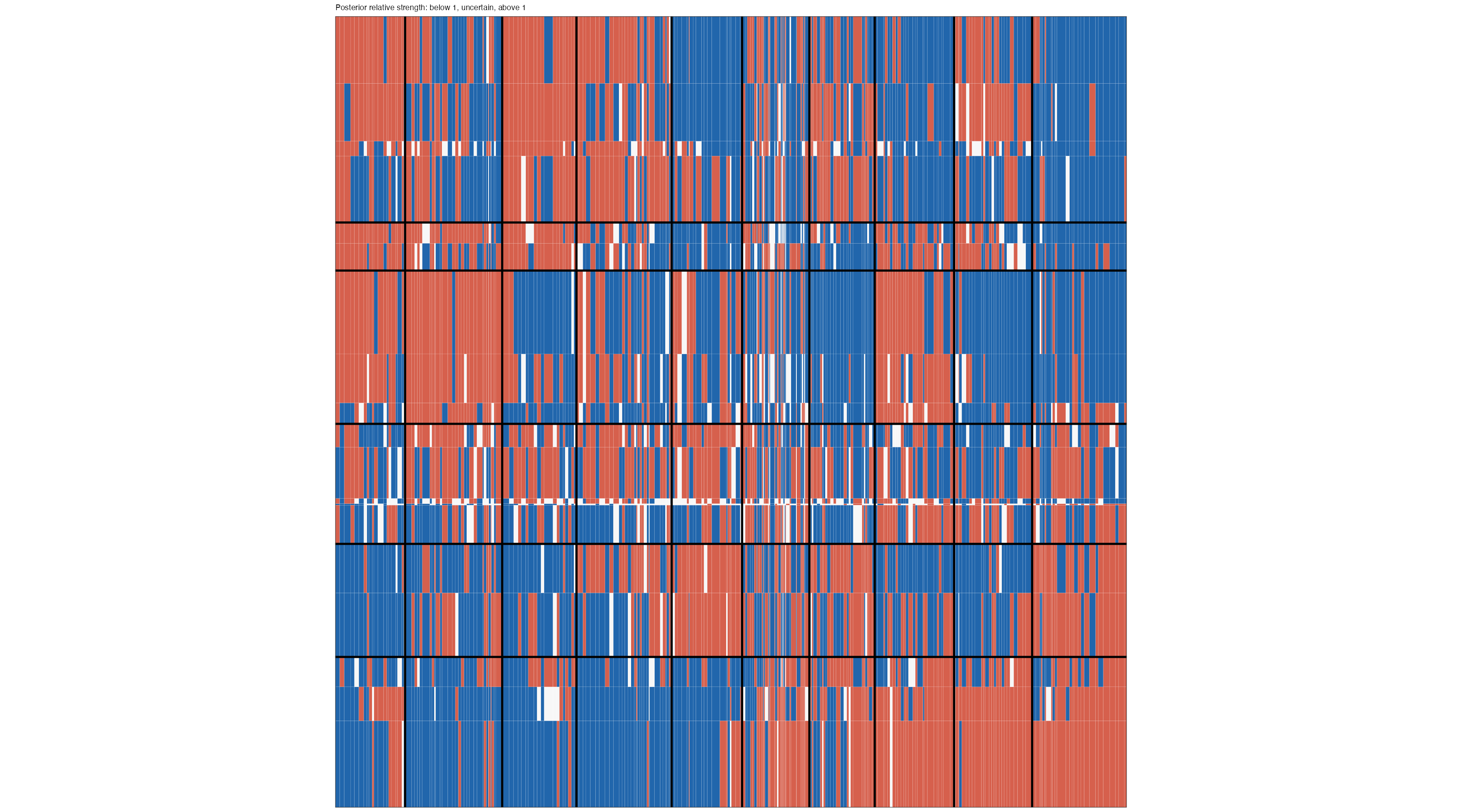}
  \caption{Posterior summary of $\widetilde\Lambda$ relative to the unit
  row-normalization reference, conditional on the $19\times259$ modal-frame
  VI partition. A cell is blue when
  $P(\widetilde\lambda_{ck}>1\mid\text{data}, C=19, K=259)\leq0.2$ and red
  when that probability is at least 0.8; otherwise it is white. Rows
  (assessor clusters) and columns (item blocks) are jointly ordered, and
  black lines mark the corresponding contiguous groups.}
  \label{fig:tcga-lambda-full}
\end{figure}

The identifiability normalization described in
Section~\ref{sec:identifiability} allows us to interpret the values of
$\widetilde\Lambda$ relative to the geometric mean within each assessor-cluster
row: an entry above 1 is above that row-wise reference, and an entry below 1
is below it \citep{newman_ranking_2022}. Figure~\ref{fig:tcga-lambda-full}
uses this same unit reference: conditional on the $19\times259$ reporting
frame, each cell summarizes the posterior probability that
$\widetilde\lambda_{ck}>1$. \suppsectionref{sec:supp-identifiability}{3} of the online Supplementary Material
gives the exact construction.

\subsection{GSEA from the joint-model gene partition}
\label{subsec:tcga-gene-enrichment}

We use gene-set enrichment analysis (GSEA) to relate the inferred sample
clusters to known biological pathways. For each sample cluster, the analysis assesses
whether genes from a pathway tend to receive higher model-based ranks than the
remaining genes.

The ordering is obtained from the full posterior distribution. In each of the
8,000 retained draws, genes inherit the midrank of their block after blocks are
ordered by their estimated strength within a sample cluster. For each VI sample
cluster, we then average the gene-level rankings of its member samples across
all retained draws. In a given draw, each
sample contributes the ranking associated with its sampled cluster assignment; gene-block labels are immaterial once strengths
have been expanded to gene-level ranks. The score used for GSEA consequently reflects uncertainty
in both the sample clustering and the gene partition, while treating genes in
the same block as tied. 

The shared gene partition makes the sample-cluster summaries directly comparable: a
gene block has the same membership in every sample cluster, but its prominence
can vary between clusters. This extends the rank-based pan-cancer analysis of
\citet{vitelli2023transcriptomic} by estimating common gene programmes and
their cluster-specific importance within one joint model.

\begin{figure}[htbp]
  \centering
  \includegraphics[width=0.9\textwidth]{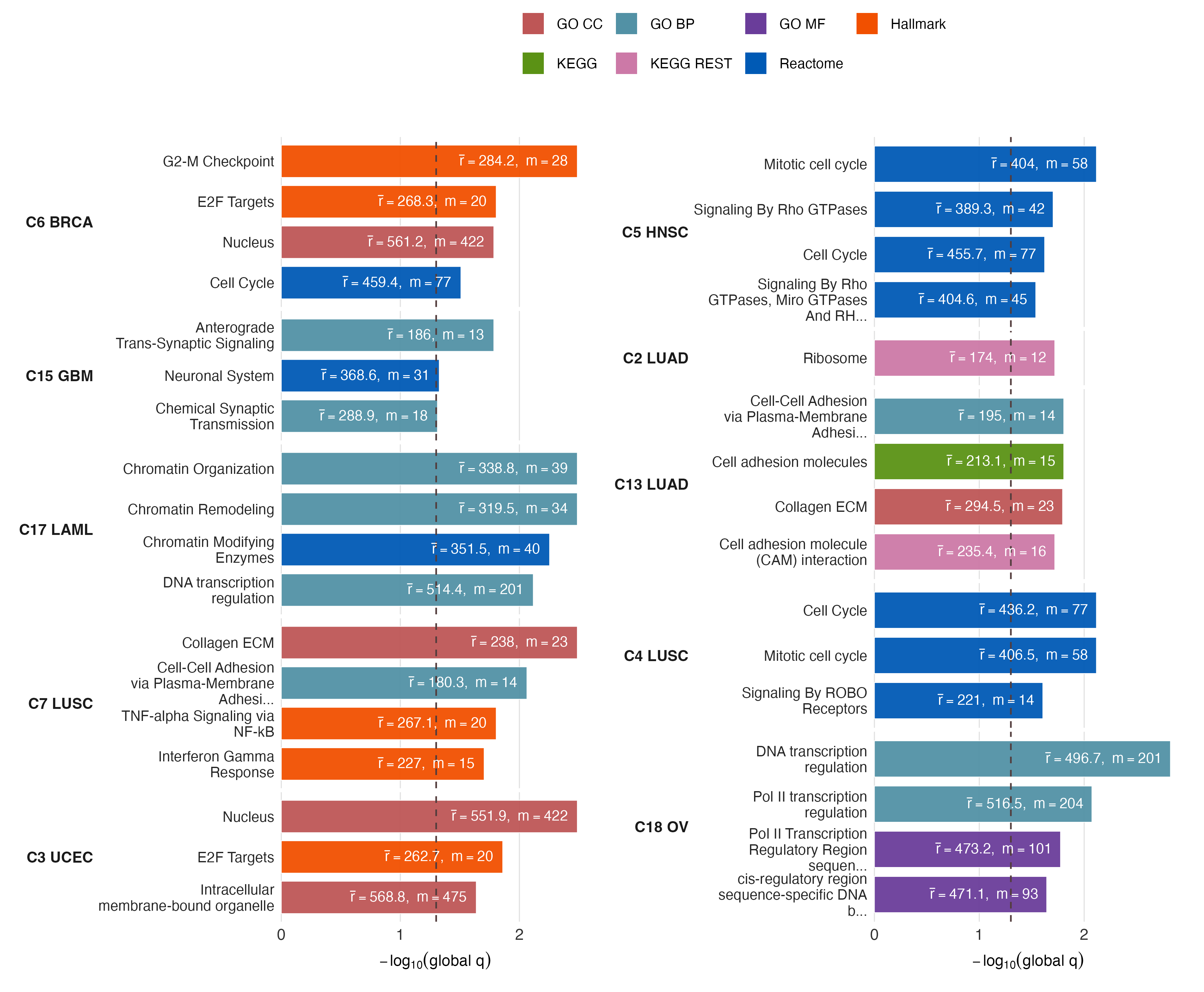}
  \caption{Rank-based GSEA results with global BH-adjusted $q<0.05$.
  Bars report $-\log_{10}(q)$, and the dashed vertical line marks $q=0.05$.
  Text labels give the mean posterior rank $\bar r$ of genes in the set and
  the number $m$ of set genes represented in the 1,247-gene analysis
  universe.}
  \label{fig:tcga-joint-toplist-barplot}
\end{figure}

For each sample cluster, we apply the one-sided rank-based competitive test in
\texttt{limma::geneSetTest} \citep{Ritchie2015}. The test compares the ranks
of genes in a set with those of all other analysed genes. We control the false
discovery rate with Benjamini--Hochberg adjustment across all sample-cluster--gene-set
pairs. Gene sets are mapped using HGNC symbols and aliases and are drawn from
the MSigDB Hallmark collection \citep{liberzon2015}, Gene Ontology
\citep{go2023}, Reactome \citep{reactome2022}, and KEGG \citep{kegg2021}.
Annotation releases and filtering are documented in \suppsectionref{sec:supp-enrichment}{7}. Of 67,944
tests, 50 have global $q<0.05$; Figure~\ref{fig:tcga-joint-toplist-barplot}
summarizes these results.

The smallest global $q$-value is for C18 OV and the broad GO term
\emph{regulation of DNA-templated transcription} (GO:0006355;
$\bar r=496.7$, $q=0.002$), which is most useful as a broad descriptive
signal. The analysis also distinguishes sample clusters with similar tissue labels:
C13 LUAD shows cell-adhesion and collagen extracellular-matrix signals,
whereas C2 LUAD shows ribosomal and interferon-gamma-response signals. C6
BRCA is enriched for cell-cycle, G2--M-checkpoint, and E2F-target annotations,
consistent with known molecular heterogeneity in BRCA
\citep{Parker2009Supervised}. The LAML and GBM clusters show chromatin-related
and neuronal-system signals, respectively.

\section{Discussion}

We proposed a Bayesian Plackett--Luce latent block model for ranked data that jointly models assessor clusters and item blocks. Items in the same block share a common strength within each assessor cluster. This gives a compact $C \times K$ representation of the ranking structure and provides a direct co-clustering formulation for ranked data. The simulation study tested the recovery performance as separation, ranking depth, and group balance varied. The sensitivity study further showed that assessor recovery can remain stable as within-block item variation increases. In the TCGA pan-cancer application, the posterior supports tissue-aligned sample structure together with a shared gene-block representation of recurrent and cluster-specific expression programmes.

The main limitation of the model is that a single item partition is shared by all assessor clusters, and items within a block have equal strength;
both assumptions can be restrictive when assessor clusters rank the items differently
or when substantial within-block heterogeneity remains. In low-information
settings, the Gnedin and Gamma priors can also have appreciable influence on
the occupied counts and relative strengths. The exact-frame TCGA strength
summary is conditional on the modal $19\times259$ dimensions, whereas the
enrichment analysis averages over all retained dimensions. The displayed VI
partitions and strength matrix are therefore reporting snapshots of a broader
posterior distribution, not substitutes for its partition uncertainty. Natural
extensions include partially shared item partitions and explicit within-block
variation, possibly using nonparametric partially exchangeable priors.

\vspace{1cm}

\begingroup
\normalsize
\setstretch{1}

\noindent\textbf{Funding.} This work was supported by Taighde Éireann --
Research Ireland, under grant numbers (18/CRT/6049) and (23/RC/13506) at
the Research Centre - Rinn Artificial Intelligence.

\noindent\textbf{Disclosure statement.} The authors report no competing
interests.

\noindent\textbf{Declaration of generative AI use.} OpenAI Codex was used for
language editing and code review. The authors remain responsible for all
scientific content, analyses, and conclusions.

\noindent\textbf{Data and computational materials availability.} TCGA source
data are publicly available from the Genomic Data Commons Data Portal. A snapshot of the code accompanies the submission and the reproduction support page is available at \href{https://github.com/laposanti/Reproducibility-Support-PL-LBM}{https://github.com/laposanti/Reproducibility-Support-PL-LBM}
\endgroup

\begingroup
\setstretch{1}
\bibliography{reference1.bib}
\endgroup

\clearpage
\section*{Supplementary Appendix}
\addcontentsline{toc}{section}{Supplementary Appendix}
\setcounter{section}{0}
\setcounter{equation}{0}
\setcounter{figure}{0}
\setcounter{table}{0}
\setcounter{algorithm}{0}
\renewcommand{\thesection}{S\arabic{section}}
\renewcommand{\thesubsection}{\thesection.\arabic{subsection}}
\renewcommand{\thesubsubsection}{\thesubsection.\arabic{subsubsection}}
\renewcommand{\theequation}{S.\arabic{equation}}
\renewcommand{\thefigure}{S\arabic{figure}}
\renewcommand{\thetable}{S\arabic{table}}
\renewcommand{\thealgorithm}{S\arabic{algorithm}}
\renewcommand{\theHsection}{S.\arabic{section}}
\providecommand{\theHalgorithm}{}
\renewcommand{\theHalgorithm}{S.\arabic{algorithm}}

\newcommand{\mainalgorithmref}[2]{%
  \ifdefined\ArxivCombined Algorithm~\ref{#1}\else Algorithm~#2 of the main text\fi}
\newcommand{\mainsectionref}[2]{%
  \ifdefined\ArxivCombined Section~\ref{#1}\else Section~#2 of the main text\fi}
\newcommand{\mainequationref}[2]{%
  \ifdefined\ArxivCombined Equation~\eqref{#1}\else Equation~#2 of the main text\fi}

This \ifdefined\ArxivCombined supplementary appendix\else online
Supplementary Material\fi{} provides the technical details supporting
the main manuscript. Sections~\ref{sec:full-conditionals}--\ref{sec:split-merge}
give the posterior-computation and split--merge derivations.
Sections~\ref{sec:supp-identifiability}--\ref{sec:supp-enrichment} record the identifiability,
posterior-summary, simulation, TCGA, and enrichment details needed to
interpret the reported analysis.

As in the main text, \emph{cluster} refers to an assessor (or sample) group
and \emph{block} to an item (or gene) group; standard terms such as
\emph{co-clustering} retain their usual two-way meaning.

\ifdefined\ArxivCombined\else
\tableofcontents
\clearpage
\fi

\section{Posterior computation details}\label{sec:full-conditionals}

This section collects the full conditional distributions used in the
Gibbs sampler described in \mainalgorithmref{alg:compact-mcmc}{1}. All
conditionals follow from the augmented posterior and the collapsed
model described in \mainsectionref{sec:computation}{4}. Throughout,
$\Ga(a,b)$ denotes a Gamma distribution with shape $a$ and rate $b$.

\subsection{Augmented likelihood}

For a ranking $\rho^{(\ell)}=(\rho_1^{(\ell)},\ldots,\rho_{m_\ell}^{(\ell)})$,
the Plackett--Luce likelihood can be augmented with independent exponential
variables $Z_{\ell t}$ using the identity
\[
  \frac{1}{\sum_{j\in A_t^{(\ell)}}\lambda_{w_\ell,x_j}}
  =
  \int_0^\infty
  \exp\!\left\{-z\sum_{j\in A_t^{(\ell)}}\lambda_{w_\ell,x_j}\right\}
  \,dz.
\]
The augmented joint density is proportional to
\[
  \prod_{\ell=1}^L\prod_{t=1}^{m_\ell}
  \lambda_{w_\ell,x_{\rho_t^{(\ell)}}}
  \exp\!\left\{
    -Z_{\ell t}\sum_{j\in A_t^{(\ell)}}\lambda_{w_\ell,x_j}
  \right\}.
\]
Integrating over $Z_{\ell t}$ recovers the original Plackett--Luce
denominator at each ranking stage. This result first appears in \citet{caronEfficientBayesianInference2012}.

\subsection{Collapsed posterior}

Let $w_{ck}$ count observed selections from item block $k$ made by assessors
in assessor cluster $c$, and let $S_{ck}$ denote the corresponding augmented exposure.
With independent $\lambda_{ck}\sim\Ga(a,b)$ priors, integration over
$\Lambda$ gives the collapsed factor
\[
  \prod_{c=1}^C\prod_{k=1}^K
  \frac{b^a}{\Gamma(a)}
  \frac{\Gamma(a+w_{ck})}{(b+S_{ck})^{a+w_{ck}}}.
\]
Combining this term with the two Gnedin partition priors gives the collapsed
target used for assessor and item allocation updates.

\subsection[Latent augmentation variables Z]{Latent augmentation variables $\mathbf Z$}

Given $(\mathbf w,\mathbf x,\Lambda)$, the latent variables are
conditionally independent. For assessor $\ell$ at stage $t$, the rate is
the total worth of all items still available in the choice set
$A_t^{(\ell)}$:
\[
  Z_{\ell t}\mid \mathbf w,\mathbf x,\Lambda,\boldsymbol\rho
  \sim
  \mathrm{Exp}\!\left(\sum_{j\in A_t^{(\ell)}} \lambda_{w_\ell,x_j}\right),
  \qquad \ell=1,\dots,L,\;\; t=1,\dots,m_\ell.
\]
Writing
\[
  c_{t,k}^{(\ell)} = \#\{j \in A_t^{(\ell)} : x_j = k\},
\]
the exponential rate simplifies to
\[
  \sum_{j\in A_t^{(\ell)}} \lambda_{w_\ell,x_j}
  =
  \sum_{k=1}^{K} c_{t,k}^{(\ell)}\,\lambda_{w_\ell,k}.
\]
For example, if the available set at step $t$ contains two items from
block $1$ and one item from block $3$, then
$c_{t,1}^{(\ell)}=2$, $c_{t,3}^{(\ell)}=1$, and the rate becomes
$2\lambda_{w_\ell,1}+\lambda_{w_\ell,3}$.

\subsection[Assessor allocations w (collapsed)]{Assessor allocations $\mathbf w$ (collapsed)}

For assessor $\ell$, define its item-block-wise contributions
\begin{align}
  w_{\ell k}
  &=
  \sum_{t=1}^{m_\ell}\ind\!\bigl(x_{\rho_t^{(\ell)}}=k\bigr),
  \label{eq:wlk-supp}\\
  S_{\ell k}
  &=
  \sum_{t=1}^{m_\ell} Z_{\ell t}\sum_{j\in A_t^{(\ell)}} \ind(x_j=k).
  \label{eq:Slk-supp}
\end{align}
Here $w_{\ell k}$ counts how many observed selections made by assessor
$\ell$ came from item block $k$, while $S_{\ell k}$ is the matching
exposure term built from the latent variables. As a simple example, if
assessor $\ell$ places items from blocks $(2,1,2)$ in the observed
positions, then $w_{\ell 1}=1$ and $w_{\ell 2}=2$. The quantity
$S_{\ell k}$ uses the same block labels, but it sums the relevant
$Z_{\ell t}$ contributions over the corresponding availability sets.

Let $w_{-\ell,ck}$ and $S_{-\ell,ck}$ denote the item-block sufficient
statistics with assessor $\ell$ removed. Let $C_{-\ell}$ be the number of
occupied assessor clusters after removing assessor $\ell$, and let
$n_{-\ell,c}^{(w)}$ be the size of cluster $c$ in $\mathbf w_{-\ell}$.

The collapsed conditional has two ingredients: a Gnedin predictive term
that depends on current cluster sizes, and an integrated likelihood term
that measures how well assessor $\ell$ fits the candidate cluster.

For an existing cluster $c=1,\dots,C_{-\ell}$:
\begin{align}
  p(w_\ell=c \mid \mathbf w_{-\ell},\mathbf x,\mathbf Z,\boldsymbol\rho)
  &\propto
  \underbrace{\bigl(n_{-\ell,c}^{(w)}+1\bigr)\,
  \bigl(L-1-C_{-\ell}+\gamma_w\bigr)}_{\text{Gnedin predictive}}
  \notag\\
  &\quad{}\times
  \prod_{k=1}^{K}
  \frac{\Gamma(a+w_{-\ell,ck}+w_{\ell k})}{\Gamma(a+w_{-\ell,ck})}
  \frac{(b+S_{-\ell,ck})^{a+w_{-\ell,ck}}}
  {(b+S_{-\ell,ck}+S_{\ell k})^{a+w_{-\ell,ck}+w_{\ell k}}}.
\end{align}

For a new cluster $c=C_{-\ell}+1$:
\begin{align}
  p(w_\ell=C_{-\ell}+1
  \mid \mathbf w_{-\ell},\mathbf x,\mathbf Z,\boldsymbol\rho)
  &\propto
  \underbrace{C_{-\ell}(C_{-\ell}-\gamma_w)}_{\text{Gnedin new}}
  \prod_{k=1}^{K}
  \frac{b^a}{\Gamma(a)}
  \frac{\Gamma(a+w_{\ell k})}{(b+S_{\ell k})^{a+w_{\ell k}}}.
\end{align}

\subsection[Item allocations x (collapsed)]{Item allocations $\mathbf x$ (collapsed)}

The item update is the same idea with the roles reversed. For item $i$,
define its contributions within each assessor cluster
\begin{align}
  w_{ic}
  &=
  \sum_{\ell=1}^{L}\ind(w_\ell=c)\sum_{t=1}^{m_\ell}\ind\!\bigl(\rho_t^{(\ell)}=i\bigr),
  \label{eq:wic-supp}\\
  S_{ic}
  &=
  \sum_{\ell=1}^{L}\ind(w_\ell=c)\sum_{t=1}^{m_\ell} Z_{\ell
  t}\ind\!\bigl(i\in A_t^{(\ell)}\bigr).
  \label{eq:Sic-supp}
\end{align}
Here $w_{ic}$ counts how many times item $i$ is selected by assessors in
cluster $c$, while $S_{ic}$ records the corresponding exposure of item
$i$ in the augmented representation. For example, if item $i$ is chosen
twice by assessors currently assigned to cluster $1$ and once by an assessor
in cluster $2$, then $w_{i1}=2$ and $w_{i2}=1$.

Let $w_{-i,ck}$ and $S_{-i,ck}$ denote the item-block sufficient
statistics with item $i$ removed. Let $K_{-i}$ be the number of occupied item
blocks after removing item $i$, and let $n_{-i,k}^{(x)}$ be the size of item block
$k$ in $\mathbf x_{-i}$.

For an existing block $k=1,\dots,K_{-i}$:
\begin{align}
  p(x_i=k \mid \mathbf x_{-i},\mathbf w,\mathbf Z,\boldsymbol\rho)
  &\propto
  \underbrace{\bigl(n_{-i,k}^{(x)}+1\bigr)\,
  \bigl(n-1-K_{-i}+\gamma_x\bigr)}_{\text{Gnedin predictive}}
  \notag\\
  &\quad{}\times
  \prod_{c=1}^{C}
  \frac{\Gamma(a+w_{-i,ck}+w_{ic})}{\Gamma(a+w_{-i,ck})}
  \frac{(b+S_{-i,ck})^{a+w_{-i,ck}}}
  {(b+S_{-i,ck}+S_{ic})^{a+w_{-i,ck}+w_{ic}}}.
\end{align}

For a new block $k=K_{-i}+1$:
\begin{align}
  p(x_i=K_{-i}+1
  \mid \mathbf x_{-i},\mathbf w,\mathbf Z,\boldsymbol\rho)
  &\propto
  \underbrace{K_{-i}(K_{-i}-\gamma_x)}_{\text{Gnedin new}}
  \prod_{c=1}^{C}
  \frac{b^a}{\Gamma(a)}
  \frac{\Gamma(a+w_{ic})}{(b+S_{ic})^{a+w_{ic}}}.
\end{align}

\subsection[Worth parameters Lambda]{Worth parameters $\Lambda$}

Once $(\mathbf w,\mathbf x,\mathbf Z,\boldsymbol\rho)$ are fixed, each
item-block worth updates independently:
\[
  \lambda_{ck}\mid \mathbf w,\mathbf x,\mathbf Z,\boldsymbol\rho
  \sim
  \Ga(a+w_{ck},\, b+S_{ck}),
  \qquad c=1,\dots,C,\;\; k=1,\dots,K,
\]
where
\begin{align}
  w_{ck}
  &=
  \sum_{\ell=1}^{L}\ind(w_\ell=c)\sum_{t=1}^{m_\ell}
  \ind\!\bigl(x_{\rho_t^{(\ell)}}=k\bigr),
  \\
  S_{ck}
  &=
  \sum_{\ell=1}^{L}\ind(w_\ell=c)\sum_{t=1}^{m_\ell}
  Z_{\ell t}\sum_{j\in A_t^{(\ell)}} \ind(x_j=k).
\end{align}

\paragraph{Remark on normalization.}
The Gamma law above is the exact full conditional for the
\emph{unconstrained} augmented strengths. After each Gamma update, the
sampler deterministically decomposes every row into its geometric mean
$s_c>0$ and the normalized strengths
$\tilde\lambda_{ck}=\lambda_{ck}/s_c$. It stores both components, so that
$s_c\tilde\lambda_{ck}$ exactly reconstructs the unconstrained draw used in
the next latent-time update. The normalization is therefore part of the
sampler parameterization, whereas the Gamma full conditional continues to
apply to the reconstructed unconstrained strengths.

\subsection{Computational complexity}

Let $m=\max_\ell m_\ell$. With cached availability counts and sufficient
statistics, updating the augmentation variables costs $O(LmK)$ per sweep.
A full assessor sweep costs $O(LCK)$, a full item sweep costs $O(nCK)$, and
refreshing $\Lambda$ costs $O(CK)$. The leading per-iteration cost is therefore
\[
  O(LmK + LCK + nCK + CK),
\]
plus the cost of any accepted or rejected split--merge proposals. The dominant
memory objects are the rankings, the allocation vectors, the $C\times K$
strength matrix, and cached item-block-availability/sufficient-statistic arrays.

\section{Split--merge updates}\label{sec:split-merge}

Local allocation sweeps change one label at a time. They can therefore be slow
to separate a heterogeneous occupied group or to combine two redundant groups.
The split--merge update makes a coordinated partition move: it either divides
one occupied group into two or combines two occupied groups into one. It is a
Metropolis--Hastings adaptation of the restricted-Gibbs construction of
\citet{jain_neal_2004}, using the collapsed Plackett--Luce item-block likelihood
and the finite-type Gnedin partition prior \citep{gnedin_species_2010}.

Each proposal has four steps. First, we select two anchors: anchors in the
same group initiate a split, while anchors in different groups initiate a
merge. Second, a restricted sequential scan constructs the proposed allocation
of the affected entities. Third, after integrating out the affected item-block
strengths, we compare the likelihood and Gnedin-prior contributions of the two
partitions. Finally, the Metropolis--Hastings correction accounts for the
probability of making, and reversing, the proposal. The proposal may use the
current augmented-data statistics to seek more useful anchors, but the exact
forward and reverse probabilities ensure that this changes efficiency only,
not the posterior target.

We describe both assessor and item moves with generic notation. Write
$\mathbf y=(y_1,\ldots,y_N)$ for a partition of $N$ entities into $H$ occupied
groups, with $n_h=\#\{r:y_r=h\}$. For an assessor move,
$\mathbf y=\mathbf w$, $N=L$, $H=C$, the Gnedin parameter is $\gamma_w$, and
the profile has one coordinate per item block. For an item move,
$\mathbf y=\mathbf x$, $N=n$, $H=K$, the Gnedin parameter is $\gamma_x$, and
the profile has one coordinate per assessor cluster. We use $d$ for this
generic coordinate.

Let $\mathbf y$ be the current partition and $\mathbf y^\star$ the proposed
one. Let $p_{\mathrm{SM}}(\mathbf y\mid\mathcal R)$ be the full conditional
for the partition being updated after the latent strengths have been
integrated out. The remaining state is
$\mathcal R=(\mathbf x,\mathbf Z,\boldsymbol\rho)$ for an assessor move and
$\mathcal R=(\mathbf w,\mathbf Z,\boldsymbol\rho)$ for an item move; fixed
hyperparameters are suppressed. If $T(\mathbf y^\star\mid\mathbf y)$ denotes
the proposal probability, the move is accepted with probability
\begin{equation}\label{eq:supp-sm-mh}
  \alpha(\mathbf y,\mathbf y^\star)
  =
  \min\left\{
  1,\,
  \frac{p_{\mathrm{SM}}(\mathbf y^\star\mid\mathcal R)}
       {p_{\mathrm{SM}}(\mathbf y\mid\mathcal R)}
  \frac{T(\mathbf y\mid\mathbf y^\star)}
       {T(\mathbf y^\star\mid\mathbf y)}
  \right\}.
\end{equation}
The first ratio asks whether the candidate partition better explains the
current augmented data and accords with the partition prior. The second ratio
corrects for any asymmetry in the proposal. The direction
$\mathbf y\to\mathbf y^\star$ is the attempted move; the opposite direction
appears only in this correction.

Collapsing the strength parameters makes the target comparison both simpler
and local: a split or merge changes only one or two groups. For a subset $A$
of entities, let $w_{Ad}$ and $S_{Ad}$ be its aggregated count and exposure
statistics in coordinate $d$. Integrating the corresponding strength
$\lambda_d$ against its $\Ga(a,b)$ prior gives
\[
  \int_0^\infty
  \frac{b^a}{\Gamma(a)}
  \lambda_d^{a+w_{Ad}-1}\exp\{-(b+S_{Ad})\lambda_d\}\,d\lambda_d
  =
  \frac{b^a}{\Gamma(a)}
  \frac{\Gamma(a+w_{Ad})}{(b+S_{Ad})^{a+w_{Ad}}}.
\]
Multiplying this contribution over columns gives the collapsed, or marginal, full conditional of
treating $A$ as one group:
\begin{equation}\label{eq:supp-sm-marginal}
  \log \mathcal M(A)
  =
  \sum_d
  \left\{
    a\log b-\log\Gamma(a)
    +\log\Gamma(a+w_{Ad})
    -(a+w_{Ad})\log(b+S_{Ad})
  \right\}.
\end{equation}
All unchanged groups cancel from the first ratio in
\eqref{eq:supp-sm-mh}. The restricted scan below therefore needs only the
incremental score
$\log\mathcal M(A\cup\{r\})-\log\mathcal M(A)$ when considering where to
place one entity.

\subsection{Anchor pairs}

Anchors determine both the direction and the starting point of a global move.
Each attempt selects an ordered pair of distinct entities $(i,j)$. If they are
in the same group, they seed two proposed daughter groups and the move is a
split; if they are in different groups, their groups are the candidates for a
merge. A useful split pair has contrasting ranking evidence, whereas a useful
merge pair comes from groups that can plausibly share one strength profile.
These are only proposal-design criteria: they do not add a modelling
assumption or biological interpretation.

We compare three ways to choose anchors. The \emph{uniform proposal} is the
neutral baseline: every ordered pair of distinct entities has probability
\[
  q_{\mathrm{pair}}(\mathbf y;i,j)=\{N(N-1)\}^{-1}.
\]
It can nevertheless attempt substantially more of one move type than the
other. The \emph{balanced proposal} first chooses between a split and a merge,
assigning probability $1/2$ to each available type (and probability one when
only one is possible), and then samples uniformly within that type. Writing
\[
  N_{\mathrm{same}}(\mathbf y)
  =\sum_{h=1}^{H} n_h(n_h-1),
  \qquad
  N_{\mathrm{diff}}(\mathbf y)
  =N(N-1)-N_{\mathrm{same}}(\mathbf y),
\]
the balanced proposal probability is
\[
  q_{\mathrm{pair}}(\mathbf y;i,j)
  =
  \begin{cases}
  p_{\mathrm{split}}/N_{\mathrm{same}}(\mathbf y),
    & y_i=y_j,\\
  p_{\mathrm{merge}}/N_{\mathrm{diff}}(\mathbf y),
    & y_i\ne y_j,
  \end{cases}
\]
where $p_{\mathrm{split}}$ and $p_{\mathrm{merge}}$ are $1/2$ when both move
types are possible and one otherwise.

Whatever anchor rule is used, its probability must be recorded because the
reverse move may be easier or harder to propose than the forward move. The
pair-selection contribution to the proposal ratio in \eqref{eq:supp-sm-mh} is
\[
  \frac{q_{\mathrm{pair}}(\mathbf y^\star;i,j)}
       {q_{\mathrm{pair}}(\mathbf y;i,j)}.
\]
The denominator is the chance of selecting $(i,j)$ in the current partition;
the numerator is the chance of selecting the same ordered pair after the move,
when attempting to reverse it. They are equal for uniform anchors. They can
differ for balanced anchors because a split or merge changes the number of
same-group and different-group pairs.

The third rule, the \emph{data-informed proposal}, retains the balanced choice
of move type but targets anchors within the chosen direction. It uses only the
current augmented-data count and exposure statistics $(w_{rd},S_{rd})$;
``data-informed'' therefore does not mean that external covariates or an
additional likelihood are introduced. The split and merge scores are
deliberately different: before a split there are no daughter groups to score,
so we look for separated entity profiles; before a merge the two groups are
fully specified, so we score their proposed union directly. All resulting,
state-dependent probabilities are evaluated in both directions. The targeting
therefore changes where the sampler proposes moves, not the posterior it
targets.

This is an adaptation of the Jain--Neal restricted split--merge construction
\citep{jain_neal_2004}; related proposal-guided merge--split methods are
discussed by
\citet{bouchard_cote_particle_gibbs_2015,peixoto_merge_split_2020,luo_shrivastava_2019}.
The particular scores and tuning constants below are implementation choices,
not a separately named algorithm.

For a split, we need a measure of whether two entities have different
\emph{patterns} across the current columns, rather than whether one column
simply has larger numerical variation than another. We therefore construct a
robustly standardised strength profile and compare profiles on this common
scale. For entity $r$ and current column $d$, define
\[
  u_{rd}=\log\frac{a+w_{rd}}{b+S_{rd}}.
\]
This is the log of the current Gamma-posterior mean strength. In an item
update, $r$ is an item and $d$ indexes assessor clusters; in an assessor
update, $r$ is an assessor and $d$ indexes item blocks.

Let $u_{\cdot d}$ denote the vector of these values across all entities in
column $d$. We put every profile coordinate on a common robust scale using
\[
  \widetilde u_{rd}
  =
  \frac{u_{rd}-\operatorname{median}(u_{\cdot d})}
       {\operatorname{MAD}(u_{\cdot d})},
  \qquad
  \operatorname{MAD}(u_{\cdot d})
  =
  \operatorname{median}
  \left\lvert u_{\cdot d}-\operatorname{median}(u_{\cdot d})\right\rvert.
\]
The MAD is the median absolute deviation: a robust measure of the typical
spread around the column median. If its value is zero, the implementation uses
the column standard deviation instead, and uses one only if that too is zero.
This is a robust z-score standardisation: the median supplies the origin and
the MAD supplies the unit of measurement. Centering alone cancels from a
pairwise Euclidean distance; the division by the MAD is what prevents a
naturally more variable column from dominating the comparison.

Let $D$ be the number of columns. For two candidate anchors $r$ and $s$, define
\[
  \delta_{rs}
  =\left\{D^{-1}\sum_{d=1}^{D}
      (\widetilde u_{rd}-\widetilde u_{sd})^2\right\}^{1/2},
  \qquad
  d_{rs}
  =\frac{\delta_{rs}}
      {\operatorname{median}_{p<q}(\delta_{pq})},
\]
where the final median is over all unordered pairs in the current partition.
Thus, $\delta_{rs}$ is the root-mean-square Euclidean distance between two
standardised profiles, and $d_{rs}$ is that distance relative to a typical
pair. The final rescaling makes the median pairwise distance one.

For example, in an item update for the TCGA analysis, $r$ and $s$ are genes
and $d$ indexes the current sample clusters. Two genes currently in the same
item block are useful split anchors when their profiles differ: for instance,
one may be selected relatively often by samples in a lung-tumour cluster and
relatively rarely by samples in a breast-tumour cluster, with the reverse
pattern for the other gene. In an assessor update, $r$ and $s$ are tumour
samples and $d$ indexes item blocks. Two samples that respectively give high
relative prominence to an immune-related block and to a cell-cycle block are
analogously plausible split anchors.

Equivalently, each column receives its own ruler: zero is its median and one
unit is its typical spread. Conditional on a split, we spend more proposals on
same-group pairs with larger profile distance. An unordered pair $(r,s)$
therefore receives weight
\[
  \exp\{\min(\kappa_{\mathrm{s}}d_{rs},B)\},
  \qquad B=40.
\]
For a merge, the candidate groups are already known. We can consequently use
the collapsed log target change $\Delta_{hh'}$ from
Section~\ref{subsec:supp-target-ratios}, including both the likelihood and
Gnedin-prior terms, and assign group pair $(h,h')$ weight
$\exp(\kappa_{\mathrm{m}}\Delta_{hh'})$. In either case, the selected
unordered pair is ordered at random; for a merge, one member of each selected
group is sampled uniformly. The normalising constants, member-selection
probabilities, and ordering probability are all retained for the forward and
reverse proposal probabilities in \eqref{eq:supp-sm-mh}.

The tuning choices regulate how strongly the proposal concentrates on the
apparently most promising anchors. The non-negative split scale
$\kappa_{\mathrm{s}}$ controls the preference for separated profiles:
$\kappa_{\mathrm{s}}=0$ is uniform among eligible split pairs, whereas a
larger value concentrates mass on the most separated pairs. Similarly,
$\kappa_{\mathrm{m}}$ tempers the preference for a merge with a favourable
collapsed target change. We used $\kappa_{\mathrm{s}}=2$ and
$\kappa_{\mathrm{m}}=0.5$ throughout the reported experiments. These are
proposal-tuning choices, not model parameters or universal defaults. A small
matched-state calibration guided the moderate targeting: at the fixed middle
TCGA state, 7 of 300 attempts were accepted with $(2,0.5)$, compared with 3
with $(1,0.5)$ and none with balanced anchors. This diagnostic helps choose a
working proposal but cannot establish improved global mixing or rank settings
generally.

The cap $B=40$ is a separate safeguard on the \emph{log} split weight. It
limits the influence of an exceptionally distant pair, so that the anchor
distribution does not become effectively deterministic; the largest possible
ratio of unnormalised split weights is $\exp(40)$. With
$\kappa_{\mathrm{s}}=2$, the cap matters only when $d_{rs}\geq20$, meaning a
pair is at least 20 times the median profile distance after standardisation.
It was inactive in all three fixed-state TCGA checks below: the largest
within-group distances were 1.99, 1.92, and 3.02 in the low-, middle-, and
high-$K$ states. Thus any cap above 6.1 would yield exactly the same
split-anchor probabilities in those checks. If the cap is reached in another
application, lowering $B$ makes the proposal more diffuse and raising it makes
it more concentrated. Either choice leaves the posterior target unchanged
because the exact proposal probability remains in \eqref{eq:supp-sm-mh}. We
did not conduct a separate full-chain sensitivity analysis for $B$, because it
was inactive in the calibration and reported fixed-state comparisons.

Table~\ref{tab:supp-informed-acceptance} asks the deliberately limited
question of whether targeting the anchors can produce acceptable item moves at
the same fixed TCGA state. Every attempt starts from the same partition and
the same augmented waiting times, and an accepted move is discarded before the
next attempt. Uniform and balanced anchors did not accept a move in these
tests, whereas the data-informed rule accepted moves at the middle and high
occupied-item-block states. This is a fixed-state proposal diagnostic, not evidence
that a full chain mixes across posterior modes or that the pattern generalises
beyond these states.

\begin{table}[htbp]
  \centering
  \caption{Matched-state acceptance of item split--merge proposals on the
  full TCGA top-500 data. Entries are accepted/attempted moves, with rates in
  parentheses; the final column is the percentage-point improvement over
  either baseline, which coincide in these tests. The likelihood-informed rule uses $\kappa_{\mathrm{s}}=2$ and
  $\kappa_{\mathrm{m}}=0.5$. Each row fixes the partition and one draw of the
  augmented waiting times throughout the comparison.}
  \label{tab:supp-informed-acceptance}
  \small
  \begin{tabular}{lrrrrrr}
    \toprule
    State & $K$ & Attempts & Uniform & Balanced & Data-informed & Gain (pp) \\
    \midrule
    Low  & 211 & 200 & 0/200 (0\%) & 0/200 (0\%) & 0/200 (0\%) & 0.00 \\
    Middle & 229 & 300 & 0/300 (0\%) & 0/300 (0\%) & 7/300 (2.33\%) & 2.33 \\
    High & 248 & 200 & 0/200 (0\%) & 0/200 (0\%) & 21/200 (10.50\%) & 10.50 \\
    \midrule
    Pooled & & 700 & 0/700 (0\%) & 0/700 (0\%) & 28/700 (4.00\%) & 4.00 \\
    \bottomrule
  \end{tabular}
\end{table}

The zero in the low-state row does not mean that split--merge moves are
unavailable. In a separate 2,000-attempt diagnostic at this same state, the
balanced direction choice generated 1,025 splits and 975 merges. Three splits
were accepted (0.293\%) and no merges were accepted. Typical merges were still
closer to acceptance: their median log acceptance ratio was $-9.17$, compared
with $-138.30$ for splits. Although the median collapsed target change for a
merge was only $-0.81$, selecting its reverse split anchors imposed a median
log penalty of $-8.36$, yielding a mean acceptance probability of 0.020\%.
The rare accepted low-state moves therefore arose from an unusually favourable
upper tail of split proposals. At the middle and high states in
Table~\ref{tab:supp-informed-acceptance}, merges were more effective than
splits (5/146 versus 2/154, and 21/106 versus 0/94, respectively). These
fixed-state diagnostics are consistent with the proposal finding compatible
merges in a fine partition while retaining a route to occasional splits in a
coarser partition. Demonstrating cross-mode mixing instead requires the
separate multi-start chain diagnostics reported below.

\subsection{Restricted scan}

The anchors identify the groups that may change, but they do not yet specify a
candidate partition. The restricted scan constructs that candidate while
leaving every other group fixed. It uses exactly one recorded sequential pass
so that the proposal probability is available for the Metropolis--Hastings
correction; an unrecorded warm-up pass would make that probability unknown.

For a split, only the other members of the anchor group can move. Let
\[
  \mathcal S=\{r:y_r=y_i\}\setminus\{i,j\}.
\]
The anchors initialise temporary groups $A$ and $B$. We visit each entity in
$\mathcal S$ once and assign it to one of them. The data-informed rule visits
first the entities for which the two anchors give most different profile
distances, $|d_{ri}-d_{rj}|$; the uniform and balanced rules use index order.
The visit order and every allocation probability are part of the proposal.

At each visit, the allocation rule should favour both a larger temporary group
under the Gnedin predictive weight and the group that better accommodates the
entity's collapsed likelihood contribution. If $A$ and $B$ contain the anchors
and all earlier allocations, define
\begin{align*}
  g_A(r)
  &=
  (|A|+1)\,
  \exp\{\log\mathcal M(A\cup\{r\})-\log\mathcal M(A)\},\\
  g_B(r)
  &=
  (|B|+1)\,
  \exp\{\log\mathcal M(B\cup\{r\})-\log\mathcal M(B)\}.
\end{align*}
The scan assigns $r$ to $A$ with probability
\begin{equation}\label{eq:supp-sm-restricted-prob}
  p_A(r)=\frac{g_A(r)}{g_A(r)+g_B(r)}.
\end{equation}
The probability of assigning $r$ to $B$ is $1-p_A(r)$. The factors
$|A|+1$ and $|B|+1$ are Gnedin predictive weights used to make a useful
proposal; the exact Gnedin prior is accounted for separately in the target
ratio below. Recording the selected side at every visit gives the scan
probability: an allocation to $A$ contributes $p_A(r)$ and an allocation to
$B$ contributes $1-p_A(r)$. Thus
\[
  \log q_{\mathrm{scan}}
  =
  \sum_{r\in\mathcal S}
  \log p\{r\hbox{ is assigned to the group selected in the scan}\}.
\]

For a merge, the candidate partition is deterministic once the anchors are
chosen: all members of their two groups are combined. There is no forward scan,
but the reverse move is a split. Its proposal probability is therefore the
probability that the same restricted scan would restore every entity to its
pre-merge side. This reverse-scan probability is the $q_{\mathrm{scan}}$ term
used in the merge acceptance ratio below.

\subsection{Target ratios}\label{subsec:supp-target-ratios}

Once the scan has produced a candidate, the first Metropolis--Hastings ratio
in \eqref{eq:supp-sm-mh} asks whether the split or merge improves the
collapsed posterior target. Only the affected groups need be compared. Let
$A$ and $B$ denote the two groups on the split side of the move: in a split,
they are the proposed daughter groups; in a merge, they are the current
groups. The likelihood comparison follows immediately from $\mathcal M$.
For a split, one group $A\cup B$ is replaced by $A$ and $B$, so
\[
  \log R_{\mathrm{lik}}^{\mathrm{split}}
  =
  \log\mathcal M(A)+\log\mathcal M(B)-\log\mathcal M(A\cup B).
\]
For a merge, the same comparison is reversed:
\[
  \log R_{\mathrm{lik}}^{\mathrm{merge}}
  =
  \log\mathcal M(A\cup B)-\log\mathcal M(A)-\log\mathcal M(B).
\]

The partition-prior contribution is obtained by evaluating the Gnedin EPPF in
\mainequationref{eq:gnedin-eppf-corrected}{(8)} at the current and proposed
partitions. We denote the resulting prior ratios for a split and merge by
$R_{\GN}^{\mathrm{split}}$ and $R_{\GN}^{\mathrm{merge}}$, respectively.
Together with the likelihood ratios above, these give the collapsed target
ratio
$p_{\mathrm{SM}}(\mathbf y^\star\mid\mathcal R)/
p_{\mathrm{SM}}(\mathbf y\mid\mathcal R)$ in \eqref{eq:supp-sm-mh}.

\subsection{Acceptance probabilities}

The acceptance ratio combines the collapsed target comparison with the cost of
constructing the candidate partition. A split uses the stochastic restricted
scan in the forward direction, so its scan probability appears as a penalty.
The pair-selection correction is always the reverse anchor probability minus
the forward anchor probability. Thus the log acceptance ratio for a split is
\begin{equation}\label{eq:supp-sm-log-accept-split}
  \log\alpha_{\mathrm{split}}
  =
  \log R_{\mathrm{lik}}^{\mathrm{split}}
  +\log R_{\GN}^{\mathrm{split}}
  -\log q_{\mathrm{scan}}
  +\log q_{\mathrm{pair}}(\mathbf y^\star;i,j)
  -\log q_{\mathrm{pair}}(\mathbf y;i,j).
\end{equation}
For a merge, the proposed allocation is deterministic but the reverse move is
a split. The scan probability therefore appears with the opposite sign:
\begin{equation}\label{eq:supp-sm-log-accept-merge}
  \log\alpha_{\mathrm{merge}}
  =
  \log R_{\mathrm{lik}}^{\mathrm{merge}}
  +\log R_{\GN}^{\mathrm{merge}}
  +\log q_{\mathrm{scan}}
  +\log q_{\mathrm{pair}}(\mathbf y^\star;i,j)
  -\log q_{\mathrm{pair}}(\mathbf y;i,j).
\end{equation}
The proposed move is accepted with probability
$\min\{1,\exp(\log\alpha)\}$. These two expressions differ only in which
direction contains the stochastic scan.

\subsection{Placement in the sampler}

The split--merge kernel supplements rather than replaces the local collapsed
Gibbs sweeps. Local sweeps efficiently refine allocations one entity at a time;
the occasional split--merge step supplies a coordinated route to a different
number of occupied groups. The algorithm below shows their order within one
iteration. We write $\mathbf s$ for the removed row geometric means, so that
$\Lambda=\operatorname{diag}(\mathbf s)\widetilde\Lambda$.

\begin{algorithm}[H]
\caption{Joint PL-LBM sampler with split--merge moves}
\label{alg:supp-gibbs}
\begin{spacing}{1.0}
\begin{algorithmic}[1]
\Require Rankings $\boldsymbol\rho$, hyperparameters
$(a,b,\gamma_w,\gamma_x)$, split--merge counts
$n_{\mathrm{SM}}^{(w)},n_{\mathrm{SM}}^{(x)}$, and an initial state
$(\mathbf w,\mathbf x,\widetilde\Lambda,\mathbf s)$
\Ensure Posterior draws of $(\mathbf w,\mathbf x,\widetilde\Lambda)$
\For{$t=1,\ldots,T$}
  \State Set $\Lambda=\operatorname{diag}(\mathbf s)\widetilde\Lambda$ and
  sample the exponential auxiliaries $\mathbf Z$ from their full conditionals
  \For{each configured assessor-allocation sweep}
    \State Sweep $\mathbf w$ using collapsed Gibbs probabilities
    \State Draw $\Lambda$ from its Gamma full conditional and set
    $(\widetilde\Lambda,\mathbf s)$ by row normalization
  \EndFor
  \For{$r=1,\ldots,n_{\mathrm{SM}}^{(w)}$}
    \State Propose a collapsed split--merge update of $\mathbf w$
    \State If accepted, draw a dimension-compatible $\Lambda$ from its Gamma
    full conditional and set $(\widetilde\Lambda,\mathbf s)$
  \EndFor
  \For{each configured item-allocation sweep}
    \State Sweep $\mathbf x$ using collapsed Gibbs probabilities
    \State Draw $\Lambda$ from its Gamma full conditional and set
    $(\widetilde\Lambda,\mathbf s)$ by row normalization
  \EndFor
  \If{$K$ is not fixed}
    \For{$r=1,\ldots,n_{\mathrm{SM}}^{(x)}$}
      \State Propose a collapsed split--merge update of $\mathbf x$
      \State If accepted, draw a dimension-compatible $\Lambda$ from its Gamma
      full conditional and set $(\widetilde\Lambda,\mathbf s)$
    \EndFor
  \EndIf
  \State Store $(\mathbf w,\mathbf x,\widetilde\Lambda,\mathbf s)$
\EndFor
\end{algorithmic}
\end{spacing}
\end{algorithm}

Rejected split--merge proposals leave both the labels and strength state
unchanged. An accepted move changes the dimension and sufficient statistics of
the strength matrix, so it is immediately followed by a dimension-compatible
Gamma refresh. We row-normalise every refreshed draw for storage and retain
the positive row scales in $\mathbf s$; their product reconstructs the
unnormalised $\Lambda$ used to sample $\mathbf Z$ at the next iteration. These
conditional refreshes are ordinary Gibbs transitions for the same augmented
target and hence do not alter the stationary posterior.

\section{Identifiability and posterior summaries}\label{sec:supp-identifiability}

The main text introduces the row normalization in
\mainsectionref{sec:identifiability}{4.3}. The following discussion records
its role in the sampler and in posterior summaries.

At any ranking stage for an assessor in cluster $c$, multiplying the row by
$\alpha_c>0$ gives
\[
  \frac{\alpha_c\lambda_{c,x_{\rho_t^{(\ell)}}}}
       {\sum_{j\in A_t^{(\ell)}}\alpha_c\lambda_{c,x_j}}
  =
  \frac{\lambda_{c,x_{\rho_t^{(\ell)}}}}
       {\sum_{j\in A_t^{(\ell)}}\lambda_{c,x_j}}.
\]
For fixed partitions, the likelihood consequently depends on contrasts
between log strengths. When the item-block comparison graph is connected
within cluster $c$, the rankings identify those contrasts and leave the
row's common level free.

The unit-geometric-mean transformation in
\mainequationref{eq:norm-geomean-rewrite}{(27)} selects one symmetric
representative of each row. In log-strength coordinates, it centers the row at
zero and treats all blocks symmetrically. This is convenient because block
labels are exchangeable and split--merge moves alter block membership. For
example, $(2,8,32)$ and $(1/4,1,4)$ yield the same Plackett--Luce
probabilities. In the normalized row, the middle block equals the
geometric-mean reference, and the remaining blocks lie below and above that
reference by reciprocal factors. Values of
$\widetilde\lambda_{ck}$ therefore describe a block's relative strength
within assessor cluster $c$.

The sampler first draws unconstrained Gamma strengths and then records the
normalized row together with its positive geometric-mean scale. For fixed
$K$, $(\widetilde\Lambda,\mathbf s)$ is a one-to-one representation of
$\Lambda$, and the retained scale reconstructs the state used in the next
latent-time update. Sampling directly in the constrained coordinates would
couple the Gamma full conditionals. The displayed normalized strengths inherit
their prior from the unconstrained Gamma specification.

\subsection*{Why the row scale is retained}

Let $\ell$ belong to cluster $c$ and write the availability-set rate at stage
$t$ as
\[
  R_{\ell t}
  =
  \sum_{j\in A_t^{(\ell)}}\lambda_{c,x_j}
  =s_c\widetilde R_{\ell t},
  \qquad
  \widetilde R_{\ell t}
  =
  \sum_{j\in A_t^{(\ell)}}\widetilde\lambda_{c,x_j}.
\]
The exact latent-time update is consequently
\[
  Z_{\ell t}\mid\widetilde\Lambda,\mathbf s,\mathbf w,\mathbf x,
  \boldsymbol\rho
  \sim \operatorname{Exp}(s_c\widetilde R_{\ell t}).
\]
Omitting $s_c$ changes this rate to $\widetilde R_{\ell t}$ and changes the
latent-time means, exposure statistics, and subsequent Gamma updates. The
resulting transition targets a different augmented distribution. A scale-free
formulation is possible after transforming the latent times, although its
normalized strength rows carry a constrained, dependent prior. Retaining
$\mathbf s$ keeps the independent Gamma updates available and leaves
$\widetilde\Lambda$ available for posterior interpretation. \citet{caronEfficientBayesianInference2012}
also consider a separate parameter-expanded scale refresh. The reported sampler
retains the scale from the unconstrained Gamma draw.

Label alignment presents a separate issue. We obtain VI partitions as
label-invariant targets and apply ECR relabeling to draws with their exact
dimensions before calculating entrywise summaries. The resulting heatmap
probabilities describe whether a VI-aligned item block lies above or below the
geometric-mean reference in its assessor cluster.

\subsection{Exact-frame ECR summary of the strength matrix}

Let $\widehat{\mathbf w}_{\mathrm{VI}}$ and
$\widehat{\mathbf x}_{\mathrm{VI}}$ denote the VI estimates, with
$\widehat C_{\mathrm{VI}}=|\widehat{\mathbf w}_{\mathrm{VI}}|$ and
$\widehat K_{\mathrm{VI}}=|\widehat{\mathbf x}_{\mathrm{VI}}|$. We form the
exact-frame index set
\[
  \mathcal I_{\mathrm{VI}}
  =
  \{s:(C^{(s)},K^{(s)})
       =(\widehat C_{\mathrm{VI}},\widehat K_{\mathrm{VI}})\}.
\]
All matrices in this set have the dimension of the reported VI partitions.

For each $s\in\mathcal I_{\mathrm{VI}}$, ECR relabeling is applied separately
to the assessor and item allocations, using
$\widehat{\mathbf w}_{\mathrm{VI}}$ and
$\widehat{\mathbf x}_{\mathrm{VI}}$ as pivots
\citep{papastamoulis_artificial_2010,papastamoulis_labelswitching_2016}.
Let $\sigma_s$ and $\tau_s$ be the resulting permutations of the assessor
clusters and item blocks. The aligned normalized strength matrix is
\begin{equation}\label{eq:supp-lambda-aligned}
  \widetilde\lambda^{(s)}_{ck,\mathrm{align}}
  =
  \widetilde\lambda^{(s)}_{\sigma_s(c),\tau_s(k)},
  \qquad s\in\mathcal I_{\mathrm{VI}}.
\end{equation}
The reported point estimate is the entrywise Monte Carlo mean
\begin{equation}\label{eq:supp-lambda-ecr-mean}
  \widehat\lambda_{ck}
  =
  \frac{1}{|\mathcal I_{\mathrm{VI}}|}
  \sum_{s\in\mathcal I_{\mathrm{VI}}}
  \widetilde\lambda^{(s)}_{ck,\mathrm{align}}.
\end{equation}
Thus every averaged entry refers to the same pair of VI-aligned assessor
clusters and item blocks. For the TCGA fit, $|\mathcal I_{\mathrm{VI}}|=3{,}513$ of the 8,000
retained draws have the exact $19\times259$ frame.

\subsection{Posterior map relative to the unit reference level}
\label{subsec:supp-lambda-state-map}

The heatmap uses the same reference level as the row normalization. Because
each aligned assessor-cluster row has geometric mean one, the posterior
probability displayed for cell $(c,k)$ is
\[
  p_{ck}
  =
  \frac{1}{|\mathcal I_{\mathrm{VI}}|}
  \sum_{s\in\mathcal I_{\mathrm{VI}}}
  \ind\!\left\{
    \widetilde\lambda^{(s)}_{ck,\mathrm{align}} > 1
  \right\}.
\]
The cell is colored blue when $p_{ck}\leq0.2$, red when
$p_{ck}\geq0.8$, and white otherwise. Consequently, a red (blue) cell
indicates posterior support that the assessor-cluster-specific relative strength is
above (below) its row-wise geometric-mean reference; white represents
posterior ambiguity around one.

For display only, rows (assessor clusters) and columns (item blocks) are
jointly ordered. We discretize
the log of the ECR-aligned posterior mean matrix into five bins between its 5th
and 95th percentiles, double-center the resulting $\widehat C_{\mathrm{VI}}
\times\widehat K_{\mathrm{VI}}$ score matrix, and take its first three
singular-vector coordinates. Separate $k$-means partitions of these row and
column coordinates (six row groups and ten column groups in the TCGA plot)
define contiguous display groups; members are ordered by their first, then
second, coordinate within group. Tile heights and widths are proportional to
the corresponding VI sample-cluster and gene-block sizes, respectively, and
black lines mark the display-group boundaries. This ordering changes neither
$p_{ck}$ nor the three-state classification.

\section{Simulation design and supporting results}\label{sec:supp-simulation}

The central simulation question is whether the model can recover the two
latent partitions when the rankings contain a controlled amount of information,
and how that recovery changes when the model is deliberately misspecified.
The structured recovery and sensitivity study below is the primary evidence
for that question.  It uses a structured factorial design for signal
separation, ranking depth, and partition balance, and then asks whether
recovery remains stable when items within a true block no longer have exactly
the same strength.

The section also includes two shorter checks that establish preconditions for
interpreting the primary study.  We first verify the end-to-end implementation
on data generated from the correctly specified model with known partitions;
because several features change together in that check, it does not estimate
the effect of any one feature.  We then hold one generated data set fixed and
run dispersed chains, so that
between-chain mixing can be assessed without confusing it with variation
between simulated data sets.  Keeping these purposes separate prevents a good
recovery result across data sets from being mistaken for evidence of MCMC
mixing.  Following \citet{morris_white_crowther_2019}, chains are pooled only
within a data set and are not treated as independent simulation replicates.
The plots retain data-set-level results and emphasize the comparisons targeted
by each experiment, in line with \citet{gasparini_morris_crowther_2021}.

\subsection{End-to-end recovery check under correct specification}

Before studying how recovery changes with signal, depth, and balance, we check
that the complete implementation behaves as intended when its assumptions are
true.  This check exercises the data generator, likelihood, sampler, and
posterior summaries together: the generating partitions and occupied counts
are known, and the fitted model is the one that generated the rankings.  We
use three representative configurations, with five replicates each:
\[
  (L,n,m)\in\{(60,80,25),\ (120,160,40),\ (200,260,60)\}.
\]
For each replicate, Algorithm~\ref{alg:supp-sim-data-generation} generates the
assessor and item partitions, assigns a well-separated item-block-strength pattern,
and samples top-$m$ rankings from the induced Plackett--Luce probabilities.
We summarize recovery with the assessor and item adjusted Rand indices (ARI)
and with the estimated occupied counts $C$ and $K$.  Since dimensions,
occupied counts, ranking depth, and separation change together across the
three configurations, this is an implementation check rather than a comparison
of their individual effects.

\begin{algorithm}[htbp]
\caption{Data generator for one end-to-end recovery check scenario}
\label{alg:supp-sim-data-generation}
\begin{algorithmic}[1]
\Require $(L,n,m,C^\star,K^\star)$, partition hyperparameters
$(\gamma_w,\gamma_x)$, and log-scale separation $\delta$
\Ensure Rankings $\rho$, true partitions $(\mathbf w^\star,\mathbf x^\star)$,
and generating strengths $\Lambda^\star$
\State Sample $\mathbf w^\star$ from a Gnedin prior on $L$ assessors,
conditioned on $C^\star$ occupied assessor clusters
\State Sample $\mathbf x^\star$ from a Gnedin prior on $n$ items,
conditioned on $K^\star$ occupied item blocks
\State Construct a deterministic $C^\star\times K^\star$ log-strength
template and map it to $\Lambda^\star$
\For{$\ell=1,\ldots,L$}
  \State Set assessor-specific item strengths to
  $\lambda_{w^\star_\ell,x^\star_j}$ for $j=1,\ldots,n$
  \State Draw a top-$m$ ranking $\rho^{(\ell)}$ by Plackett--Luce sampling
\EndFor
\State \Return $(\rho,\mathbf w^\star,\mathbf x^\star,\Lambda^\star)$
\end{algorithmic}
\end{algorithm}

\begin{table}[htbp]
  \centering
  \caption{End-to-end recovery check under correct model specification. Each configuration has five replicates. Data are generated by conditioning the Gnedin partitions to attain $(C^*, K^*)$ and then assigning a well-separated $\Lambda^*$ template, and are fitted with the corresponding PL-LBM. $\hat C_{\mathrm{VI}}$ and $\hat K_{\mathrm{VI}}$ are median [minimum, maximum] across replicates; ARI is mean $\pm$ standard deviation. The table checks that the full workflow recovers known structure; it is not a comparison of isolated scale effects.}
  \label{tab:joint-sim-generative}
  \scriptsize
  \resizebox{\textwidth}{!}{%
  \begin{tabular}{lcccccc}
    \toprule
    Scenario & $L/n/m$ & $C^*/K^*$ & $\hat C_{\mathrm{VI}}$ & $\hat K_{\mathrm{VI}}$ & ARI$_w$ & ARI$_x$ \\
    \midrule
    Small & 60 / 80 / 25 & $2 / 3$ & $2\,[2,2]$ & $3\,[3,3]$ & $0.97 \pm 0.06$ & $0.98 \pm 0.02$ \\
    Medium & 120 / 160 / 40 & $3 / 4$ & $3\,[2,3]$ & $4\,[4,4]$ & $0.84 \pm 0.32$ & $1.00 \pm 0.01$ \\
    Large & 200 / 260 / 60 & $4 / 5$ & $3\,[3,3]$ & $5\,[5,5]$ & $0.91 \pm 0.09$ & $0.99 \pm 0.01$ \\
    \bottomrule
  \end{tabular}
  }
\end{table}

\begin{figure}[htbp]
  \centering
  \includegraphics[width=\textwidth]{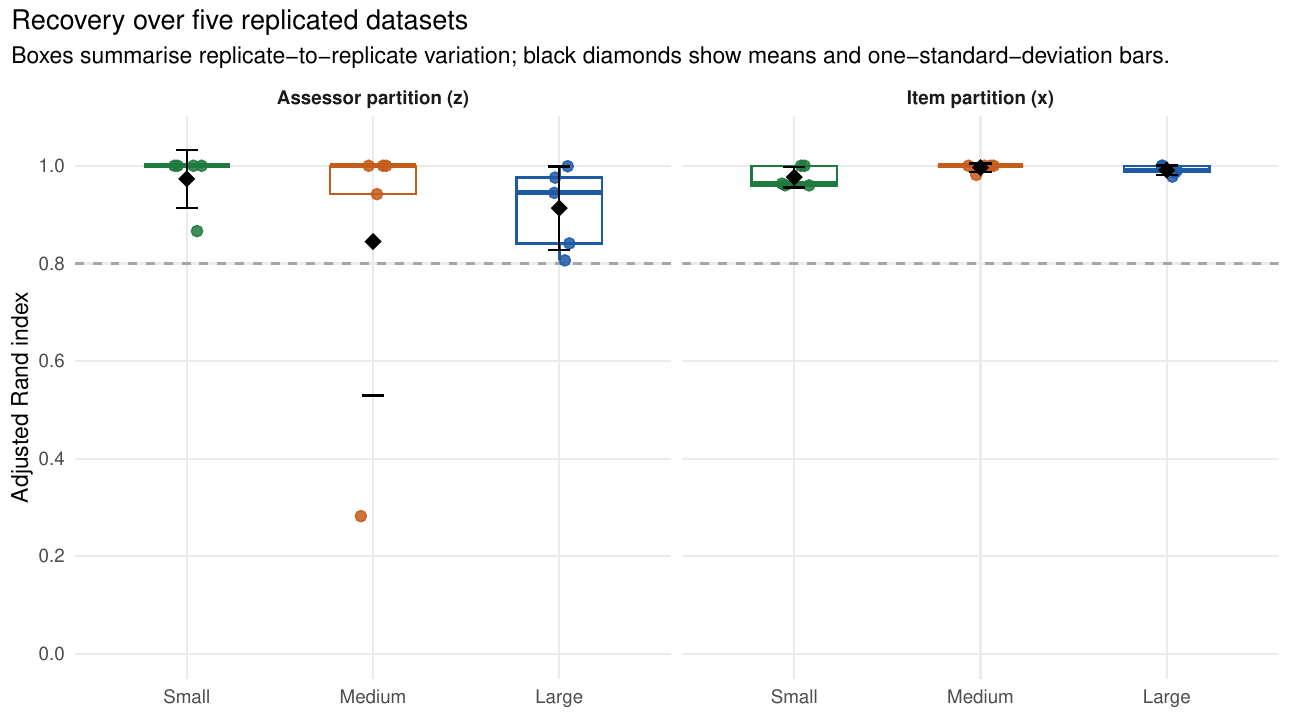}
  \caption{Partition recovery in the end-to-end check under correct model
  specification.
  Adjusted Rand index of the VI-minimizing partition for the
  assessor partition $\mathbf w$ and the item partition $\mathbf x$ across the
  three controlled generative scenarios and five replicates. Boxes summarize
  replicate-to-replicate variation; black diamonds mark the mean with
  one-standard-deviation bars. The dashed line indicates $\mathrm{ARI}=0.8$.}
  \label{fig:supp-joint-sim-recovery-uncertainty}
\end{figure}

The item partition is recovered almost completely in all three configurations,
whereas the assessor partition is more variable in the medium configuration
(Table~\ref{tab:joint-sim-generative} and
Figure~\ref{fig:supp-joint-sim-recovery-uncertainty}).  This illustrates
successful end-to-end recovery in these selected regimes, but it does not
attribute the difference to scale, depth, or separation; the primary study
below makes those contrasts explicit.

\subsection{Three-chain convergence study with informed split--merge moves}
\label{subsec:supp-convergence-simulation}

Recovery across independent data sets does not by itself show that the MCMC
sampler escapes different initialization-specific regions.  We therefore
freeze the first replicate of the medium end-to-end check scenario and ask whether
chains started from widely different numbers of occupied groups reach the same
posterior region.  The frozen data set has $L=120$, $n=160$, $m=40$,
$(C^\star,K^\star)=(3,4)$, conditioned Gnedin partitions with
$\gamma_w=\gamma_x=0.55$, and a row-normalized fixed-strength template with
log-gap $1.10$.  Holding the data fixed isolates between-chain mixing from
between-replicate variation.

We ran three independent chains from dispersed occupied counts
\[
  (C_0,K_0)\in\{(1,1),(8,12),(20,30)\}.
\]
Each chain ran for 10,000 iterations, with the first 3,000 discarded.  At each
iteration it received two random-scan allocation sweeps and two data-informed
split--merge proposals for each partition; the restricted split allocation
used one exact pass.  The forward and reverse proposal probabilities include
the informed anchor probabilities for the relevant partition.  The same
implementation is used for the joint, assessor-only, and item-only samplers.
For identifiability, each strength row is centered to have zero log-mean and
its row scale is retained as an auxiliary state variable.

\begin{figure}[htbp]
  \centering
  \includegraphics[width=\textwidth]{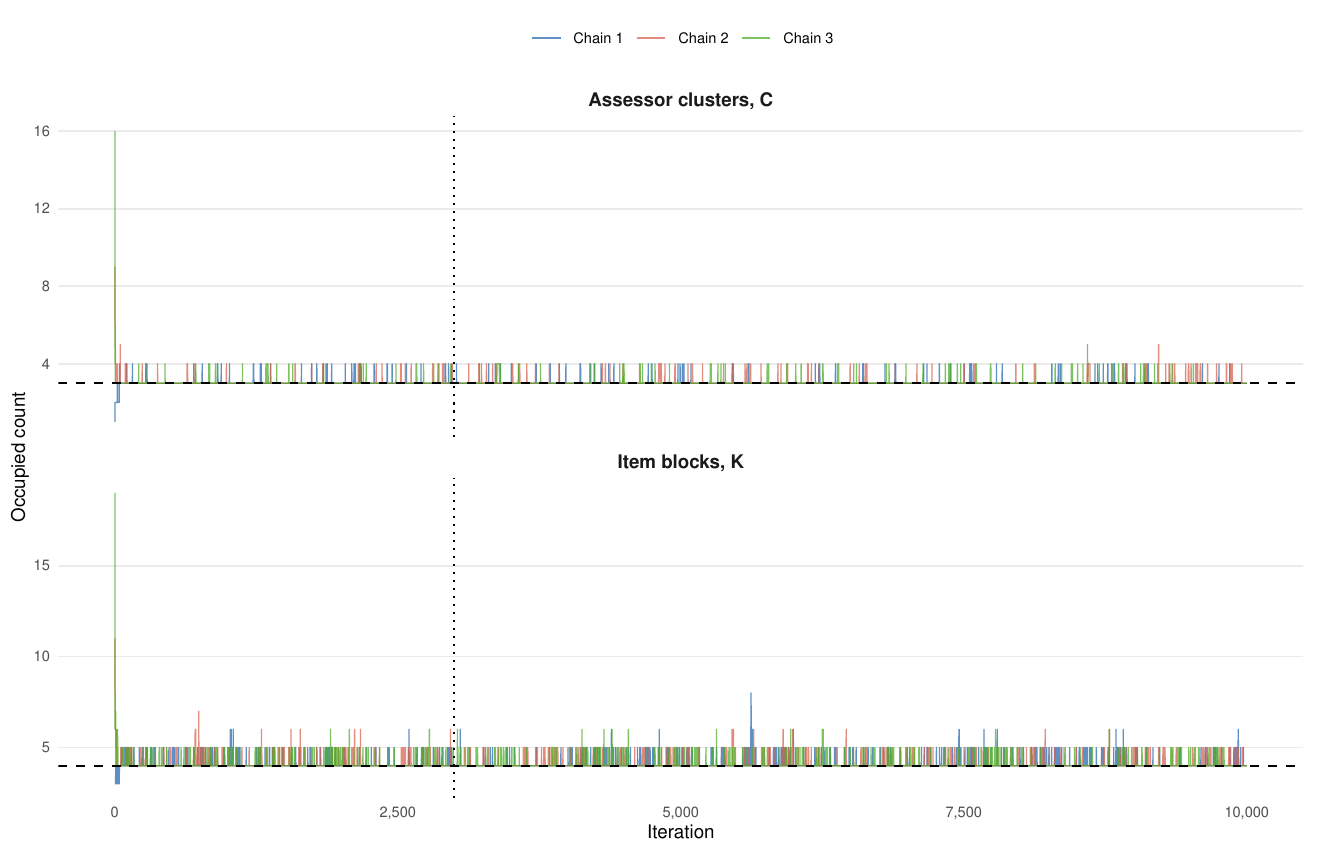}
  \caption{Three-chain traces of the occupied assessor count $C$ and item-block
  count $K$ for the frozen medium simulation replicate.  The vertical dotted
  line marks the end of burn-in and the horizontal dashed line is the true
  generating count.  The chains begin from strongly different occupied counts
  but enter the same region well before the retained period.}
  \label{fig:supp-convergence-simulation-trace}
\end{figure}

\begin{table}[htbp]
  \centering
  \caption{Three-chain convergence diagnostics for occupied assessor clusters $C$ and item blocks $K$ on the frozen medium simulation replicate. Chain summaries are ordered from underdispersed to strongly overdispersed starts. ESS is rank-normalized bulk effective sample size, and the pooled ESS and $\widehat R$ use all 21,000 retained draws.}
  \label{tab:supp-convergence-simulation}
  \begin{tabular}{lccccc}
    \toprule
    Quantity & Truth & Chain means & Chain bulk ESS & Pooled bulk ESS & $\widehat R$ \\
    \midrule
    $C$ & 3 & 3.01 / 3.02 / 3.02 & 2,422 / 1,027 / 1,045 & 3,640 & 1.0004 \\
    $K$ & 4 & 4.05 / 4.04 / 4.04 & 2,280 / 4,382 / 2,877 & 8,249 & 1.0001 \\
    \bottomrule
  \end{tabular}
\end{table}

All three chains enter the same occupied-count region well before the retained
period (Figure~\ref{fig:supp-convergence-simulation-trace}).
Rank-normalized $\widehat R$ is 1.0004 for $C$ and 1.0001 for $K$; pooled bulk
ESS is 3,640 and 8,249, respectively, from 21,000 retained draws
(Table~\ref{tab:supp-convergence-simulation}).  The informed split--merge
steps accepted 469 of 60,000 assessor proposals (0.78\%) and 1,795 of 60,000
item proposals (2.99\%).  In this well-identified setting the global moves
are selective, but their accepted moves supplement the local allocation
updates.  This is a fixed-data check rather than a general convergence proof;
it supports satisfactory occupied-count mixing for this controlled replicate.
The corresponding TCGA diagnostics are reported separately in
Section~\ref{sec:supp-tcga-results}.

\subsection{Structured recovery and sensitivity study}
\label{subsec:supp-structured-simulation}

The primary study asks how much separation and ranking depth are needed for
reliable recovery, and whether a small latent group is harder to recover than
the overall partition.  It therefore holds the problem size fixed at
$L=120$ assessors, $n=160$ items, $C^\star=3$ assessor clusters, and
$K^\star=4$ item blocks, while varying three features that govern available
information: the adjacent log-strength gap $\delta$, the observed ranking
depth $m$, and partition balance.  A larger $\delta$ makes neighboring item blocks
more distinct; a larger $m$ reveals more of each assessor's ranking.  We use
$\delta=0.50,0.70,1.10$ and $m=16,64$.  The balanced and imbalanced conditions
respectively include equal-sized groups and a smallest assessor cluster of 15
and item block of 16.  Thus, the balance contrast reveals whether good global
recovery masks difficulty with a scientifically small subgroup.  The
low-signal, shallow-ranking combinations define the intended low-information
boundary.  A pilot at $\delta=0.35$ was still more severely prior-dominated,
so it motivated $\delta=0.50$ as the lower boundary of the primary study.
Using explicit generative contrasts to control difficulty follows recent
latent-block simulation studies \citep{frisch_leger_grandvalet_2021,riverain_fossier_nadif_2022}.

Within each generated data set, labels are randomly permuted subject to the
specified group sizes: $(40,40,40)$ and $(40,40,40,40)$ in the balanced
condition, and $(60,45,15)$ and $(64,48,32,16)$ in the unbalanced condition.
Balance is therefore controlled by construction rather than confounded with
chance variation in group sizes.  The fitting model uses Gnedin priors; only
the end-to-end check above uses the prior-conditioned partition generator.

Each correctly specified condition has eight independently generated data
sets, each fitted from two dispersed initializations.  A chain has 4,000
iterations with 1,200 burn-in iterations; each iteration makes five assessor
and two item split--merge proposals using data-informed anchors and exact
one-pass restricted allocation.  Before producing a data-set-level summary,
we screen the two chains using rank-normalized $\widehat R$: a data set with
$\widehat R>1.01$ is reported as not meeting the common-budget screen, rather
than having its chains pooled.  Qualifying chains are then pooled to compute
posterior similarity matrices, point partitions, and occupied-count
probabilities.  This makes the distinction between data-set recovery and the
finite-budget MCMC check visible.  In weak-signal conditions, posterior mass
may genuinely be spread across several partitions and occupied counts; that
uncertainty is part of the information-boundary result, not a reason to force
a single answer.

We use complementary recovery summaries because no single number answers all
parts of the question: assessor and item ARI measure recovery on each side,
CARI measures recovery of their joint assessor--item structure, and best-matched F1
focuses on the smallest true group.  We also report exact recovery and
posterior probability of the true occupied counts.  The rank-normalized
diagnostic and bulk ESS follow \citet{vehtari_gelman_simpson_2021}.  The goal
is to characterize recovery and Monte Carlo variation across generated data
sets, not nominal interval coverage.

The linked sensitivity experiment asks whether the primary conclusions depend
on the exact common-strength assumption within an item block.  It retains the
moderate-signal, deep-ranking setting ($\delta=0.70$, $m=64$), then adds
centered item-level log-strength deviations with standard deviation 0.2 or 0.4
within every true item block.  Thus the item-block mean is preserved while individual
items deviate from it.  The rankings remain Plackett--Luce, but the fitted
common-strength item-block model is now deliberately misspecified.  We use eight
independently generated data sets per setting and balance condition.  The
homogeneous and heterogeneous data sets share settings but not random-number
draws, so the resulting curves describe sensitivity across conditions rather
than paired effects.

\begin{table}[htbp]
  \centering
  \caption{Primary recovery study under the correctly specified PL-LBM. Rows vary signal separation $\delta$, ranking depth $m$, and partition balance while holding problem size fixed. Performance is summarized only for independently generated data sets whose two chains passed the common-budget convergence screen; ``Converged'' gives that number out of eight. F1 is computed for the smallest true assessor cluster or item block after matching estimated and true groups. The final two columns average the posterior probability assigned to the generating occupied counts.}
  \label{tab:appendix-information-recovery}
  \scriptsize
  \resizebox{\textwidth}{!}{%
  \begin{tabular}{llllrrrrrr}
    \toprule
    $\delta$ & $m$ & Partition balance & Converged & ARI$_w$ & ARI$_x$ & F1$_{w,\min}$ & F1$_{x,\min}$ & $\overline{P(C=C^\star\mid\rho)}$ & $\overline{P(K=K^\star\mid\rho)}$ \\
    \midrule
    $\delta=0.50$ & $m=16$ & Balanced & 1/8 & 0.52 & 0.38 & 0.80 & 0.58 & 0.04 & 0.21 \\
    $\delta=0.70$ & $m=16$ & Balanced & 6/8 & 0.95 & 0.76 & 0.99 & 0.84 & 0.83 & 0.38 \\
    $\delta=1.10$ & $m=16$ & Balanced & 8/8 & 1.00 & 0.91 & 1.00 & 0.93 & 0.97 & 0.80 \\
    $\delta=0.50$ & $m=64$ & Balanced & 5/8 & 1.00 & 0.94 & 1.00 & 0.96 & 0.86 & 0.42 \\
    $\delta=0.70$ & $m=64$ & Balanced & 7/8 & 1.00 & 0.99 & 1.00 & 0.99 & 0.97 & 0.86 \\
    $\delta=1.10$ & $m=64$ & Balanced & 8/8 & 1.00 & 1.00 & 1.00 & 1.00 & 1.00 & 0.99 \\
    $\delta=0.50$ & $m=16$ & Unbalanced & 0/8 & -- & -- & -- & -- & -- & -- \\
    $\delta=0.70$ & $m=16$ & Unbalanced & 4/8 & 0.80 & 0.51 & 0.90 & 0.96 & 0.68 & 0.30 \\
    $\delta=1.10$ & $m=16$ & Unbalanced & 8/8 & 0.98 & 0.84 & 0.99 & 1.00 & 0.94 & 0.80 \\
    $\delta=0.50$ & $m=64$ & Unbalanced & 4/8 & 0.94 & 0.86 & 1.00 & 1.00 & 0.81 & 0.49 \\
    $\delta=0.70$ & $m=64$ & Unbalanced & 5/8 & 1.00 & 0.97 & 1.00 & 1.00 & 0.97 & 0.85 \\
    $\delta=1.10$ & $m=64$ & Unbalanced & 7/8 & 1.00 & 1.00 & 1.00 & 1.00 & 1.00 & 0.98 \\
    \bottomrule
  \end{tabular}
  }
\end{table}
\begin{table}[htbp]
  \centering
  \caption{Sensitivity to deliberate within-block misspecification at $\delta=0.70$ and $m=64$. The generating configurations add centered item-level log-strength deviations while preserving geometric block means, whereas the fitted model still assumes a common strength within each block. Comparisons are across matched settings rather than paired data sets; ``Converged'' gives the number passing the common-budget screen.}
  \label{tab:appendix-heterogeneity-robustness}
  \scriptsize
  \resizebox{\textwidth}{!}{%
  \begin{tabular}{lllrrrrr}
    \toprule
    Partition balance & Generating strengths & Converged & ARI$_w$ & ARI$_x$ & $\hat C_{\mathrm{VI}}=C^\star$ & $\hat K_{\mathrm{VI}}=K^\star$ & $\overline{P(K=K^\star\mid\rho)}$ \\
    \midrule
    Balanced & Common strength within blocks & 7/8 & 1.00 & 0.99 & 1.00 & 1.00 & 0.86 \\
    Balanced & Item-specific variation (SD 0.2) & 6/8 & 1.00 & 0.81 & 1.00 & 0.00 & 0.00 \\
    Balanced & Item-specific variation (SD 0.4) & 7/8 & 1.00 & 0.38 & 1.00 & 0.00 & 0.00 \\
    Unbalanced & Common strength within blocks & 5/8 & 1.00 & 0.97 & 1.00 & 1.00 & 0.85 \\
    Unbalanced & Item-specific variation (SD 0.2) & 8/8 & 1.00 & 0.84 & 1.00 & 0.12 & 0.00 \\
    Unbalanced & Item-specific variation (SD 0.4) & 6/8 & 1.00 & 0.31 & 1.00 & 0.00 & 0.00 \\
    \bottomrule
  \end{tabular}
  }
\end{table}
\begin{table}[htbp]
  \centering
  \caption{Sampler diagnostics across all data sets in the structured recovery and sensitivity study under the common 4,000-iteration budget. A data set is flagged when rank-normalized $\widehat R>1.01$, which is why only qualifying pairs are pooled in the recovery summaries. Acceptance is the median split--merge acceptance proportion; chain ARI compares the two chain-specific VI partitions and PSM MAD is their mean absolute posterior-similarity difference.}
  \label{tab:appendix-structured-diagnostics}
  \resizebox{\textwidth}{!}{%
  \begin{tabular}{lrrrrrr}
    \toprule
    Partition & Data sets & $\widehat R>1.01$ & Median bulk ESS & Median acceptance & Median chain ARI & Median PSM MAD \\
    \midrule
    Assessors & 128 & 14 & 2033 & 0.018 & 1.000 & 0.000 \\
    Items & 128 & 34 & 457 & 0.154 & 0.956 & 0.003 \\
    \bottomrule
  \end{tabular}
  }
\end{table}

\begin{figure}[htbp]
  \centering
  \includegraphics[width=\textwidth]{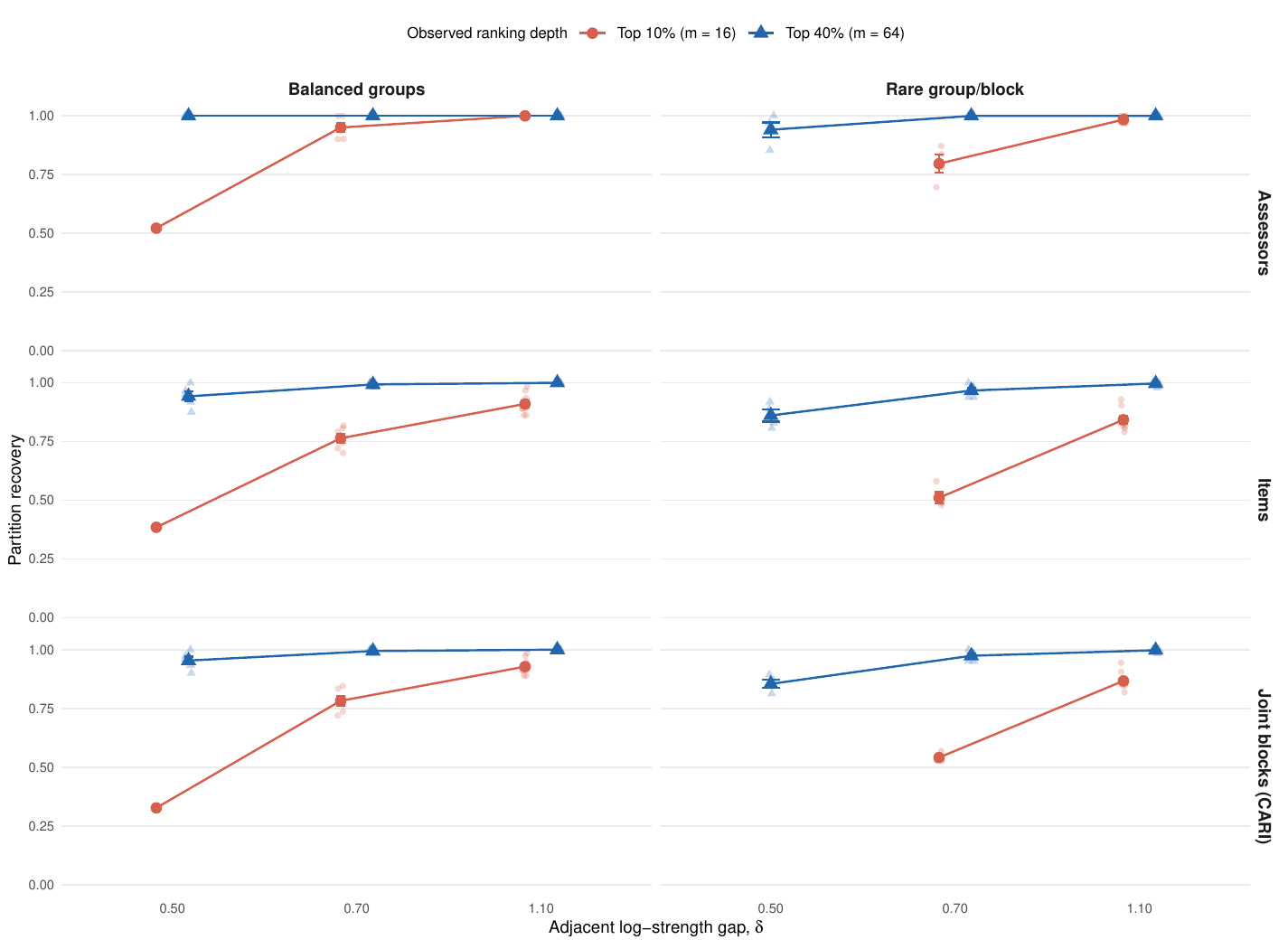}
  \caption{Recovery profiles for the correctly specified structured study.
  Faint points show independently generated data sets passing the convergence
  screen; solid points and bars are means and one Monte Carlo standard error.
  The numbers passing are reported in Table~\ref{tab:appendix-information-recovery}.
  CARI complements the two
  side-specific ARIs with a joint assessor--item measure.}
  \label{fig:supp-structured-recovery}
\end{figure}

\begin{figure}[htbp]
  \centering
  \includegraphics[width=0.9\textwidth]{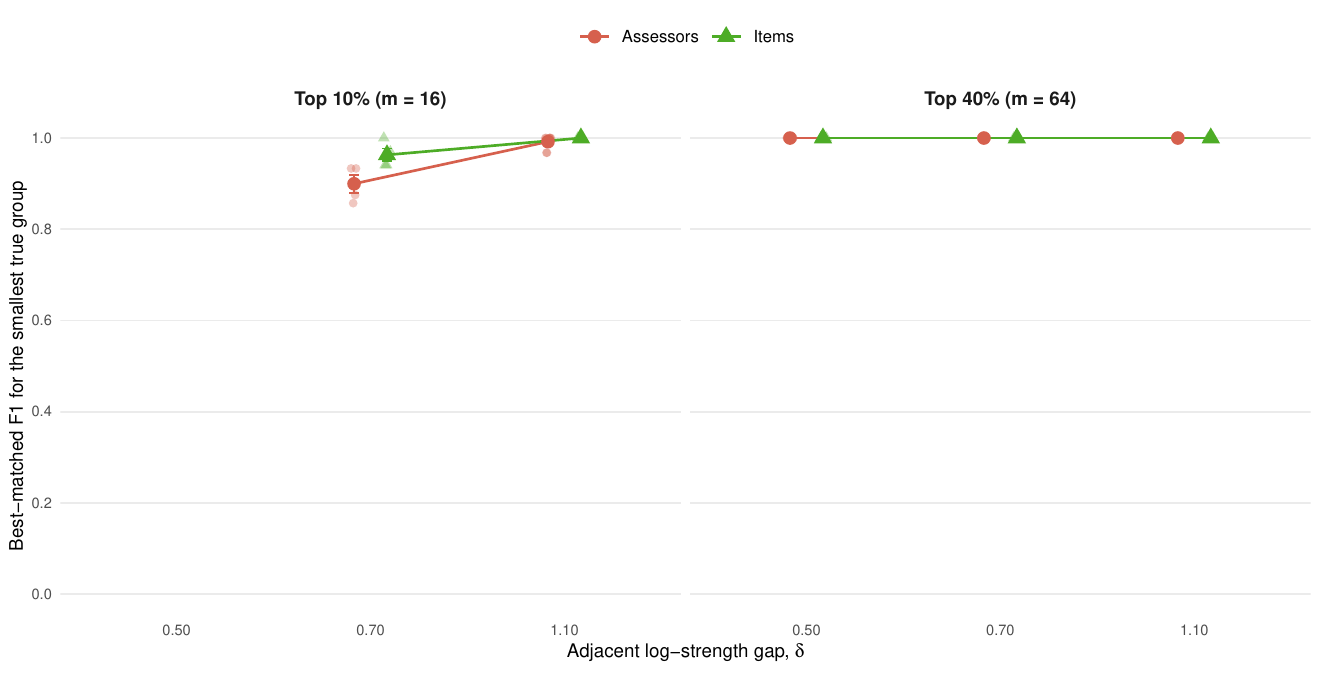}
  \caption{Recovery of the smallest assessor cluster and item block in the
  imbalanced scenarios among data sets passing the convergence screen.
  Best-matched F1 complements global ARI by penalizing both merging a rare
  group with a large group and fragmenting it.}
  \label{fig:supp-structured-rare}
\end{figure}

\paragraph{Recovery under the fitted model.}
Table~\ref{tab:appendix-information-recovery} and
Figure~\ref{fig:supp-structured-recovery} show a clear information gradient:
among data sets passing the common-budget screen, increasing either separation
$\delta$ or ranking depth $m$ improves recovery of both partitions and of the
occupied counts.  The rare-group view in
Figure~\ref{fig:supp-structured-rare} adds an important qualification: global
ARI can remain high while recovery of the smallest group deteriorates as
signal or ranking depth declines.  Overall, 90 of 128 independently generated
data sets passed the common-budget screen.  This includes 31 of 32
strong-signal data sets, but only one of the 16 weak-signal,
shallow-ranking data sets.  At weak signal, increasing depth from $m=16$ to
$m=64$ raised the number passing from 1 to 9 out of 16 across the two balance
conditions.  The absence of a pooled summary for the weak, shallow,
unbalanced regime therefore says that its two finite chains did not establish a
common occupied-count distribution; it does not say that a unique true
partition should have been recoverable from such limited information.

\begin{figure}[htbp]
  \centering
  \includegraphics[width=\textwidth]{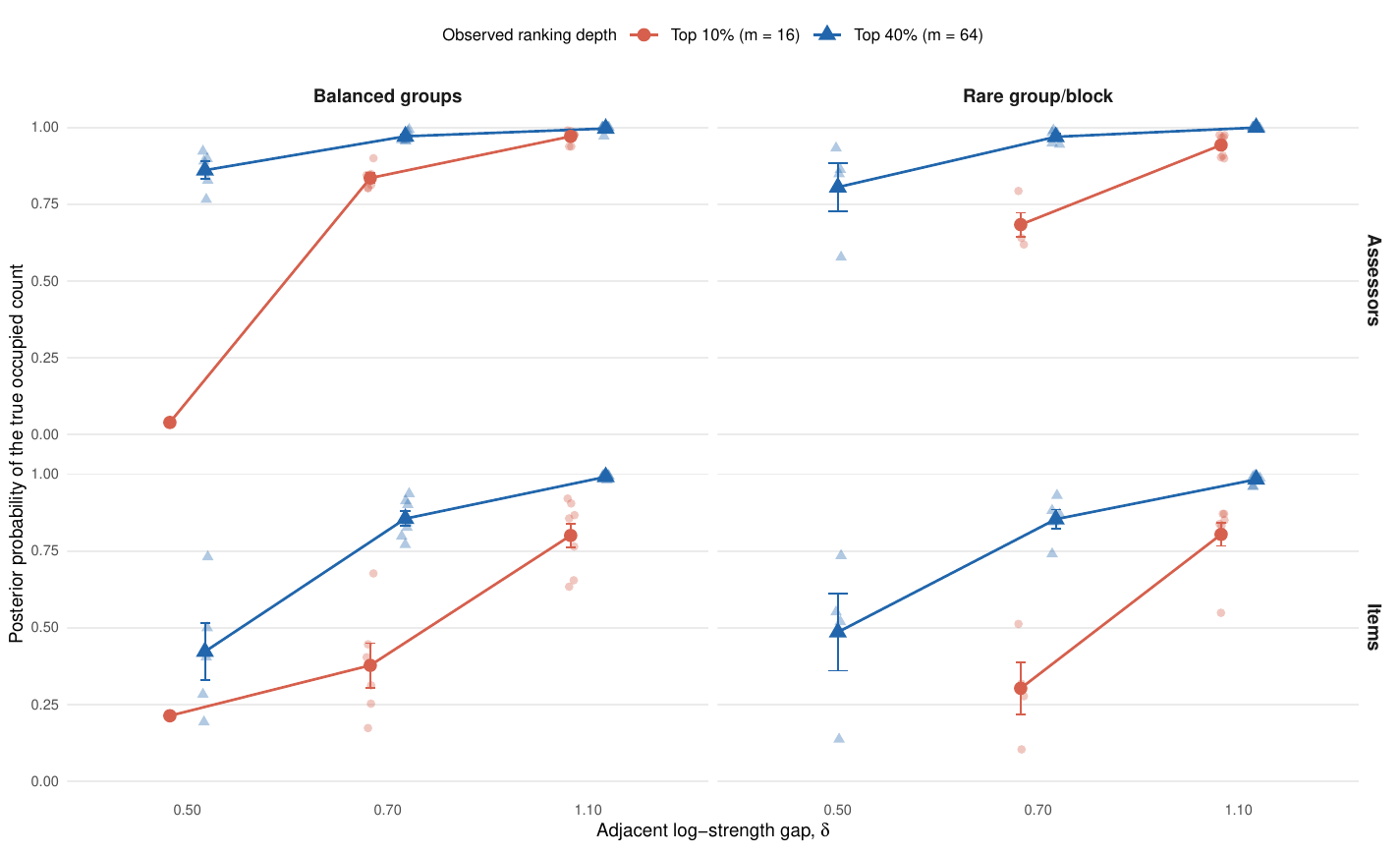}
  \caption{Data-set-level posterior probability assigned to the true occupied
  assessor and item counts for data sets passing the convergence screen.
  Chains are pooled within each qualifying data set before the probabilities
  are calculated.}
  \label{fig:supp-structured-counts}
\end{figure}

For the moderate-signal, deep-ranking conditions, mean item ARI was 0.99 for
balanced and 0.97 for unbalanced partitions among qualifying data sets, and
the corresponding mean posterior probabilities of the true item count were
0.86 and 0.85.  Together with the count profiles in
Figure~\ref{fig:supp-structured-counts}, these results show that a small group
does not prevent recovery once the rankings contain sufficient information,
but it makes the low-information boundary more demanding.

\begin{figure}[htbp]
  \centering
  \includegraphics[width=0.9\textwidth]{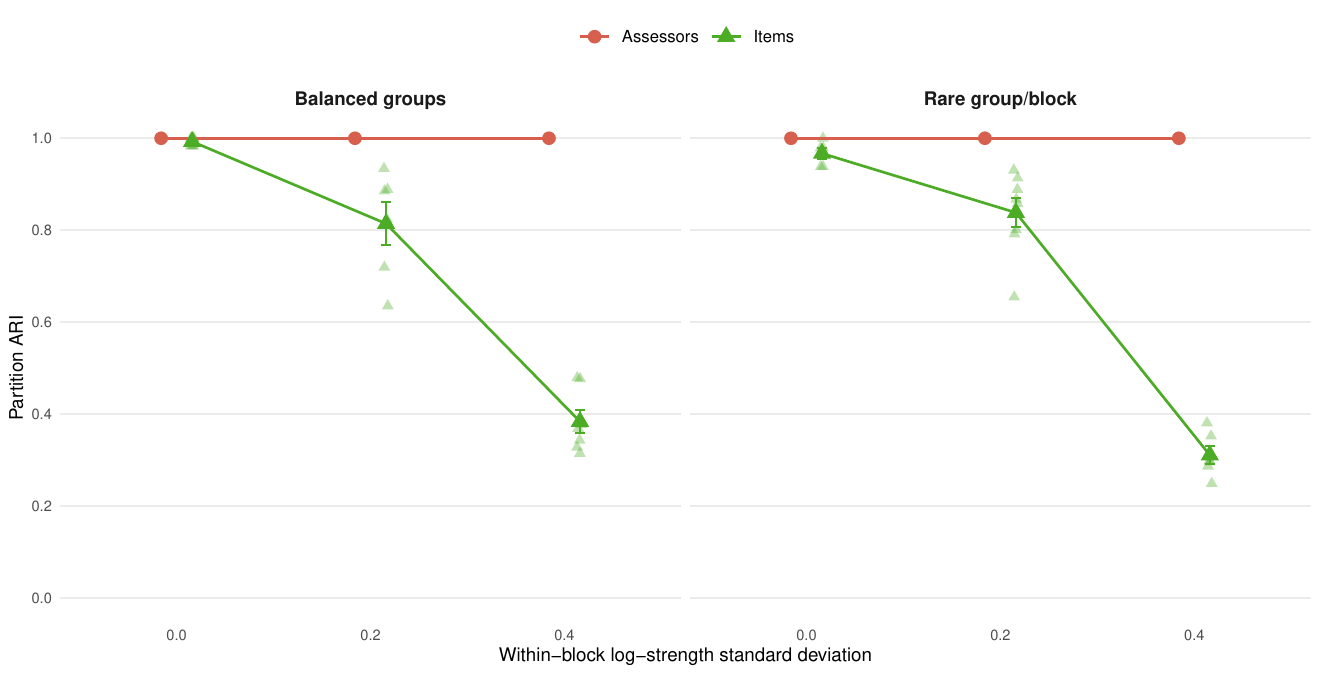}
  \caption{Sensitivity of partition recovery to centered within-block
  item-level variation in matched moderate-signal, deep-ranking settings.
  Replicate points are shown together with means and one Monte Carlo standard
  error.}
  \label{fig:supp-structured-robustness}
\end{figure}

\paragraph{Robustness to within-block variation.}
Table~\ref{tab:appendix-heterogeneity-robustness} and
Figure~\ref{fig:supp-structured-robustness} isolate the consequence of
violating the common-strength assumption.  The assessor partition remains
stable (ARI $=1.00$), whereas item ARI declines from 0.99 to 0.81 and 0.38 in
the balanced setting and from 0.97 to 0.84 and 0.31 in the unbalanced setting
as the perturbation standard deviation increases from 0 to 0.2 and 0.4.
Posterior probability on $K=K^\star$ is near zero for either nonzero
perturbation.  In other words, the fitted model represents continuous
within-block item variation by introducing additional occupied item blocks,
rather than recovering the generating common-block count.

\section{TCGA data and model fitting}\label{sec:supp-tcga-data}

The TCGA analysis uses $L=2{,}617$ tumor samples and a fixed universe of
$n=1{,}247$ selected genes. Each sample contributes a top-$m$ partial ranking
with $m=500$ retained ranks; the remaining genes are treated as unranked items
still present in the choice set. The joint, assessor-only, and item-only models
are run for 10,000 iterations with 2,000 burn-in iterations, retaining 8,000
posterior draws. The joint fit uses $\gamma_w=0.5$, $\gamma_x=0.99$, and
$\lambda_{ck}\sim\Ga(50,\exp\{\psi(50)\})$.

The manuscript reports sample-level conditional WAIC with one top-500 ranking
as the pointwise unit, computed from 400 evenly spaced post-burn-in draws from
each fit. The joint fit has lower conditional WAIC than the
assessor-cluster-only fit, while the item-block-only fit is less favorable.
The joint model uses 4,921 PL strength parameters, compared with 14,964 for the
assessor-only fit, and therefore combines improved conditional repeat-ranking
fit with a shared, lower-dimensional gene-block representation.

\section{Additional TCGA results and diagnostics}\label{sec:supp-tcga-results}

All dimension-specific TCGA displays use one frozen relabeling map obtained by
conditioning the VI summary on the joint posterior-modal dimension. Under this map, the two
large pure KIRC clusters are C11 and C12 in the table, composition plot, and
posterior-similarity outputs. Across retained draws, $C$ lies almost entirely
on 19 and 20, while $K$ spans 257--261 and has mode 259. The exact-frame ECR
summary provides a common $19\times259$ representation of the strength matrix
from 3,513 compatible draws; the occupied-count and split--merge diagnostics
are reported below.

\begin{figure}[htbp]
  \centering
  \includegraphics[width=0.48\textwidth]{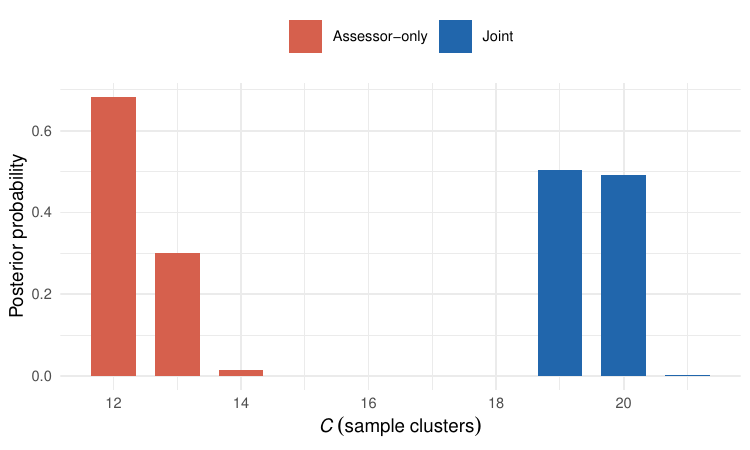}\hfill
  \includegraphics[width=0.48\textwidth]{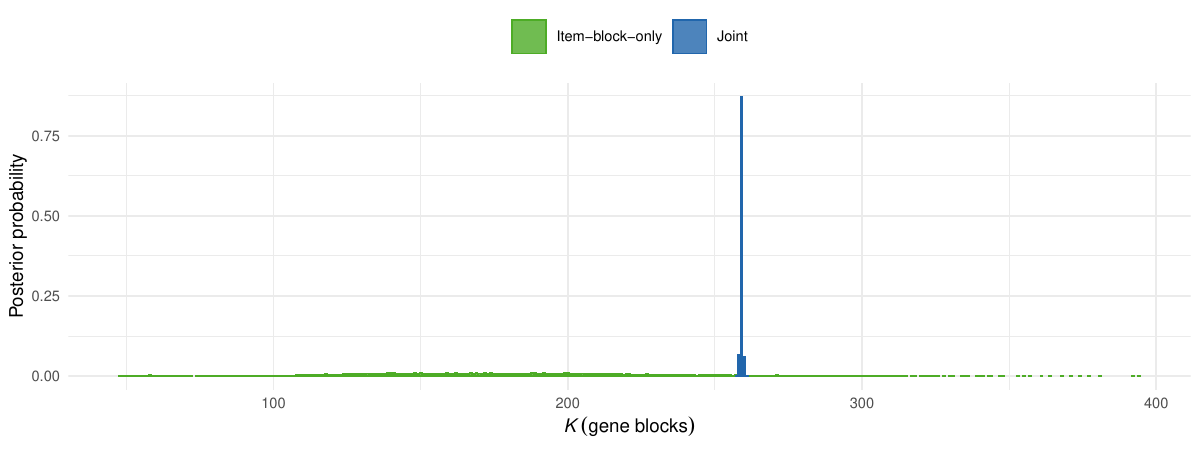}
  \caption{Marginal posterior distributions of the number of sample clusters
  $C$ (left) and gene blocks $K$ (right) for all three fitted models. In the
  joint model, $50.5\%$ of post-burn-in draws have $C=19$ and $49.2\%$ have
  $C=20$; $K$ spans 257--261, with $87.2\%$ at its mode $K=259$ and
  $\mathrm{ESS}(K)\approx245$.}
  \label{fig:supp-tcga-posterior-CK-marginals}
\end{figure}

\begin{figure}[htbp]
  \centering
  \includegraphics[width=0.62\textwidth]{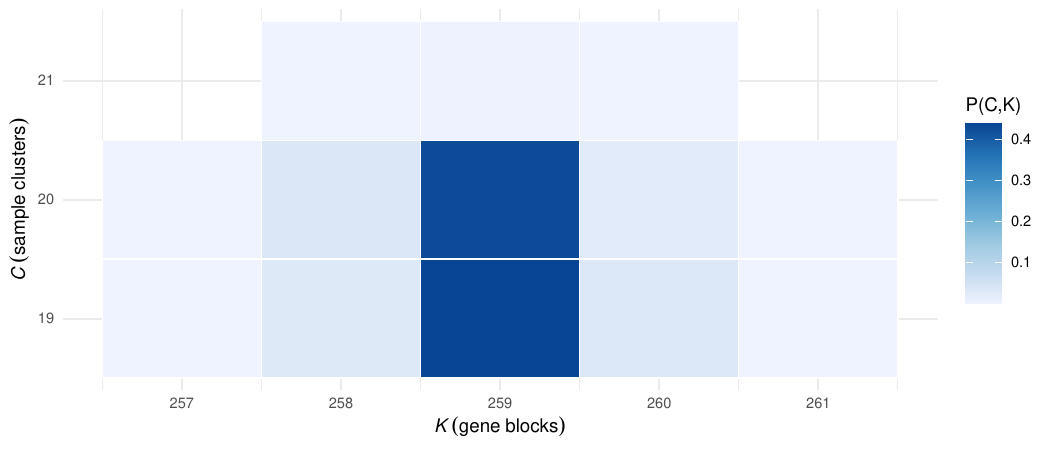}
  \caption{Joint posterior $P(C,K\mid\text{data})$ for the joint PL-LBM,
  shown over the range of $(C,K)$ values visited by the chain. The principal
  posterior mass lies at $C\in\{19,20\}$ and $K=259$.}
  \label{fig:supp-tcga-posterior-CK}
\end{figure}

\begin{figure}[htbp]
  \centering
  \includegraphics[width=0.62\textwidth]{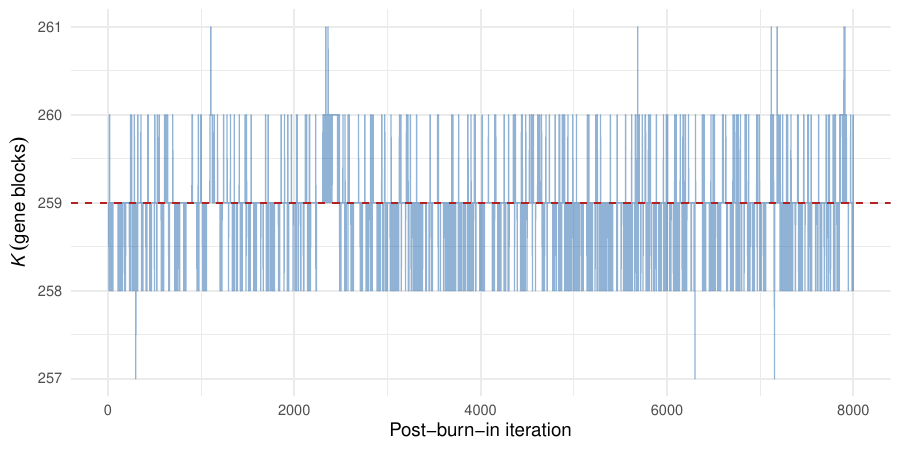}
  \caption{Post-burn-in trace of the occupied gene-block count $K$ for the
  joint PL-LBM. The dashed red line marks the posterior mode.}
  \label{fig:supp-tcga-mixing}
\end{figure}

The stored occupied-count effective sample sizes are approximately 15 for $C$
and 245 for $K$. The joint sampler accepts 31 of 50,000 assessor split--merge
proposals (0.062\%) and 101 of 20,000 item split--merge proposals (0.505\%).

\begin{longtable}{rrllr}
  \caption{Posterior summaries for the $\widehat C_{\mathrm{VI}}=19$ sample clusters under the joint
    Plackett--Luce latent block model on the TCGA pan-cancer data
    ($L=2,617$ samples, $n=1,247$ genes, $m=500$ retained ranks per sample).
    Size is the number of samples assigned to the cluster in the point-estimate partition $\hat{\mathbf{w}}$.
    The first tissue column reports the cancer type with the largest posterior mean share within cluster $c$;
    the second reports the runner-up when its share exceeds 10\%.
    The strongest gene block is the gene block with the largest posterior mean strength in that sample cluster.
    Since genes inside a fitted block are exchangeable, no arbitrary data-order gene representatives are reported.}
  \label{tab:tcga-cluster-summary}\\
  \toprule
  $c$ & Size & Most prevalent tissue & Second most prevalent tissue & Strongest block \\
  \midrule
  \endfirsthead
  \toprule
  $c$ & Size & Most prevalent tissue & Second most prevalent tissue & Strongest block \\
  \midrule
  \endhead
  \midrule \multicolumn{5}{r}{\emph{continued on next page}}\\
  \endfoot
  \bottomrule
  \endlastfoot
  1 & 69 & BLCA (72\%) & UCEC (12\%) & $k_{115}$ (7 genes) \\
  2 & 90 & LUAD (78\%) & LUSC (13\%) & $k_{105}$ (10 genes) \\
  3 & 49 & UCEC (55\%) & BLCA (12\%) & $k_{39}$ (7 genes) \\
  4 & 162 & LUSC (70\%) & HNSC (21\%) & $k_{20}$ (10 genes) \\
  5 & 276 & HNSC (88\%) & --- & $k_{20}$ (10 genes) \\
  6 & 77 & BRCA (94\%) & --- & $k_{39}$ (7 genes) \\
  7 & 69 & LUSC (35\%) & HNSC (26\%) & $k_{208}$ (4 genes) \\
  8 & 212 & BRCA (100\%) & --- & $k_{196}$ (5 genes) \\
  9 & 170 & BRCA (100\%) & --- & $k_{196}$ (5 genes) \\
  10 & 5 & KIRC (60\%) & BRCA (20\%) & $k_{37}$ (14 genes) \\
  11 & 287 & KIRC (100\%) & --- & $k_{37}$ (14 genes) \\
  12 & 112 & KIRC (100\%) & --- & $k_{37}$ (14 genes) \\
  13 & 128 & LUAD (100\%) & --- & $k_{105}$ (10 genes) \\
  14 & 17 & LUAD (100\%) & --- & $k_{105}$ (10 genes) \\
  15 & 99 & GBM (100\%) & --- & $k_{67}$ (4 genes) \\
  16 & 222 & COAD (70\%) & READ (29\%) & $k_{35}$ (4 genes) \\
  17 & 162 & LAML (100\%) & --- & $k_{174}$ (14 genes) \\
  18 & 220 & OV (96\%) & --- & $k_{92}$ (3 genes) \\
  19 & 191 & UCEC (100\%) & --- & $k_{92}$ (3 genes) \\
\end{longtable}

\begin{figure}[htbp]
  \centering
  \includegraphics[width=0.85\textwidth]{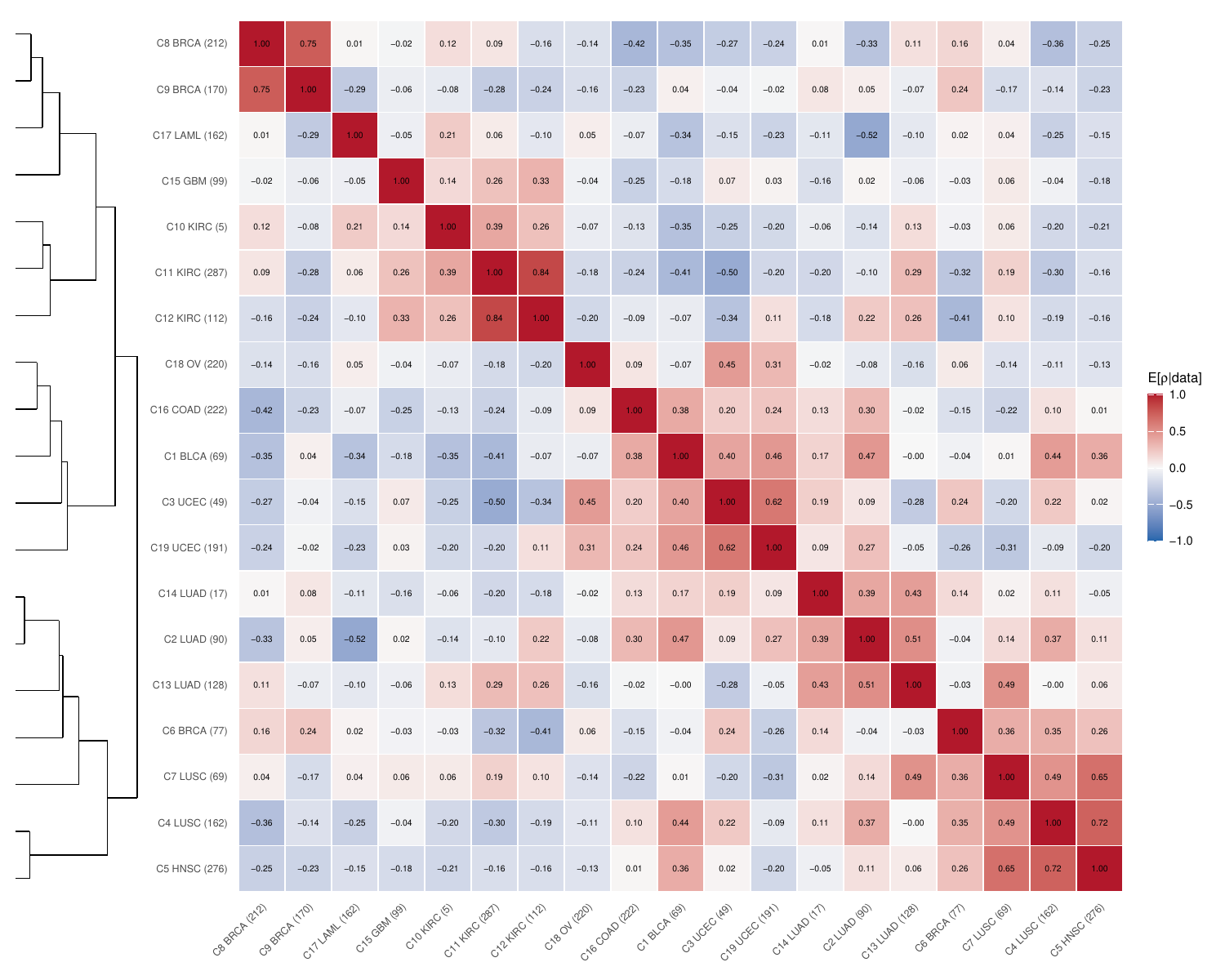}
  \caption{Posterior mean Pearson correlation of log-strength profiles among
  the 19 sample clusters, averaged over the 3,513 exact-frame draws
  label-matched to $\hat{\mathbf w}$ and $\hat{\mathbf x}$. Warm colors mark
  concordant gene-priority patterns; cool colors mark discordant ones. Rows and
  columns are reordered by Ward HC on $1-\widehat\rho$.}
  \label{fig:supp-tcga-cluster-similarity}
\end{figure}

\begin{figure}[htbp]
  \centering
  \includegraphics[width=0.85\textwidth]{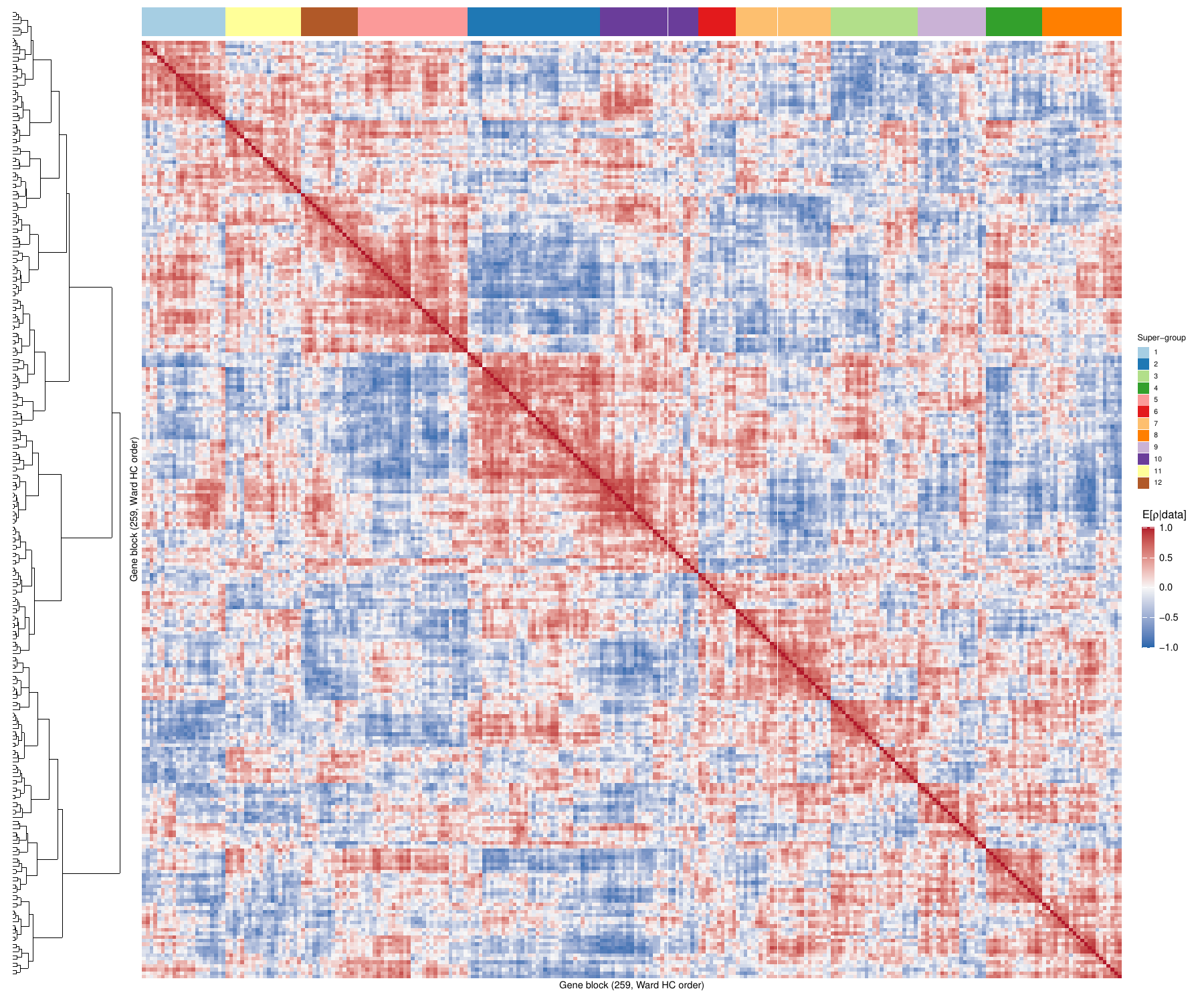}
  \caption{Posterior mean Pearson correlation of log-strength profiles among
  the 259 gene blocks, averaged over the 3,513 exact-frame draws. Rows and
  columns are reordered by Ward HC on $1-\widehat\rho_{kk'}$; the strip above
  the heatmap marks the 12 super-groups from the dendrogram cut.}
  \label{fig:supp-tcga-block-similarity}
\end{figure}

\section{Gene set enrichment analysis}\label{sec:supp-enrichment}

The reported GSEA prioritizes genes by posterior midrank summaries derived from
the joint PL-LBM and applies a one-sided rank-based competitive test to the
full gene ordering. The implementation uses \texttt{limma::geneSetTest}
\citep{Ritchie2015}; no significance threshold or top-list cutoff is applied
before testing. Multiplicity is controlled by global Benjamini--Hochberg
adjusted $q$-values across all sample-cluster and gene-set pairs.

For posterior draw $t$, let $K^{(t)}$ be the number of gene blocks, let
$x_i^{(t)}$ be the block containing gene $i$, and let
\[
  n_k^{(t)} = |\{i:x_i^{(t)}=k\}|
\]
be the draw-level block size. Let $I_c=\{\ell:\widehat z_\ell=c\}$ be the
samples in VI sample cluster $c$, with $N_c=|I_c|$. For every draw, the share of
those samples allocated to sampled assessor cluster $j$ is
\[
  w_{cj}^{(t)}
  =
  \frac{1}{N_c}\sum_{\ell\in I_c}\mathbb{I}\{z_\ell^{(t)}=j\}.
\]
For sampled assessor cluster $j$, the blocks are ordered by decreasing
$\lambda^{(t)}_{j,k}$. For block $k$, define the number of genes in strictly
stronger blocks as
\[
  H_{jk}^{(t)}
  =
  \sum_{h:\lambda^{(t)}_{j,h}>
           \lambda^{(t)}_{j,k}}
  n_h^{(t)} .
\]
The gene-level midrank assigned to every gene in block $k$ is
\[
  \widetilde r_{jk}^{(t)}
  =
  H_{jk}^{(t)}
  +
  \frac{n_k^{(t)}+1}{2}.
\]
The sample-cluster-specific posterior score is then
\[
  s_{ic}
  =
  \frac{1}{T}\sum_{t=1}^T
  \sum_{j=1}^{C^{(t)}} w_{cj}^{(t)}
  \frac{n - \widetilde r^{(t)}_{j,x_i^{(t)}} + 1/2}{n},
\]
where $n=1{,}247$ is the fixed TCGA gene universe size. Larger values indicate
stronger sample-cluster-specific up-weighting. The fixed gene-rank scale makes scores
comparable across posterior draws with different numbers of blocks. All genes
in a draw-level block receive the same midrank; accounting for $n_k^{(t)}$
places that tied block at the center of its occupied rank interval rather than
assigning every member the rank of the interval's leading boundary. The
computation uses all 8,000 posterior draws, not only the 3,513 draws in the
exact $19\times259$ frame. No relabeling is needed: under a permutation of
sample-cluster labels, the weights $w_{cj}^{(t)}$ and the corresponding rows
of $\Lambda^{(t)}$ are permuted together. If a VI sample cluster is split across
sampled assessor clusters in a draw, each part contributes according to its share of
the VI-sample-cluster members. The mean largest such share is 0.988.

For each sample cluster, genes are ordered by $s_{ic}$. Each curated gene set
is tested for a shift toward the top of this complete ordering using
\texttt{geneSetTest} with a one-sided alternative and rank-only statistics.
After intersecting the annotations with the 1,247-gene TCGA universe and
retaining sets with 3--500 mapped genes, the analysis contains 3,576 sets from
seven collections. Across 67,944 tests, 50 have global BH-adjusted $q<0.05$.

\begin{table}[htbp]
  \centering
  \caption{Gene-set libraries retained for the TCGA joint PL-LBM enrichment screen after intersecting with the 1,247-gene universe and applying the set-size filter.}
  \label{tab:tcga-gene-set-libraries}
  \small
  \begin{tabular}{lr}
    \toprule
    Collection & Gene sets \\
    \midrule
    GO\_Biological\_Process\_2025 & 1687 \\
    GO\_Cellular\_Component\_2025 &  175 \\
    GO\_Molecular\_Function\_2025 &  272 \\
    KEGG\_2021\_Human &  230 \\
    KEGG\_REST &  274 \\
    MSigDB\_Hallmark\_2020 &   49 \\
    Reactome\_2022 &  889 \\
    \bottomrule
  \end{tabular}
\end{table}

\begin{table}[htbp]
  \centering
  \caption{Top rank-based GSEA results for sample-cluster-specific posterior midrank scores from the joint PL-LBM, using limma::geneSetTest with global BH correction across all sample-cluster--gene-set pairs.}
  \label{tab:tcga-joint-vitelli-gsea}
  \scriptsize
  \begin{tabularx}{\textwidth}{llXrrr}
    \toprule
    Cluster & Collection & Gene set & Set size & Mean rank & global $q$ \\
    \midrule
    C18 OV & GO BP & Regulation of DNA-templated Transcription (GO:0006355) & 201 & 496.7 & 0.002 \\
    C7 LUSC & GO CC & Collagen-Containing Extracellular Matrix (GO:0062023) & 23 & 238.0 & 0.003 \\
    C3 UCEC & GO CC & Nucleus (GO:0005634) & 422 & 551.9 & 0.003 \\
    C6 BRCA & Hallmark & G2-M Checkpoint & 28 & 284.2 & 0.003 \\
    C17 LAML & GO BP & Chromatin Organization (GO:0006325) & 39 & 338.8 & 0.003 \\
    C17 LAML & GO BP & Chromatin Remodeling (GO:0006338) & 34 & 319.5 & 0.003 \\
    C17 LAML & Reactome & Chromatin Modifying Enzymes R-HSA-3247509 & 40 & 351.5 & 0.006 \\
    C5 HNSC & Reactome & Cell Cycle, Mitotic R-HSA-69278 & 58 & 404.0 & 0.008 \\
    C4 LUSC & Reactome & Cell Cycle R-HSA-1640170 & 77 & 436.2 & 0.008 \\
    C17 LAML & GO BP & Regulation of DNA-templated Transcription (GO:0006355) & 201 & 514.4 & 0.008 \\
    C4 LUSC & Reactome & Cell Cycle, Mitotic R-HSA-69278 & 58 & 406.5 & 0.008 \\
    C18 OV & GO BP & Regulation of Transcription by RNA Polymerase II (GO:0006357) & 204 & 516.5 & 0.008 \\
    \bottomrule
  \end{tabularx}
\end{table}

\begin{figure}[htbp]
  \centering
  \includegraphics[width=\textwidth]{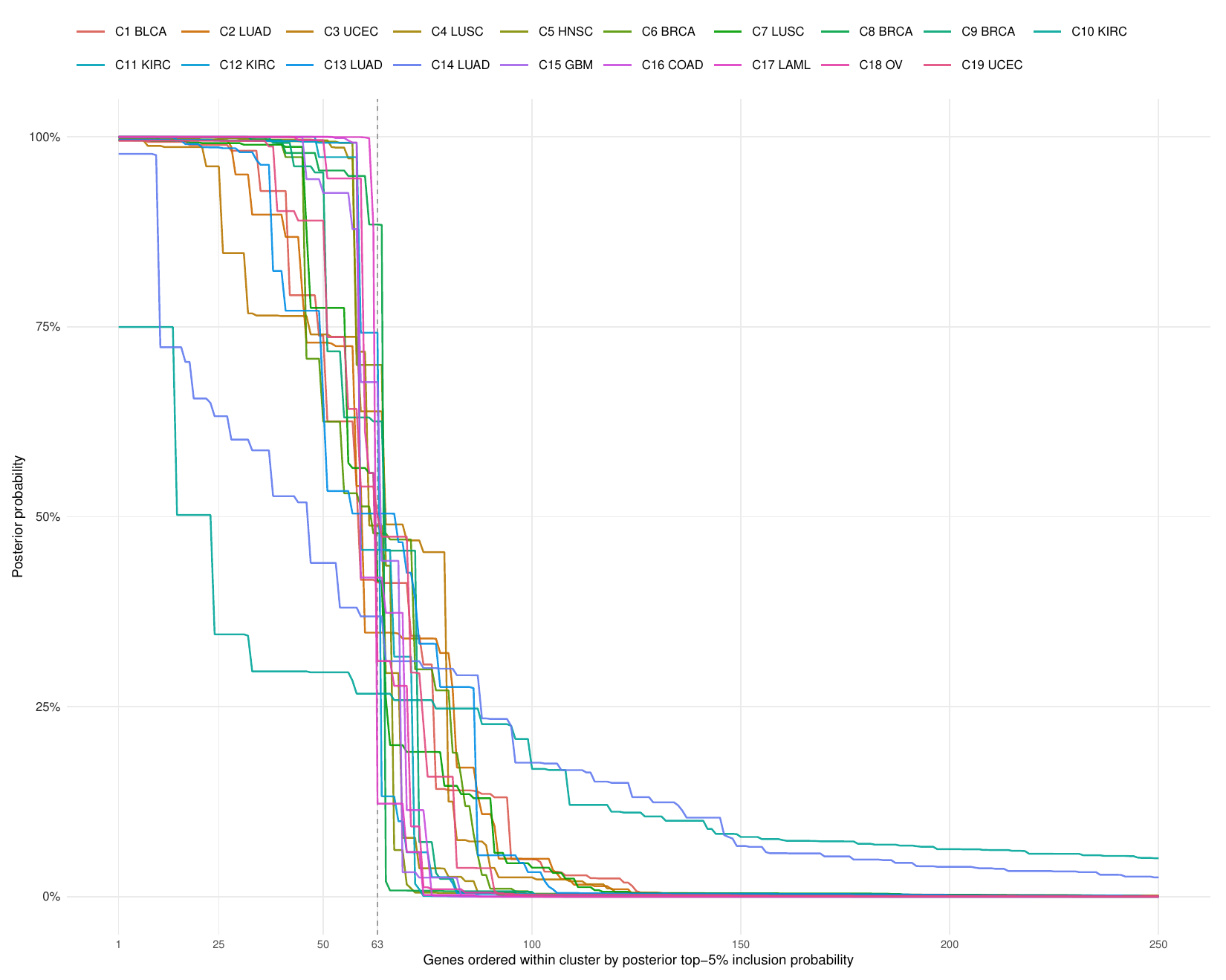}
  \caption{Posterior probability that each gene is in the top 5\% of the
  sample-cluster-specific induced gene ranking. Within each sample cluster, genes are
  sorted from largest to smallest posterior inclusion probability. The dashed
  vertical line marks rank 63, corresponding to 5\% of the 1,247-gene TCGA
  universe. Sharp drops indicate sample clusters where the posterior supports a
  comparatively crisp high-priority gene list.}
  \label{fig:supp-tcga-top5-probability}
\end{figure}

\end{document}